\documentclass[aps,prb,notitlepage,nofootinbib]{revtex4-2}
\usepackage{graphicx}
\usepackage[dvipsnames]{xcolor}
\usepackage{amsfonts}
\usepackage{amssymb}
\usepackage{amsmath}
\usepackage{mathtools}
\usepackage{afterpage}
\usepackage[mathscr]{euscript}
\usepackage{braket}
\usepackage{subfigure}
\usepackage{multirow}
\usepackage{float}
\usepackage{qcircuit}
\usepackage{enumitem}
\usepackage{color}   
\usepackage{hyperref}
\hypersetup{
    colorlinks=true, 
    linktoc=all,     
    linkcolor=blue,  
}

\usepackage[utf8]{inputenc} 
\usepackage{yfonts}
\usepackage{dsfont}
\usepackage{cancel}
\usepackage[scr=boondox]{mathalpha}

\usepackage{leftidx}
\usepackage{epstopdf}
\usepackage{booktabs}
\usepackage{pst-all}
\usepackage[off]{auto-pst-pdf}

\usepackage[bottom]{footmisc}

\newcommand{\act}[2]{{}^{{\bf #1}} #2}
\newcommand{\acts}[3]{{}^{{\bf #1}_{#2}} #3}

\newcommand\restr[2]{{
  \left.\kern-\nulldelimiterspace 
  #1 
  \vphantom{\big|} 
  \right|_{#2} 
  }}

\begin{document}
	\def\U{\mathrm{U}(1)}
	\def\SU{\mathrm{SU}}
	\def\PSU{\mathrm{PSU}}
	\def\SO{\mathrm{SO}}
	\def\Pin{\mathrm{Pin}}
	\def\Spin{\mathrm{Spin}}
	\def\H{\mathcal{H}}
	\def\F{\mathcal{F}}
	\def\E{\mathbb{E}}
	\def\TT{\mathsf{T}}
	\def\A{\mathcal{A}}
	\def\L{\mathcal{L}}
	\def\C{\mathcal{C}}
	
\newcommand{\Z}{\mathbb Z}
\newcommand{\R}{\mathbb R}
    \renewcommand{\L}{\mathcal L}
	\renewcommand{\H}{\mathcal H}
	
\newcommand{\commentrk}[1]{\textcolor{purple}{[RK: #1]}}
	   
\newcommand{\commentdb}[1]{\textcolor{red}{[DB: #1]}}

\newcommand{\commentst}[1]{\textcolor{olive}{[ST: #1]}}
\newcommand{\commentmb}[1]{\textcolor{red}{[MB: #1]}}

\newcommand\curvearrowbotright{\rotatebox[origin=c]{180}{$\curvearrowleft$}}
\newcommand\curvearrowbotleft{\rotatebox[origin=c]{180}{$\curvearrowright$}}

\def\Sq{\mathop{\mathrm{Sq}}\nolimits}
\newcommand{\mathbbm}[1]{\text{\usefont{U}{bbm}{m}{n}#1}}

\def\Sq{\mathop{\mathrm{Sq}}\nolimits}

        \title{(3+1)D path integral state sums on curved U(1) bundles \\ and U(1) anomalies of (2+1)D topological phases
        }
     \author{Ryohei Kobayashi}
     \author{Maissam Barkeshli}
     \affiliation{Condensed Matter Theory Center and Joint Quantum Institute, Department of Physics, University of Maryland, College Park, Maryland 20472 USA}
    
    \begin{abstract}
    Given the algebraic data characterizing any (2+1)D bosonic or fermionic topological order with a global symmetry group $G = \mathrm{U}(1) \rtimes H$,  we construct a (3+1)D topologically invariant path integral in the presence of a curved background $G$ gauge field, as an exact combinatorial state sum. Specifically, the U(1) component of the $G$ gauge field can have a non-trivial second Chern class, extending previous work that was restricted to flat $G$ bundles. Our construction expresses the U(1) gauge field in terms of a Villain formulation on the triangulation, which includes a $1$-form $\mathbb{R}$ gauge field and $2$-form $\Z$ gauge field. We develop a new graphical calculus for anyons interacting with ``Villain symmetry defects", associated with the $1$-form and $2$-form background gauge fields. This graphical calculus is used to define the (3+1)D path integral, which can describe either a bosonic or fermionic symmetry-protected topological (SPT) phase. For example, we can construct the topological path integral on curved U(1) bundles for the (3+1)D fermionic topological insulator in class AII and topological superconductor in class AIII given appropriate (2+1)D fermionic symmetry fractionalization data; these then give invariants of 4-manifolds with Spin$^c$ or Pin$^c$ structures and their generalizations. The (3+1)D path integrals define anomaly indicators for the (2+1)D topological orders; in the case of Abelian (2+1)D topological orders, we derive by explicit computation all of the mixed U(1) anomaly indicator formulas proposed by Lapa and Levin. We also propose a Spin$^c$ generalization of the Gauss-Milgram sum, valid for super-modular categories. 
    \end{abstract}

        \maketitle
        \tableofcontents
        
\section{Introduction}        

In the past several years, a comprehensive algebraic framework has emerged to completely characterize symmetry fractionalization in both bosonic and fermionic (2+1)D topological phases of matter in general \cite{barkeshli2019,bulmashSymmFrac,aasen21ferm}. In some cases, a given pattern of symmetry fractionalization may be anomalous, in the sense that it cannot occur in a purely (2+1)D system, but it can occur at the (2+1)D surface of a (3+1)D symmetry-protected topological (SPT) state. In the language of quantum field theory, the (2+1)D theory possesses a 't Hooft anomaly for the global symmetry group. A basic question then is to compute the anomaly given the algebraic data that characterizes the (2+1)D topological order; that is, to determine which (3+1)D SPT can host the given (2+1)D topological phase at its surface. Since it has also been understood in recent years that the Lieb-Schultz-Mattis theorem and its generalizations can be rephrased in terms of mixed anomalies between translation symmetry and an on-site symmetry \cite{cheng2016lsm, ChoManifestation, JianLSM, MetlitskiThorngren, TanizakiSU3}, these questions have far-reaching implications for placing constraints on the possible (2+1)D topological orders that can emerge from a given microscopic model. 

Recently it has been understood that one can in general compute anomalies of (2+1)D topological orders as follows \cite{barkeshli2019tr,bulmash2020,tata2021anomalies}. One can use the algebraic data that defines the (2+1)D topological order and its possibly anomalous symmetry fractionalization pattern for a symmetry group $G$ as input into the construction of a (3+1)D topological path integral defined on a principal $G$ bundle. This (3+1)D path integral can be constructed in terms of an exact combinatorial sum over a finite set of labelings of a triangulated 4-manifold in the presence of a background $G$ gauge field. This construction defines an invertible topological quantum field theory (TQFT) with $G$ symmetry, as appropriate for describing (3+1)D SPT states, and can be viewed as a symmetry-enriched generalization of the Crane-Yetter-Walker-Wang constructions \cite{crane1993,walker2012}. 

If the input is a fermionic topological order, i.e. a super-modular category, with fermionic symmetry fractionalization, then the topological path integral depends on a generalized spin structure as well, and defines an invertible spin TQFT with a fermionic symmetry group $G_f$ \cite{tata2021anomalies}. 

The state sum path integral is quite powerful, as it can also be used to derive a ground state wave function for the (3+1)D system and an exactly solvable parent Hamiltonian. As such it provides a comprehensive, exact description of the (3+1)D system although this perspective has only been worked out in some cases \cite{walker2012,williamson2017,bulmash2020}.

A significant limitation of previously known constructions is that they only give topological path integrals that can be defined on \it flat \rm $G$-bundles. When $G$ contains a continuous component, it is known that the path integral on curved $G$ bundles is quite useful, and potentially necessary, to fully characterize a (3+1)D SPT. For example, the famous (3+1)D topological insulator in Class AII \cite{qi2010RMP,hasan2010} can be characterized by its U(1) electromagnetic response, which contains a theta term at $\Theta = \pi$, and one needs to introduce a non-trivial field strength for this term to be non-trivial. In fact, all TQFT state sum constructions so far have only be defined for flat $G$ bundles, and it is generally assumed that one must always restrict to flat $G$ bundles. 

In this paper, we show that when the symmetry $G$ contains a $\U$ subgroup, one can generalize the previous constructions to define a (3+1)D topological path integral on a \it curved \rm bundle. More specifically, $\U$ bundles over (3+1)D manifolds can be completely characterized by their second Chern number, in addition to the holonomies around non-contractible cycles. Here we show that one can define a topological path integral for any $\U$ bundle, including any non-zero second Chern number. In the case of fermionic topological phases, our construction allows us to define topological path integrals on 4-manifolds with Spin$^c$ structures, again with non-trivial Chern numbers.

Our construction proceeds by utilizing a Villain formulation for the background $\U$ gauge field. We view a $1$-form $\U$ gauge field in terms of a $1$-form $\mathbb{R}$ gauge field and a $2$-form $\mathbb{Z}$ gauge field. We then introduce a new kind of diagrammatic formalism, which we refer to as a ``Villain" graphical calculus, which describes anyons interacting with symmetry defects associated with the $1$-form $\mathbb{R}$ gauge field and $2$-form $\mathbb{Z}$ gauge field. We show how this new graphical calculus can be used to consistently define the (3+1)D topological path integral on curved $\U$ bundles. 

As an important application of our results, our constructions allow a derivation of the anomaly indicator formulas presented in \cite{lapa2019}, which we explicitly demonstrate in the case of Abelian topological phases. Our results also provide a Spin$^c$ generalization of the Gauss-Milgram sum for super-modular categories, which provides a simple explicit formula to determine the chiral central charge $c_- \text{ mod } 1$ of a super-modular category with $\U^f$ symmetry fractionalization just from the modular data and fractional charge assignments. 

\subsection{Summary of main results}
\label{summarySec}

We consider both bosonic and fermionic (2+1)D topological orders with a global symmetry that involves U(1). 
In the bosonic cases, the global symmetry group has the form $G=\U\rtimes H$. We allow $H$ to contain anti-unitary symmetry group elements, and $H$ can act on U(1) by charge conjugation.

In the fermionic cases, the system has a fermionic global symmetry group $G_f$. $G_f$ contains a fermion parity symmetry $\Z_2^f$, and the group that acts on bosonic operators is expressed as $G_b=G_f/\Z_2^f$. We take the bosonic symmetry group to be of the form $G_b=\U\rtimes H$. In general, $G_f$ is a nontrivial extension of $G_b$ by $\Z_2^f$, whose extension is specified by a 2-cocycle $\omega_2\in Z^2(BG_b,\Z_2)$. When $G_f$ involves a nontrivial extension of $\U$ by $\Z_2^f$, the fermion carries charge under the extended U(1) symmetry $\Z_2^f\to \U^f\to\U$. More generally, the existence of a non-trivial $\omega_2$ can be understood as the physical fermion carrying fractional quantum numbers under $G_b$. In particular, the above global symmetry group contains the Altland-Zirnbauer symmetry classes A, AI, AII and AIII that involve U(1) symmetry. The specific 2-cocycle $\omega_2(\mathbf{g},\mathbf{h})$ is determined by the action of $G_f$ on local fermion operators in the microscopic theory~\cite{bulmashSymmFrac,aasen21ferm}.

In this paper, we assume that we are given the data of a modular or super-modular fusion category~\cite{bruillard2017a,bruillard2017b,bonderson2018} $\mathcal{C}$ in the bosonic or fermionic case respectively. In addition to $\mathcal{C}$, we are given the data that specifies the symmetry fractionalization \cite{barkeshli2019,bulmashSymmFrac,aasen21ferm}. This data consists of a group homomorphism $[\rho]:H\to \mathrm{Aut}(\mathcal{C})$, which specifies how the symmetry $H$ permutes the anyons, and a set of U(1) phases $\eta_a: G \times G \rightarrow \U$ for each anyon $a$, which must obey certain consistency equations and which are defined up to certain gauge transformations. We review this data in Appendix~\ref{app:symmfrac}. The data $\{\eta_a\}$ in particular specify the $\U$ fractional charges $\{ Q_a \}$ of each anyon $a$, that specifies how the $\U$ symmetry is fractionalized. 

In this paper, given $\mathcal{C}$ and the symmetry fractionalization data, we show how to construct a (3+1)D topologically invariant path integral coupled to a background gauge field with non-trivial Chern class. In the fermionic case, the path integral depends on both a $G_b$ background gauge field and a twisted spin structure $\xi$. 

The U(1) gauge field is expressed in a Villain form, as $({h},{c})$, which is a pair of a 1-form $\R$ gauge field $h$ and a 2-form $\Z$ gauge field $c$. The integral of $c$ over a closed 2-cycle gives the first Chern number of the $U(1)$ gauge field evaluated on that 2-cycle.  

The state sum path integral is given by a symmetry-enriched generalization of the Crane-Yetter-Walker-Wang construction, which utilizes the diagrammatic calculus of (2+1)D topological order in the presence of symmetry defects \cite{bulmash2020,tata2021anomalies}. In order to define the path integral in the presence of the Villain background gauge fields, we introduce a new graphical calculus that characterizes the interaction between the anyons and the ``Villain symmetry defects," which include codimension-1 and codimension-2 defects. This new graphical calculus is regarded as encoding the symmetry action of 0-form $\R$ and 1-form $\Z$ symmetry on the (2+1)D anyon system, and allows us to define a (3+1)D topologically invariant theory coupled with the generic Villain U(1) gauge field.

\subsubsection{Construction of topological path integrals}

Concretely, in the bosonic cases our (3+1)D path integral takes the form of $Z(M^4,( h, c), A_H)$, where $M^4$ is a closed 4-manifold, $(h, c)$ is the Villain gauge field subject to the action of $H$, and $A_H$ is a flat $H$ gauge field. In the fermionic cases, the path integral is given in the form of $Z(M^4,( h, c), A_H, \xi_{\mathcal{G}})$ with the additional dependence on twisted spin/pin structure $\xi_{\mathcal{G}}$ referred to as a $\mathcal{G}_f$-structure in this paper. The $\mathcal{G}_f$-structure of a manifold $M^4$ is specified by a choice of a 1-cochain $\xi_{\mathcal{G}}$ that satisfies
\begin{align}
            \delta \xi_\mathcal{G} = w_2 + w_1^2 + 2{c}+A_H^* \omega_H, 
        \end{align}
where $w_k\in H^k(M^4,\Z_2)$ is the $k^{\mathrm{th}}$ Stiefel-Whitney class, and $\omega_H \in H^2(BH,\Z_2)$ characterizes the group extension by $\Z_2^f$ restricted to $H$. Here we view the background $H$ gauge field as a map $A_H: M^4 \rightarrow BH$, where $BH$ is the classifying space of $H$, and $A_H^*$ denotes the pullback from $H^2(BH, \Z_2)$ to $H^2(M^4, \Z_2)$. For example, in the case that $H$ is trivial, $\xi_{\mathcal{G}}$ is a Spin$^c$ structure on oriented manifolds. When $H =\Z_2^{\mathbf{T}}$ with $\Z_2^{\mathbf{T}}$ a time-reversal symmetry and $\omega_{H}$ is trivial, $\xi_{\mathcal{G}}$ becomes a Pin$^c$ structure or Pin$^{\tilde{c}}_{-}$ structure depending on whether $\Z_2^{\mathbf{T}}$ conjugates the U(1).

The fermionic path integral $Z(M^4,( h,  c), A_H,\xi_{\mathcal{G}})$ is constructed by first taking the super-modular category with $G_b$ symmetry, and constructing a (3+1)D bosonic path integral using the new graphical calculus involving Villain U(1) background gauge fields $(h,  c)$. This gives a (3+1)D path integral $Z_b(M^4,( h,  c), A_H)$ for a bosonic topological phase which contains a single non-trivial point-like excitation, which is a fermion. We then condense this fermion, following the prescription of Refs.~\cite{Gaiotto:2015zta, tata2021anomalies}.

The fermion condensation is performed by first considering a $\Z_2$ 2-form symmetry of the theory $Z_b(M^4,( h,  c), A_H)$ generated by a Wilson line of a fermion. We turn on the background gauge field $f_3\in Z^3(M^4,\Z_2)$ of the 2-form symmetry, yielding a path integral $Z_b(M^4,( h,  c), A_H, f_3)$. In the presence of the 3-form background gauge field $f_3$, the path integral turns out to possess a 't Hooft anomaly of the $\Z_2$ 2-form symmetry characterized by a (4+1)D response action
\begin{align}
    e^{i S_{5,b}}=(-1)^{\int \Sq^2(f_3)+(2{c}+A_H^*\omega_H)\cup f_3}
    \label{eq:S5bintro}
\end{align}
where $\Sq^2$ is a cohomology operation called Steenrod square. Due to the 't Hooft anomaly, $Z_b$ is no longer topologically invariant in the presence of the 3-form background gauge field $f_3$.

In order to obtain a topologically invariant path integral by condensing the fermion, we need to appropriately compensate for these anomalies by including an additional factor. The full fermionic path integral then takes the form:
\begin{align}
            Z(M^4, ( h,  c), A_H, \xi_{\mathcal{G}}) = \frac{1}{\sqrt{|H^2(M^4,\Z_2)|}} \sum_{[f_3] \in H^3(M^4,\Z_2)} Z_b(M^4, ( h,  c), A_H, f_3) z_c( M^4, f_3, \xi_{\mathcal{G}}) ,
\end{align}
where $z_c( M^4, f_3, \xi_{\mathcal{G}})$ is a fermionic theory coupled with ${\mathcal{G}}_f$-structure $\xi_{\mathcal{G}}$, which has the same 't Hooft anomaly as that of $Z_b$ characterized by the response action Eq.~\eqref{eq:S5bintro}. By stacking the bosonic path integral $Z_b$ with $z_c$, we obtain a fermionic theory free of 't Hooft anomaly with respect to the diagonal 2-form $\Z_2$ symmetry. We can gauge the diagonal 2-form symmetry by summing over distinct $[f_3]$ classes, resulting in a fermionic theory $Z$. This gauging process implements fermion condensation within the path integral framework.

The definition of the fermionic theory $z_c( M^4, f_3, \xi_{\mathcal{G}})$ is based on the path integral over Grassmann variables that decorate the triangulation of a manifold; for orientable manifolds this is given by the construction of Gu and Wen \cite{gu2014, Gaiotto:2015zta}, while the generalization to non-orientable manifolds was given in \cite{Kobayashi2019pin}. The full generalization for $\xi_{\mathcal{G}}$ was obtained in~\cite{tata2021anomalies}.

In both bosonic and fermionic cases, the path integral $Z$ is invariant under re-triangulations and also invariant under gauge transformations of background gauge fields and $\mathcal{G}_f$-structure, and thus gives a topological invariant of a manifold $M^4$ with a given $\mathcal{G}_f$-structure and background gauge field with non-trivial Chern class. 

We argue that the path integral $Z$ always defines an invertible topological quantum field theory (TQFT), and therefore describes a (3+1)D SPT. This implies that the path integral should give a $\U$ phase which is a smooth bordism invariant. More precisely, in the bosonic case, our path integrals define elements of the Pontryagin dual of the smooth bordism group of 4-manifolds equipped with (possibly curved) background $G$ gauge fields. In the fermionic case, the bordism groups are equipped with (possibly curved) background $G_b$ gauge fields and twisted spin structures (i.e. $\mathcal{G}_f$ structures). 

When the input (2+1)D topological order characterized by the (super)-modular category $\mathcal{C}$ is Abelian, we can explicitly perform the computation of the path integral $Z$ on various 4-manifolds, for both bosonic and fermionic theories. These path integrals give anomaly indicator formulas for a given topological order $\mathcal{C}$ with a global symmetry, that indicates the presence or absence of 't Hooft anomaly of the given (2+1)D topological order.

Below we list the results of the computations for Abelian topological order $\mathcal{C}$ as follows. For the Abelian topological order, the results shown below produce all of the anomaly indicator formulas proposed by Lapa and Levin in~\cite{lapa2019}. In the case with the time-reversal symmetry, the indicators were derived in~\cite{Chenjie2021mirror} based on a completely different method where the time-reversal is replaced by a crystalline reflection symmetry.


    \subsubsection{Abelian bosonic topological phases with $G = \U$} 
    
    Let $\mathcal{C}$ be an Abelian unitary modular tensor category (UMTC) with $G=\U$ symmetry. We consider a path integral on $\mathbb{CP}^2$ with nontrivial Chern class $\int F/(2\pi)=C\in\Z$ evaluated on a fundamental 2-cycle that generates $H_2(\mathbb{CP}^2,\Z)=\Z$. The partition function is given by~\footnote{The partition function is a number given by evaluating a path integral on a closed manifold. In this paper, we sometimes use the terminologies ``partition function'' and ``path integral'' interchangeably.}
    \begin{align}
        Z(\mathbb{CP}^2,(h,c))=\frac{1}{\mathcal{D}}\sum_{a\in\mathcal{C}}e^{-2\pi iQ_a\cdot C}\theta_a,
        \label{eq:CP2bosonintro}
    \end{align}
    where $Q_a\in\R/\Z$ is a fractional U(1) charge of an anyon $a$. 
    
    The above theory is conjectured to give the bosonic invertible TQFT with U(1) symmetry. The invertible TQFT is classified up to isomorphisms as the Pontryagin dual of the 4D oriented bordism group $\Omega_4^{\mathrm{SO}}(B\U)=\Z\times\Z$, given by $\mathrm{Hom}(\Omega_4^{\mathrm{SO}}(B\U),\U)=\U\times\U$. 
    The $\U\times\U$ corresponds to the two continuous theta terms given by
    \begin{align}
        S=\frac{\Theta}{8\pi^2}\int F\wedge F+\frac{2\pi c_-}{192\pi^2}\int\mathrm{Tr}(R\wedge R),
        \label{eq:thetaterm}
    \end{align}
    The theta parameter $\Theta$ has $4\pi$ periodicity for bosonic phases, while $c_-$ has periodicity 8.
    On a $\mathbb{CP}^2$ with the Chern number $C\in\Z$, the above action is evaluated as $e^{\frac{i\Theta}{2}C^2+\frac{2\pi i}{8}c_-}$. We can then read each parameter from the indicator formula in Eq.~\eqref{eq:CP2bosonintro} as
    \begin{align}
    \begin{split}
        e^{\frac{i\Theta}{2}}&=\frac{Z(\mathbb{CP}^2,C=1)}{Z(\mathbb{CP}^2,C=0)}=\frac{\sum_{a\in\mathcal{C}}e^{-2\pi iQ_a}\theta_a}{\sum_{a\in\mathcal{C}}\theta_a} \\
        e^{\frac{2\pi i}{8}c_-}&={Z(\mathbb{CP}^2,C=0)}=\frac{1}{\mathcal{D}}{\sum_{a\in\mathcal{C}}\theta_a}
        \end{split}
        \label{eq:indicatortheta}
    \end{align}
    The latter formula for $e^{\frac{2\pi i}{8}c_-}$ reproduces the well-known Gauss-Milgram formula for chiral central charge $c_-$ of a UMTC $\mathcal{C}$, which corresponds to the framing anomaly characterized by the 4D response action in the second term of Eq.~\eqref{eq:thetaterm}.
    
    Note that here the $\U \times \U$ classification does not classify invertible topological phases of matter, which correspond to \it deformation classes \rm of invertible TQFTs; therefore the formulas above do not correspond to 't Hooft anomalies of the (2+1)D phase, which are conceptually distinct from the framing anomaly. 
    
    The $\Theta$ term also determines the Hall conductivity of the (2+1)D surface to be $\sigma_H = \frac{1}{2\pi} \frac{\Theta}{2\pi} \text{ mod } 2$, since the boundary of the electromagnetic theta term~\eqref{eq:thetaterm} reduces locally to the response action $\frac{\Theta}{8\pi^2}A\mathrm{d}A$, which gives the Hall response $\sigma_H$ on the boundary. Thus, defining $\overline{\sigma}_H = 2\pi \sigma_H$, we have
    \begin{align}
        e^{i \Theta/2} = e^{i \pi \overline{\sigma}_H} = \theta_v^* = \frac{\sum_{a\in\mathcal{C}}e^{-2\pi iQ_a}\theta_a}{\sum_{a\in\mathcal{C}}\theta_a},
        \label{eq:thetavison}
    \end{align}
    where $\theta_v$ is the topological twist of the vison, which is the anyon induced by $2\pi$ flux insertion (see Section~\ref{subsec:U1frac} for a more precise definition of the vison).

    We also consider the partition function on $S^2\times S^2$ with nontrivial Chern numbers $C_{\alpha}, C_{\beta}\in\Z$ evaluated on each fundamental 2-cycle that generates $H_2(S^2\times S^2,\Z)=\Z\times\Z$. The partition function is given by
    \begin{align}
\begin{split}
    Z(S^2\times S^2,(h,c))=\frac{1}{\mathcal{D}}\sum_{a,b\in \mathcal{C}}e^{-2\pi i Q_a\cdot C_\beta}e^{-2\pi i Q_b\cdot C_\alpha}S^*_{{a},b}=(\theta_v^*)^{2C_{\alpha}C_{\beta}}.
    \end{split}
    \label{eq:S2S2}
\end{align}

The topological action in Eq.~\eqref{eq:thetaterm} evaluates on this $S^2\times S^2$ as $e^{i\Theta C_{\alpha}C_{\beta}}$. We hence have
\begin{align}
    e^{i\Theta C_{\alpha}C_{\beta}}=(\theta_v^*)^{2C_{\alpha}C_{\beta}},
\end{align}
which is consistent with Eq.~\eqref{eq:thetavison}.

    \subsubsection{Abelian bosonic topological phases with $G = \U \rtimes \Z_2^{\bf T}$}
    
    For the symmetry group $G = \U\rtimes\Z_2^{\mathbf{T}}$, the classification of (3+1)D invertible TQFT is given by $(\Z_2)^3$~\cite{kapustin2014c}. These are partially accounted for by the four choices of
    $(\Theta, c_-)$ that are compatible with time-reversal invariance: $(\Theta, c_-) = (0,0), (2\pi, 0), (0,4),(2\pi,4)$. The formulas given above for $(\Theta, c_-)$ therefore give anomaly indicator formulas to diagnose the 't Hooft anomalies of (2+1)D topological order that arise at the surface of these bosonic topological insulators. 
    The third factor of $\Z_2$ is distinguished by the path integral on $\mathbb{RP}^4$ \cite{barkeshli2019tr}. 
    
    

    \subsubsection{Abelian bosonic topological phases with $G = \U \times \Z_2^{\bf T}$}

Let $\mathcal{C}$ be an Abelian UMTC with $G=\U\times\Z_2^{\mathbf{T}}$ symmetry. We consider a partition function on $S^2\times \mathbb{RP}^2$ with the nontrivial Chern number $\int_{S^2}F/(2\pi)=C\in\Z$ evaluated on $S^2$. We assume that the time-reversal symmetry $\Z_2^{\mathbf{T}}$ does not permute the label of anyons. The partition function is then given by
\begin{align}
\begin{split}
    Z(S^2\times\mathbb{RP}^2,(h,c),A_{H})=\frac{1}{\mathcal{D}}
    \sum_{a,b\in\mathcal{C}} e^{-2\pi i Q_bC}\eta_a^{\mathbf{T}}S_{a,b}
    \end{split}
    \label{eq:S2RP2intro}
\end{align}
where we define $\eta_a^{\mathbf{T}}:= \eta_a(\mathbf{T},\mathbf{T})$, which determines whether the anyon $a$ carries a local Kramers degeneracy. 

(3+1)D invertible TQFTs with $G=\U\times\Z_2^{\mathbf{T}}$ symmetry are classified by $(\Z_2)^4$~\cite{kapustin2014c}, and the partition function on $S^2\times\mathbb{RP}^2$ diagnoses one of these four topological terms given by
\begin{align}
    e^{iS}=(-1)^{\int w_1^2\frac{F}{2\pi}}.
\end{align}
The formula Eq.~\eqref{eq:S2RP2intro} with $C=1$ indicates whether a given symmetry fractionalization class for the (2+1)D topological order has this mixed 't Hooft anomaly. The other three topological terms are 
\begin{align}
    (-1)^{\int w_1^4},\quad (-1)^{\int\left(\frac{F}{2\pi}\right)^2 }, \quad (-1)^{\int w_2^2},
\end{align}
where the action $(-1)^{\int w_1^4}$ is detected by the partition function on $\mathbb{RP}^4$~\cite{barkeshli2019tr}, and the rest are diagnosed by the indicators $e^{\frac{i\Theta}{2}}, e^{\frac{2\pi i}{8}c_-}$ obtained in Eq.~\eqref{eq:indicatortheta}.

    \subsubsection{Abelian fermionic topological phases with $G_f = \U^f$ and Spin$^c$ Gauss-Milgram formula}     
    
     Let $\mathcal{C}$ be an Abelian super-modular tensor category with $G_b=\U$ symmetry. We consider a partition function $Z$ of a fermionic theory on $\mathbb{CP}^2$ equipped with a Spin$^c$ structure $\xi_{\mathcal{G}}$, with nontrivial Chern class $\int F/(2\pi)=C\in\Z+1/2$ evaluated on a fundamental 2-cycle that generates $H_2(\mathbb{CP}^2,\Z)=\Z$. The partition function is given by
    \begin{align}
        Z(\mathbb{CP}^2,(h,c),\xi_{\mathcal{G}})=\frac{1}{\sqrt{2}\mathcal{D}}\sum_{a\in\mathcal{C}}e^{-2\pi iQ_a\cdot C}\theta_a,
        \label{eq:CP2intro}
    \end{align}
    where $Q_a\in\R/(2\Z)$ is a fractional U(1) charge of an anyon $a$. 
    
    (3+1)D invertible Spin$^c$ TQFTs are classified by $\U\times \U$ up to isomorphisms, which is the Pontryagin dual of the bordism group $\Omega_{4}^{\Spin^c}=\Z\times\Z$.
    The $\U\times\U$ corresponds to the two continuous theta terms in the topological effective action ~\cite{Seiberg2016Gapped}
    \begin{align}
    S = \Theta_1 I_1+\Theta_2 I_2,
    \end{align}
    \begin{align}
        I_1=\int\left(-\frac{1}{192\pi^2}\mathrm{Tr}(R\wedge R)+\frac{1}{8\pi^2}F\wedge F\right),\quad I_2=\frac{1}{\pi^2}\int F\wedge F,
        \label{eq:thetatermspinc}
    \end{align}
    with the Dirac quantization condition $\int F/(2\pi)=\int w_2/2$ on closed oriented 2-cycles.
    Both $I_1$ and $I_2$ are integers on Spin$^c$ manifolds, so $\Theta_1$ and $\Theta_2$ have periodicity $2\pi$. On $\mathbb{CP}^2$ with the Chern number $C$, the above response action evaluates to $\exp(i\Theta_1(-\frac{1}{8}+\frac{C^2}{2})+\Theta_2 4C^2)$. We can then read off each theta parameter from the indicator formula 
    \begin{align}
            e^{i\Theta_1}=e^{-2\pi i c_-}= \frac{Z\left(\mathbb{CP}^2,C=\frac{3}{2}\right)}{Z\left(\mathbb{CP}^2,C=\frac{1}{2}\right)^9},\quad e^{i\Theta_2}= Z\left(\mathbb{CP}^2,C=\frac{1}{2}\right).
            \label{eq:indicatorsspinc}
    \end{align}
   Here, the chiral central charge $c_-$ of the Spin$^c$ topological order is given by $c_-=-\Theta_1/(2\pi)$ mod 1. 
   
   The formula for $e^{i\Theta_1}$ can be regarded as a generalization of the Gauss-Milgram formula Eq.~\eqref{eq:indicatortheta} of $c_-$ to the Spin$^c$ case. In other words, this provides a simple formula to determine the chiral central charge $c_- \text{ mod } 1$, given the data of the super-modular category with $\U^f$ fermionic symmetry fractionalization. For non-Abelian topological orders, we conjecture that the partition function on $\mathbb{CP}^2$ is given by
   \begin{align}
        Z(\mathbb{CP}^2,(h,c),\xi_{\mathcal{G}})=\frac{1}{\sqrt{2}\mathcal{D}}\sum_{a\in\mathcal{C}}d_a^2 e^{-2\pi iQ_a\cdot C}\theta_a,
        \label{eq:CP2nonabel}
    \end{align}
    and the formulas for the theta parameters are given by Eq.~\eqref{eq:indicatorsspinc}. In Appendix~\ref{app:gauss}, we explicitly compute the indicator formula for $e^{i\Theta_1}$ for several Read-Rezayi fractional quantum Hall states which are non-Abelian~\cite{readrezayi1999}, and demonstrate that it exactly gives $c_-$ mod 1.
   
   Note that even though it is known that a super-modular category alone, without any additional global symmetry, determines the chiral central charge $c_- \text{ mod } 1/2$, a simple explicit formula generalizing the Gauss-Milgram sum is not known, although it is known that it should correspond to the topological fermionic path integral evaluated on the K3 surface with a choice of spin structure~\cite{walkerSCGPtalk}. 
    

\subsubsection{Fermionic topological phases in class AII or AIII}

Here we consider general (2+1)D fermionic topological phases, which may be non-Abelian, characterized by a general super-modular category $\mathcal{C}$, in symmetry class AII or AIII. Symmetry class AII corresponds to $G_f = [\U^f \times \Z_4^{{\bf T},f}]/\Z_2$, where $\Z_4^{{\bf T}, f}$ is the non-trivial extension of $\Z_2^{\bf T}$ by $\Z_2^f$, and the mod $\Z_2$ identifies the $Z_2^f$ from the two factors. Symmetry class AIII corresponds to $G_f = \U^f \times \Z_2^{\bf T}$. The resulting TQFTs for class AII and AIII depend on a Pin$^{\tilde{c}}_+$ or Pin$^c$ structure, respectively.  

In these cases we can compute the partition function on $\mathbb{RP}^4$ with U(1) gauge field turned off, with the result
\begin{align}
\label{ZRP4}
Z(\mathbb{RP}^4,\xi_{\mathcal{G}}) &= \frac{1}{\sqrt{2}\mathcal{D}} \left(\sum_{x | x = {\,^{\bf T}}x} d_x \theta_x \eta^{\bf T}_x \pm i \sum_{x | x = {\,^{\bf T}}x \times \psi} d_x \theta_x \eta^{\bf T}_x\right)
\end{align}
Here, $\xi_{\mathcal{G}}$ denotes the Pin$^{\tilde{c}}_+$ or Pin$^c$ structure, and the choice of sign in $\pm i$ depends on the possible two choices of $\xi_{\mathcal{G}}$ structure. $\eta^{\bf T}_x$ is given by
\begin{equation} 
            \eta_a^{\bf T} := \begin{cases}
            \eta_a({\bf T},{\bf T}) & \,^{\bf T}a=a\\
            \eta_a({\bf T},{\bf T})U_{\bf T}(a,\psi;a\psi)F^{a,\psi, \psi} & \,^{\bf T}a = a \times \psi ,
            \end{cases} 
        \end{equation}
        and determines the Kramers degeneracy of the anyon $a$. 
        
        The computation of the $\mathbb{RP}^4$ partition function is essentially the same as that done in~\cite{tata2021anomalies} for Pin$^+$ structure, which corresponds to class DIII (where $G_f = \Z_4^{\mathbf{T},f}$).
        $\mathbb{RP}^4$ generates the $\Z_2$ subclass of the Pin$^{\tilde c}_+$ bordism group $\Omega_4^{\mathrm{Pin}^{\tilde c}_+}=(\Z_2)^3$, and generates the $\Z_8$ subclass of the Pin$^{c}$ bordism group $\Omega_4^{\mathrm{Pin}^{c}}=\Z_8\times\Z_2$~\cite{Freed:2016rqq, shiozaki2018fermion}. These classes correspond to the topological insulator in class AII and topological superconductor in class AIII respectively. 
        
        Therefore, given a (2+1)D fermionic topological order with a choice of $\{\eta_a^{\bf T}\}$, if the theory is compatible with class AII, Eq.~\eqref{ZRP4} will evaluate to $\pm 1$, while if the theory is compatible with class AIII, Eq.~\eqref{ZRP4} will evaluate to an 8$^{\mathrm{th}}$ root of unity. 
        
        One of the $\Z_2^3$ factors for the classification in class AII corresponds to the conventional topological band insulator, with electromagnetic response given by $\Theta_1=\pi$ in Eq.~\eqref{eq:thetatermspinc} on oriented manifolds. The other two generators have topological actions given by~\cite{wangpotter2014}
        \begin{align}
            (-1)^{\int w_1^4}, \quad (-1)^{\int w_2^2}.
        \end{align}
        Since both the topological path integral for the conventional topological band insulator and $(-1)^{\int w_1^4}$ evaluate to $-1$ on $\mathbb{RP}^4$ \cite{metlitski2015, witten2016}, $Z(\mathbb{RP}^4,\xi_{\mathcal{G}})$ cannot distinguish these two phases. The indicator that can distinguish the topological insulator is given by the $S^2\times S^2$ partition function with $\frac{1}{8\pi^2}\int FF=1$. We conjecture that the $S^2\times S^2$ partition function has the form 
        \begin{align}
            Z(S^2\times S^2,(h,c))=\frac{1}{2\mathcal{D}}\sum_{a,b\in \mathcal{C}}d_a d_b e^{-2\pi i Q_a\cdot C_\beta}e^{-2\pi i Q_b\cdot C_\alpha}S^*_{{a},b},
        \end{align}
        which is regarded as a generalization of Eq.~\eqref{eq:S2S2} to the non-Abelian and fermionic case, proposed by~\cite{lapa2019}.
        
        The (3+1)D theory that generates the $\Z_8$ subclass of class AIII is generated by a topological superconductor, which again has a $\U$ response $\Theta_1=\pi$ on oriented manifolds~\cite{metlitski2015, Seiberg2016Gapped}. The other $\Z_2$ class is generated by the bosonic action $(-1)^{\int w_2^2}$, which is diagnosed by $e^{i\Theta_2}$ in Eq.~\eqref{eq:indicatorsspinc}.

        
        \subsection{Organization of paper}

This paper is organized as follows. In Sec.~\ref{sec:villain}, we introduce the Villain formulation for the background U(1) gauge field utilized throughout this paper to describe the curved U(1) bundle.  In Sec.~\ref{sec:U1frac}, we first review the symmetry fractionalization of U(1) symmetry. We then introduce a new graphical calculus to describe the (2+1)D anyon systems coupled to the U(1) Villain gauge field, which is beyond the scope of the existing symmetry fractionalization theory. In Sec.~\ref{sec:statesumu1}, we provide the (3+1)D state sum path integral based on the Villain graphical calculus in the simplest setup for a bosonic phase with U(1) symmetry, which can describe the electromagnetic response of U(1) theta term. In Sec.~\ref{sec:u1rtimesg}, we generalize the bosonic state sum to the case with $\U\rtimes H$ symmetry, where $H$ is a symmetry group that can act on $\U$ by charge conjugation.
In Sec.~\ref{sec:condensation}, we review the fermion condensation utilized to construct a path integral of fermionic SPT phases starting with a path integral of a bosonic theory, referred to as the bosonic shadow. 
In Sec.~\ref{sec:shadow}, we construct a bosonic path integral that gives a (3+1)D fermionic SPT phase with $\U\rtimes H$ symmetry via fermion condensation. For example, the resulting fermionic theory can describe Spin$^c$ theta terms, a topological insulator in class AII and topological superconductor in class AIII. In Sec.~\ref{sec:evaluation}, we compute the partition functions of our path integrals for various manifolds, and prove the anomaly indicator formulas proposed by Lapa and Levin for the case of Abelian topological phases based on our path integral. We close with some discussion of open questions in Sec.~\ref{sec:discussions}. Many technical discussions and detailed calculations are relegated to appendices.

\section{U(1) gauge field on lattice: Villain formulation}
\label{sec:villain}
\subsection{U(1) gauge field}

Let us consider a $\U$ gauge theory on a $d$-dimensional manifold $M$ equipped with a triangulation and a branching structure, following the Villain-type formulation~\cite{Villain, Sulejmanpasic}. 
That is, we start with a non-compact $\mathbb{R}$ gauge field on each 1-simplex $h\in C^1(M,\mathbb{R})$. Then, we gauge the $\mathbb{Z}$ 1-form symmetry of the theory by introducing a 2-form $\mathbb{Z}$ gauge field $c\in Z^2(M,\mathbb{Z})$ on each 2-simplex. Here, $C^k(M,X)$ and $Z^k(M,X)$ denote a $k^{\mathrm{th}}$ cochain and cocycle with coefficient $X$ respectively. We then postulate the gauge invariance
\begin{align}
    c\to c-\delta \lambda, \quad h\to h+\lambda
    \label{eq:gaugetrans}
\end{align}
with $\lambda\in C^1(M,\mathbb{Z})$. The resulting theory realizes a $\R/\Z=\U$ gauge theory whose field strength is given by 
\begin{align}
\frac{F}{2\pi} = c + \delta h, 
\end{align}
where the field strength $F$ is normalized so that it satisfies the Dirac quantization condition $\int F=2\pi\Z$ on oriented 2-cycles.
The usual U(1) gauge transformation is realized by
\begin{align}
    c\to c, \quad h\to h+\delta\chi
\end{align}
with $\chi\in C^0(M,\mathbb{R})$. 

Therefore, a background U(1) gauge field can be described by the pair of background fields $(h,c)$, subject to the gauge transformation~\eqref{eq:gaugetrans}.

Note that one way to understand the above Villain formulation is as follows. Usually to describe a $\U$ gauge field, we define a $\U \simeq \mathbb{R}/\Z$ gauge field on links, $a_{ij} \in \mathbb{R}/\Z$. The field strength in this case would be $\delta a \in \mathbb{R}/\Z$; since this is a coboundary, the Chern number will always vanish on closed manifolds. In order to describe a manifold with non-trivial Chern number, we need to resolve a contribution to the flux per plaquette that is not a coboundary. This can be done by defining a lift $h$ of $a$ from $\mathbb{R}/\Z$ to $\mathbb{R}$, and introducing an additional degree of freedom $c \in \Z$ on each $2$-simplex, which keeps track of the integer part of the flux that is not a coboundary. The total flux is then $F = c + \delta h$. Since we are really describing a $\U$ gauge field, the only gauge invariant quantities are $\mathbb{R}/\Z$-valued holonomies through non-contractible cycles and the field strength $F$ through each plaquette. Therefore we have the gauge transformation $h \rightarrow h + \lambda$, $c \rightarrow c - \delta \lambda$ for $\lambda \in C^1(M,\Z)$, which keeps all physical quantities invariant. 

\subsection{$\U^f$ gauge field and Spin$^c$ structure}

We are also interested in fermionic theories, where the symmetry group contains a $\U^f$ factor. Here $\U^f$ means that the $\pi$ rotation in the $\U$ is associated with fermion parity. In other words, $\U^f$ is a non-trivial central extension of $\U$ by fermion parity, $\Z_2^f$. 

In this case, the fermionic path integral depends on a background $\U$ gauge field $A$ and a Spin$^c$ structure $\xi$, which is a trivialization of $w_2 + F/\pi \text{ mod 2}$: 
\begin{align}
\delta \xi = w_2 + \frac{F}{\pi} \text{ mod 2} ,
\end{align}
where $w_2\in H^2(M,\Z_2)$ is the 2nd Stiefel-Whitney class. 

If we describe the $\U$ gauge field in terms of $(h,c)$ as discussed above, then $\int_C \frac{F}{2\pi} = \int_C c$ and so we require 
\begin{align}
        \int_C c= \frac{1}{2} \int_C w_2 \quad \mod 1,
        \label{eq:spincdirac}
    \end{align}
for any oriented two-dimensional cycle $C$ in the spacetime. This is realized by the gauge fields $(h,c)$ with the field strength $F/(2\pi) = c+\delta h$, setting $h\in C^1(M,\R)$,  $2c\in\Z$ on each 2-simplex and then $[2c]=[w_2]$ mod $2$ as an element of $H^2(M,\Z_2)$. The gauge transformation is again given by~\eqref{eq:gaugetrans}.

Note that in the above, the gauge field $A$ is the gauge field for the bosonic symmetry $\U^f/\Z_2^f \simeq \U$, for which the periodicity of the $\U$ is $1/2$. One could choose a different normalization for the gauge field so that the periodicity is $1$, which would replace $F/\pi$ by $F/(2\pi)$ in the above equations. We will see that in the normalization we chose above, the fermion $\psi$ has a charge $Q_\psi = 1 \text{ mod 2}$ under the $\U$; changing the normalization would change the fermion charge to $1/2 \text{ mod 1}$.

    \section{U(1) symmetry}
    \label{sec:U1frac}
    
        In this section, we discuss a braided tensor category (BTC) in the presence of a global U(1) symmetry. See Appendix~\ref{sec:anyon} for a general review of BTC.
        The global symmetry of the BTC is described by the action of symmetry defects on its anyon data. The symmetry defect is equivalent to the background gauge field of the global symmetry inserted in its Poincar\'e dual. When a given symmetry defect corresponds to the flat background gauge field, the description of symmetry defects in the BTC is fully established within the framework of symmetry fractionalization theory developed in~\cite{barkeshli2019,bulmashSymmFrac,aasen21ferm} based on G-crossed BTC, see Appendix~\ref{app:symmfrac} for a detailed review. 
        
        We want to construct a (3+1)D state sum path integral coupled with a curved U(1) bundle. To do this, we extend the symmetry fractionalization theory to deal with the non-flat background gauge field. This gives a new diagrammatic calculus utilizing the Villain gauge fields, and enables us to construct a (3+1)D state sum path integral coupled with Villain U(1) gauge field, starting with a U(1) symmetry fractionalization data of the BTC.
        In Sec.~\ref{subsec:U1frac}, we give a brief explanation of U(1) symmetry fractionalization for flat U(1) gauge field, and then extend the framework to the non-flat U(1) gauge field in Sec.~\ref{subsec:nonflat}.
        The symmetry fractionalization data and the calculus of Villain gauge fields for $\U\rtimes H$ symmetry is developed in Appendix~\ref{app:tilde}.
    
    \subsection{U(1) symmetry fractionalization}
    \label{subsec:U1frac}
    
    \subsubsection{Bosonic case: modular tensor category}
    
    Here we briefly summarize the symmetry fractionalization data of unitary modular tensor category (UMTC) with global $G = \U$ symmetry following~\cite{barkeshli2019, cheng2016lsm}. 
    Most generally in BTC, the global symmetry acts on the anyons and the topological state space through the action of a group homomorphism
\begin{align}
[\rho] : G \rightarrow \text{Aut}(\mathcal{C}).
\end{align}
The elements of $\text{Aut}(\mathcal{C})$ in general act by permuting the anyons. Since $\text{Aut}(\mathcal{C})$ is a finite group, when $G=\U$ the map $[\rho_{\bf g}]$ must be trivial. 

\begin{figure}
    \centering
    \includegraphics[width=0.4\linewidth]{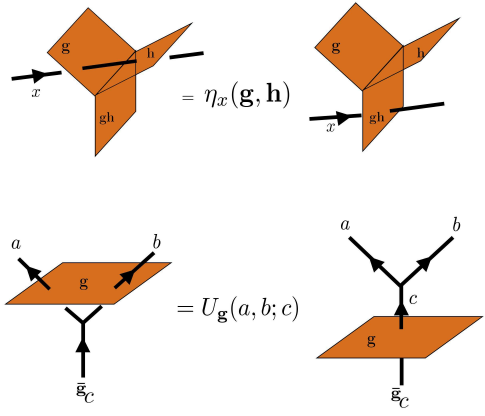}
    \caption{Anyon lines (black) passing through branch sheets (orange) and graphical definitions of the $U$ and $\eta$ symbols. }
    \label{fig:symfracmain}
\end{figure}

A given representative $\rho_{\bf g}$ defines $N_{ab}^c \times N_{ab}^c$ matrices $U_{\bf g}(a,b;c)$, which characterize the action of $\rho_{\bf g}$ on the fusion and splitting spaces in a given basis. 
Once a representative $\rho$ is specified, the symmetry fractionalization pattern is characterized by the phases $\eta_a({\bf g}, {\bf h})$, defined diagrammatically in Fig.~\ref{fig:symfracmain}. These $\eta$ and $U$'s are subject to the consistency equations

\begin{align}
\begin{split}
    {U}_{\mathbf{g}}({a},{b} ;{c}){U}_{\mathbf{h}}({a},{b} ;{c}) &= {U}_{\mathbf{gh}}({a},{b} ;{c})\frac{\eta_c({\bf g}, {\bf h})}{\eta_a({\bf g}, {\bf h})\eta_b({\bf g}, {\bf h})},
\\
  {\eta}_{a}({\bf h},{\bf k}) {\eta}_a({\bf gh}, {\bf k})
                                                             &= {\eta}_a({\bf g},{\bf h}) {\eta}_a({\bf g}, {\bf hk}).
                  \end{split}                  
                  \label{eq:consistencymain}
\end{align}

For a given set of $\{\eta, U\}$, one can define fractional charges of anyons that characterizes the U(1) symmetry fractionalization as follows. For a fixed anyon $a$, let $n$ be the smallest integer such that $a^n$ contains the identity as a fusion product. Choose a sequence of anyons $a, a^2,\dots a^n=1$ such that $a\times a^k$ contains $a^{k+1}$ as a fusion product. Then define a fractional charge $Q_a\in\R/\Z$ as \cite{bulmashSymmFrac}
\begin{align}
    e^{2\pi i Q_a}:= \prod_{m=1}^{n-1}\eta_a\left(\frac{1}{n}, \frac{m}{n}\right)U_{\frac{1}{n}}(a,a^m; a^{m+1}),
\end{align}
where the elements of $\U=\R/\Z$ is labeled by the numbers in $[0,1)$. One can check that the quantity $e^{2\pi i Q_a}$ is gauge-invariant.
Since the map $[\rho]$ is trivial, we can fix a gauge where $U = 1$, so that $\eta$ satisfies ${\eta_c({\bf g}, {\bf h})}=\eta_a({\bf g}, {\bf h})\eta_b({\bf g}, {\bf h})$ when $N^{c}_{ab}>0$. One can then see that
\begin{align}
    e^{2\pi i Q_a}e^{2\pi i Q_b}=e^{2\pi i Q_c}\quad \text{when $N_{a,b}^c\neq 0$.} \label{eq:fusionu1charge}
\end{align}

In the gauge where $U = 1$, we can write the phases $\eta_a({\bf g}, {\bf h})$ as
\begin{align}
    \eta_a({\bf g},{\bf h}) = M_{a, \mathfrak{t}({\bf g},{\bf h})} ,
\end{align}
for $\mathfrak{t} \in Z^2(B\U, \mathcal{A})$. A representative 2-cocycle $\mathfrak{t}$ is given by
\begin{align}
    \mathfrak{t}({\bf g}, {\bf h}) = v^{{\bf g} + {\bf h} - [ {\bf g} + {\bf h}]} ,
\end{align}
where $v \in \mathcal{A}$ is referred to as the vison, $\mathbf{g},\mathbf{h}\in \R/\Z$ takes the values in $[0,1)$, and $[\mathbf{g}+\mathbf{h}]$ means the sum mod 1. 

The fractional charge can be written as \cite{cheng2016lsm}
\begin{align}
    e^{2\pi i Q_a} = M_{a,v} ,
\end{align}
in terms of which we then have 
\begin{align}
    \eta_a(\mathbf{g},\mathbf{h})=e^{2\pi iQ_a(\mathbf{g}+\mathbf{h}-[\mathbf{g}+\mathbf{h}])}.
\end{align}




    
    \subsubsection{Fermionic case: super-modular tensor category}
    \label{subsubsec:spinccharge}
    
    In fermionic topological phases, we consider the fermionic $G_f=\U^f=\R/\Z$ global symmetry given by the symmetry extension
    \begin{align}
        \Z_2^f\to G_f\to G_b,
    \end{align}
    where $G_b=\U$ is the bosonic symmetry group. Note that $G_b=\R/(\Z/2)$ has periodicity 1/2, since we want the fermionic group $G_f=\R/\Z$ to have periodicity 1.
    
    In this case, the symmetry fractionalization for the anyons is formulated by considering the symmetry action of $G_b$ on the super-modular category $\mathcal{C}$.
    
    We demand, as reviewed in Appendix~\ref{app:symmfrac}, that \cite{bulmashSymmFrac,aasen21ferm}
        \begin{align}
            U_{\bf g}(\psi,\psi;1) &= 1\\
            \eta_{\psi}({\bf g}, {\bf h}) &= \omega_2({\bf g}, {\bf h})
        \end{align}
        where $[\omega_2] \in H^2(BG_b,\mathbb{Z}_2)$ is the cohomology class specifying $G_f$ as a group extension of $G_b$ by $\mathbb{Z}_2$.
    
    For fermionic symmetry fractionalization defined for super-modular categories, the first step is to pick a map
    \begin{align}
        [\rho] : G_b \rightarrow \text{Aut}_{LR}(\mathcal{C}), 
    \end{align}
    where $\text{Aut}_{LR}(\mathcal{C})$ is the group of locality-respecting auto-equivalences of $\mathcal{C}$ \cite{bulmashSymmFrac,aasen21ferm}.  
    
    Given a choice of $[\omega_2]$ and $[\rho]$, there can be obstructions to defining any consistent pattern of symmetry fractionalization. In the case of $G_f = \U^f$, one can define symmetry fractionalization only for those super-modular categories $\mathcal{C}$ for which the map $\Upsilon_\psi$, is locality-respecting \cite{bulmashSymmFrac}. (see Appendix~\ref{app:symmfrac} for a definition of $\Upsilon_\psi$ and the locality respecting condition). For the case of $\U^f$, one can also show that the map $[\rho]$ must be the identity map for those super-modular categories that admit a consistent definition of symmetry fractionalization \cite{bulmashSymmFrac}. 
    
    For super-modular categories where there is no obstruction to defining $\U^f$ symmetry fractionalization, we can define fractional charges of anyons analogously to the bosonic case. For a fixed anyon $a$, let $n$ be the smallest integer such that $a^n$ contains the identity as a fusion product. Choose a sequence of anyons $a, a^2,\dots a^n=1$ such that $a\times a^k$ contains $a^{k+1}$ as a fusion product. Then define a fractional charge $Q_a\in\R/(2\Z)$ as \cite{bulmashSymmFrac}
\begin{align}
    e^{\pi i Q_a}:= \prod_{m=1}^{n-1}\eta_a\left(\frac{1}{2n}, \frac{m}{2n}\right)U_{\frac{1}{2n}}(a,a^m; a^{m+1}),
    \label{eq:chargedefspinc}
\end{align}
where the elements of $G_b$ is labeled by numbers in $[0,1/2)$.\footnote{Note that our normalizations are different by a factor of $2$ compared with \cite{bulmashSymmFrac}. }

    Since $[\rho]$ is the identity map, there is a gauge in which we can set $U = 1$, so that 
    $\eta$ satisfies ${\eta_c({\bf g}, {\bf h})}=\eta_a({\bf g}, {\bf h})\eta_b({\bf g}, {\bf h})$ when $N^{c}_{ab}>0$. Then one can see that $e^{\pi i Q_a}$ satisfies
      \begin{align}
    e^{\pi i Q_a}e^{\pi i Q_b}=e^{\pi i Q_c}\quad \text{when $N_{ab}^c\neq 0$}.
    \label{eq:fusionchargespinc}
\end{align}  
Since $\omega_2({\bf g}, {\bf h})=e^{2\pi i (\mathbf{g}+\mathbf{h}-[\mathbf{g}+\mathbf{h}]_{1/2})}$, the fractional charge of a transparent fermion $\psi$ satisfies
\begin{align}
    Q_{\psi}=1 \mod 2.
    \label{eq:Qpsi1}
\end{align}

For fixed fractional charges $e^{\pi i Q_a}$, an example of $\eta_a(\mathbf{g},\mathbf{h})$ realizing the fractional charges $e^{2\pi i Q_a}$ is given by
\begin{align}
    \eta_a(\mathbf{g},\mathbf{h})=e^{2\pi iQ_a(\mathbf{g}+\mathbf{h}-[\mathbf{g}+\mathbf{h}]_{1/2})},
\end{align}
where $\mathbf{g},\mathbf{h}\in \R/(\Z/2)$ takes the values in $[0,1/2)$, and $[\mathbf{g}+\mathbf{h}]_{1/2}$ means the sum mod 1/2. 


    \subsection{New graphical calculus with Villain symmetry defects}
    \label{subsec:nonflat}
    Here, we provide a new diagrammatic calculus to describe the action of U(1) symmetry defects on anyons of the BTC, by utilizing the Villain formulation of the curved U(1) background gauge field. This is done by extending the diagrammatics of U(1) symmetry fractionalization to the case with non-flat gauge fields described by the Villain gauge field $(h,c)$. The new diagrammatics takes the symmetry fractionalization data of the BTC characterized by the fractional charge $Q_a\in\R/\Z$ as an input. This is utilized to define a (3+1)D state sum path integral coupled with curved U(1) gauge fields, based on a given symmetry fractionalization data for a BTC.
    
\begin{figure}[htb]
    \centering
    \includegraphics{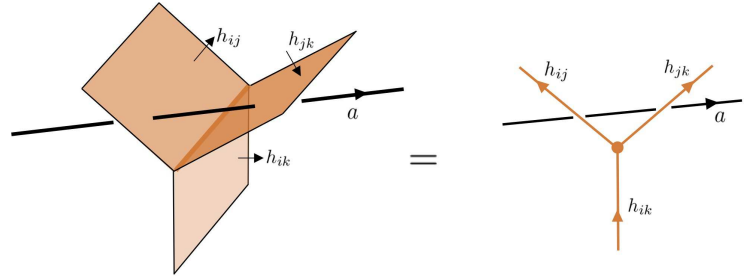}
    \caption{The $\mathbb{R}$ symmetry defect is realized as a codimension-1 plane in (2+1)D, and we describe the symmetry action on the anyons by the diagrammatics given by projecting the 3D picture onto a plane, where the $\mathbb{R}$ defects are expressed as orange lines. The orange bold dot at the junction of $\mathbb{R}$ defects indicates that $h$ is not closed in general and $\R$ defects can end at the junction.}
    \label{fig:projection}
\end{figure}

\begin{figure}[htb]
    \centering
    \includegraphics{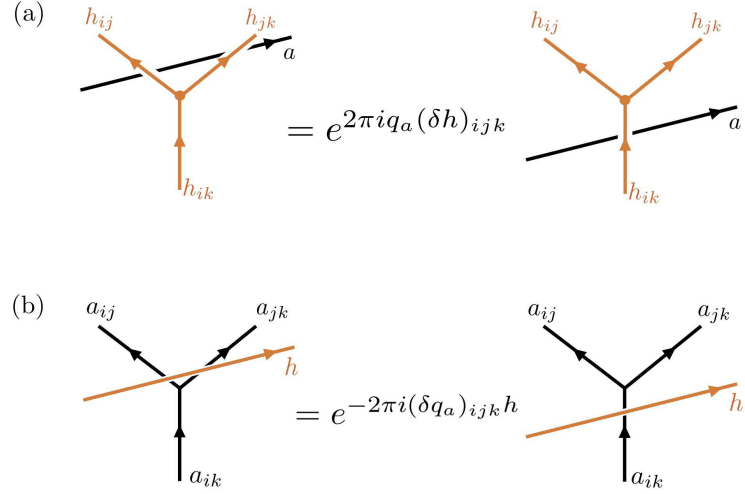}
    \caption{The graphical calculus for the $\R$ symmetry defects. (a) This figure reflects that the anyon $a$ is charged by $e^{2\pi i q_a h}$ under the $\R$ symmetry action. (b) The difference of the $\mathbb{R}$ symmetry action under fusion of anyons is expressed via the difference of $\mathbb{R}$ fractional charges $\delta q_a:= q(a_{ij})+q(a_{jk})-q(a_{ik})$.}
    \label{fig:Uetau1}
\end{figure}

    Let us consider the U(1) Villain gauge field $(h,c)\in C^1(M,\mathbb{R})\times Z^2(M,\mathbb{Z})$. Since the codimension-1 symmetry defect sheet (Poincar\'e dual of $h$) is now valued in $\mathbb{R}$, the fractional charge of anyons should also take value in $\mathbb{R}$. Let us denote the charge $q_a\in \mathbb{R}$ of an anyon $a$, which is defined by taking the $\mathbb{R}$ lift of $Q_a\in \mathbb{R}/\Z$. We describe the $\R$ symmetry action on anyons in the diagrammatics as shown in Fig.~\ref{fig:projection}, which projects the 3D configuration of defects onto a plane. The symmetry actions are then encoded in the diagrammatic calculus shown in Fig.~\ref{fig:Uetau1}. We note that the calculus shown in Fig.~\ref{fig:Uetau1} depends on a choice of the $\mathbb{R}$ lift $q_a\in \mathbb{R}$ of the fractional charge $Q_a\in \mathbb{R}/\Z$, but we will later see in Sec.~\ref{sec:statesumu1} that the resulting (3+1)D state sum topological invariant does not depend on the $\R$ lift of fractional charges.  Our (3+1)D state sum path integral hence gives a well-defined theory based on a given data of $\{Q_a\}$.
    
\begin{figure}[htb]
    \centering
    \includegraphics{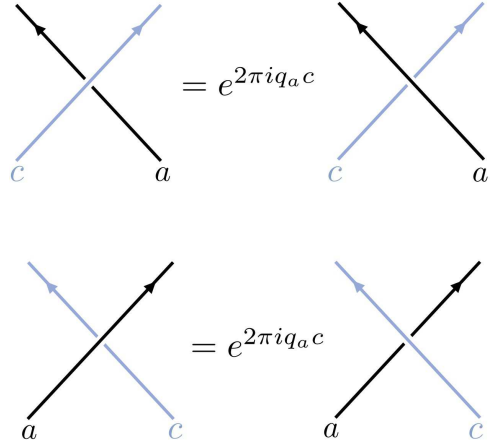}
    \caption{The action of the $\Z$ 1-form symmetry on anyons. The $c$ line acts on anyons as a 't Hooft line which emits Aharanov-Bohm phase by braiding with an anyon.}
    \label{fig:Zaction}
\end{figure}

    The 2-form field $c\in Z^2(M,\Z)$ is regarded as a defect line for a $\Z$ 1-form symmetry, which acts on an anyon by the mutual braiding, as shown in Fig.~\ref{fig:Zaction}. This corresponds to the Aharanov-Bohm phase between the unit magnetic flux carried by the defect line and the charged anyon.
    
\begin{figure}[htb]
    \centering
    \includegraphics{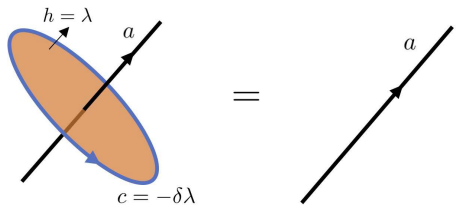}
    \caption{The local $\Z$ gauge transformation generates a bubble of symmetry defects $(h,c)=(\lambda,-\delta\lambda)$. This bubble acts trivially on anyons since the 0-form symmetry defect acts as $e^{2\pi i\lambda}$ by crossing with the end of the defect, while $c$ acts as $e^{-2\pi i\lambda}$ by braiding, and these contributions cancel out. This makes our diagrammatics consistent with the invariance under the $\Z$ gauge transformation.}
    \label{fig:gaugeinv}
\end{figure}

We have to check that the symmetry actions of the U(1) symmetry defect $(h,c)$ are independent of the gauge transformation of the $\Z$ 1-form symmetry $(h,c)\to (h+\lambda, c-\delta\lambda)$ with $\lambda\in C^1(M,\Z)$, which is regarded as redundancy to express the same configuration of U(1) background gauge field.
This puts a constraint on the fractional charges $\{q_a\}$ carried by anyons. To see this, consider a shift of the symmetry defect $h\to h+\delta\Lambda$ by $\Lambda\in C^0(M,\Z)$ using the gauge transformation $\lambda=\delta\Lambda$. Since $c$ is invariant under this transformation, the process does not generate the defect line for $c$. When the $h$ defect shifts across the fusion vertex of anyons as shown in Fig.~\ref{fig:Uetau1} (b) by this transformation, it emits a phase $e^{-2\pi i (\delta q)\delta\Lambda}$, which should always be 1. We hence require that 
\begin{align}
    e^{2\pi i q_a}e^{2\pi i q_b}=e^{2\pi i q_c}\quad \text{when $N_{a,b}^c\neq 0$,}
\label{eq:fusioncharge}
\end{align}
which is satisfied by Eq.~\eqref{eq:fusionu1charge}, since $q_a$ is $\R$ lift of $Q_a$.
We note that the symmetry actions of the defects $(h,c)$ on anyons illustrated in Figs.~\ref{fig:projection},~\ref{fig:Zaction} are consistent with the gauge invariance under $(h,c)\to (h+\lambda, c-\delta\lambda)$ with $\lambda\in C^1(M,\Z)$. To see this, consider shifting the gauge field locally by $(h,c)=(\lambda,-\delta\lambda)$, which generates a small disc of a symmetry defect $h=\lambda$ in the spacetime, bounded by the $c=-\delta\lambda$ defect line. One can see that this object acts trivially on anyons, since $h$ acts as $e^{2\pi i\lambda}$, while $c$ acts as $e^{-2\pi i\lambda}$ by braiding, and these contributions cancel out. See Fig.~\ref{fig:gaugeinv}.

 The background gauge field $(h,c)\in C^1(M,\mathbb{R})\times Z^2(M,\mathbb{Z})$ is regarded as the symmetry defect of a $\R$ 0-form symmetry and $\Z$ 1-form symmetry respectively. Our calculus is then understood as encoding the action of $\Z$ 1-form symmetry in addition to the 0-form $\R$ symmetry action, in a way respecting the gauge invariance Eq.~\eqref{eq:gaugetrans}.  We will see that this diagrammatic calculus makes it possible to define a (3+1)D state-sum topological invariant correctly coupled with the Villain U(1) gauge field.
 
 The diagrammatic calculus developed in this context is therefore beyond the usual one used to describe symmetry fractionalization \cite{barkeshli2019,bulmashSymmFrac,aasen21ferm}, which only considers flat background gauge fields for the 0-form symmetries. Note that the full formalism of G-crossed BTCs does incorporate non-flat gauge fields, corresponding to cases in which the codimension-1 symmetry defects end at a codimension-2 line in space-time; however, the precise relationship between the formalism we have developed here and the usual one for G-crossed BTCs is unclear. 

\subsubsection*{Villain symmetry defects for $\U^f$ symmetry of super-modular tensor categories}
\label{subsec:nonflatspinc}

In the case of super-modular categories with $\U^f$ symmetry, the local fermion $\psi$ carries charge $1$ under the $\U^f$ symmetry. Under the bosonic $\U = \U^f/\Z_2^f$ symmetry, we choose the convention that $\psi$ continues to carry charge $1$, so that the bosonic $\U = \R/(\Z/2)$, which is half the quantization of the $\U^f$. 

  In this case, we use the Villain gauge fields $(h,c)$ to describe the background $\U$ gauge field. The 
  $\mathbb{R}$ symmetry action of the $h$ defects on an anyon $a\in\mathcal{C}$ are again described in terms of the fractional charges $q_a\in\mathbb{R}$, which in this case is defined by lifting the U(1) fractional charges $Q_a\in\R/(2\Z)$ to $\R$.
  
  The symmetry defects act on anyons according to the diagrammatics shown in Fig.~\ref{fig:projection},~\ref{fig:Zaction}. The fractional charges are subject to the constraint
  \begin{align}
      e^{\pi i q_a}e^{\pi i q_b}=e^{\pi i q_c}\quad \text{when $N_{a,b}^c\neq 0$,}
      \label{eq:fusionchargeliftedspinc}
  \end{align}
reflecting Eq.~\eqref{eq:fusionchargespinc}.

Due to Eq.~\eqref{eq:Qpsi1}, the fractional charge of the transparent fermion $\psi$ satisfies
  \begin{align}
      q_{\psi}=1\quad \mod 2.
      \label{eq:fermionchargelifted}
  \end{align}

\section{Bosonic state sum of (3+1)D TQFT with U(1) symmetry}
\label{sec:statesumu1}

\begin{figure}[htb]
    \centering
    \includegraphics{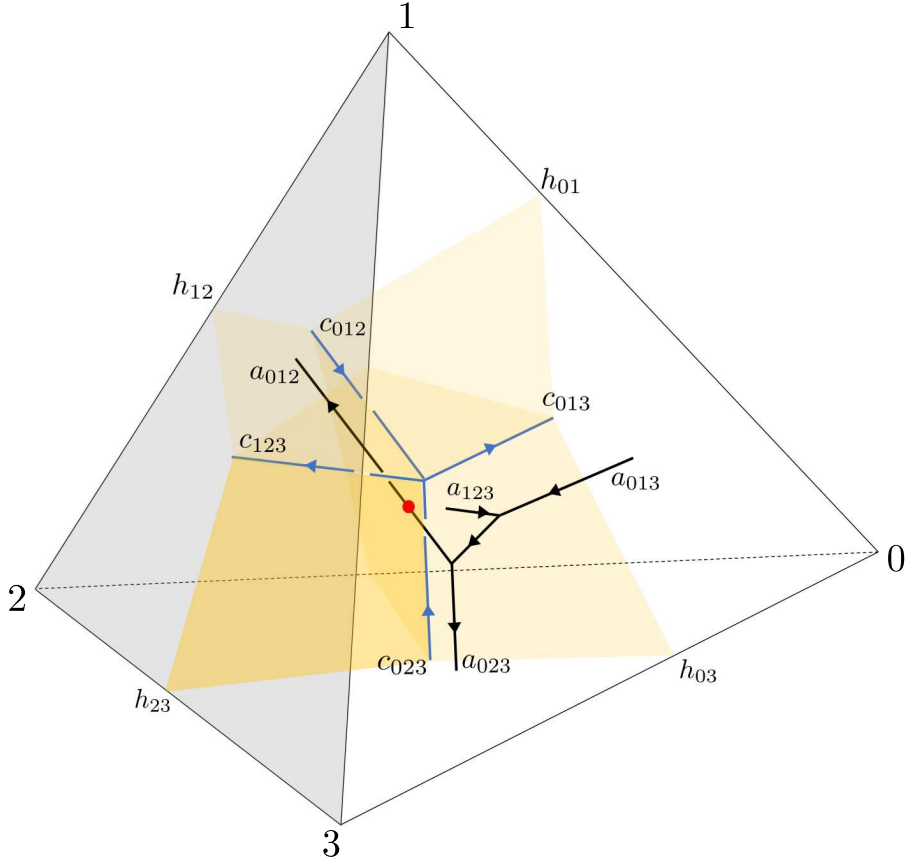}
    \caption{Anyon lines and gauge fields on a 3-simplex. $h_{ij}$ is described as a yellow surface, $c_{ijk}$ is a blue line, and anyons $a_{ijk}$ are black lines Poincar\'e dual of $2$-simplices. $a_{ijk}$ lines are dislocated from $c_{ijk}$ lines, and intersect with a $h_{23}$ surface at a red point. The directions of $c$ lines are assigned so that $c+\delta h$ becomes a gauge-invariant field strength. That is, the direction of $c$ is the same as that of $\delta h$ which is given by the ``right-handed screw'' rule on each 2-simplex, e.g., directed outwards on $\braket{123}$.}
    \label{fig:3simplexu1}
\end{figure}

In this section, we provide a path integral state sum for a (3+1)D bosonic topological phase coupled with U(1) Villain gauge field. The input data to our construction of a bosonic state sum is a unitary modular tensor category (UMTC) $\mathcal{C}$, U(1) gauge field $(h,c)$, and U(1) charge $Q_a\in\R/\Z$ assigned for each anyon $a\in\mathcal{C}$.
Given a 4-manifold $M^4$, we pick a triangulation and a branching structure (i.e. a local ordering of vertices). 
Each simplex can then be either a $+$ simplex or a $-$ simplex, depending on whether the ordering agrees with the orientation or not. 
We associate to each simplex the following data:
\begin{itemize}
    \item $1$-simplex $ij$: a link variable $h_{ij} \in \R$
    \item $2$-simplex $ijk$: a face variable $c_{ijk}\in \Z$, an anyon $a_{ijk}\in\mathcal{C}$
    \item $3$-simplex $ijkl$: an anyon $b_{ijkl}\in\mathcal{C}$ obeying certain rules
\end{itemize}
The 3-simplex data assumes $N_{ij}^k \leq 1$, and is determined as follows. Consider a particular $3$-simplex $\langle 0123\rangle$, with link variables $h_{ij}\in\R$ on its 1-simplices, anyons $a_{ijk}$ and face variables $c_{ijk}\in\Z$ on its 2-simplices. This is shown in Fig.~\ref{fig:3simplexu1}.
Then we demand that the anyon $b_{0123}$ placed on this 3-simplex obeys
\begin{equation}
N_{a_{023},a_{012}}^{b_{0123}} \neq 0 \text{ and } N_{a_{013}, a_{123}}^{b_{0123}} \neq 0
\label{eqn:3simplexRule}
\end{equation} 
where $N_{a,b}^c$ are fusion coefficients. 

\begin{figure}[htb]
    \centering
    \includegraphics{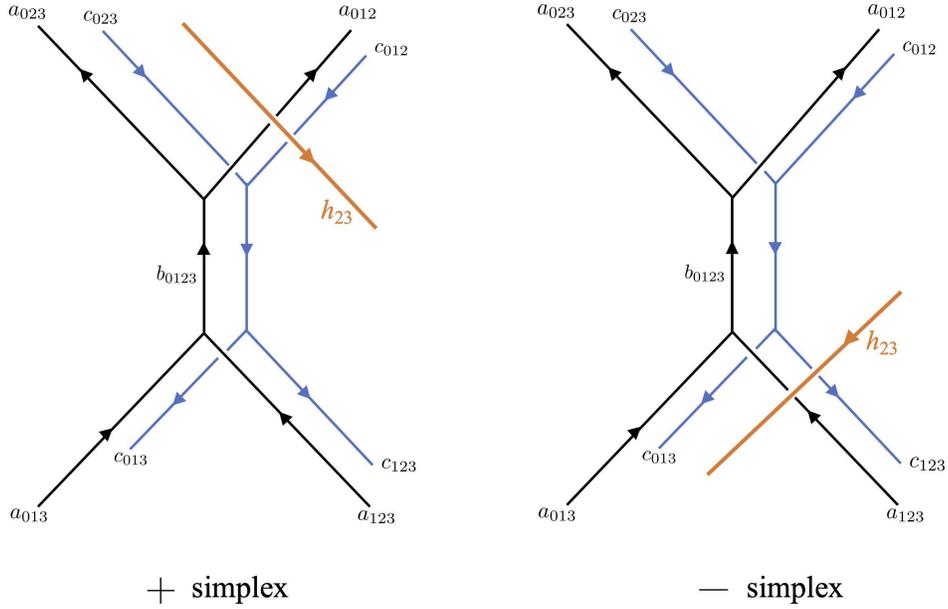}
    \caption{Graphical calculus for a 3-simplex with $+$ or $-$ orientation. The orange line denotes the symmetry defect of the 0-form $\mathbb{R}$ symmetry, while the blue line denotes the symmetry defect of the 1-form $\mathbb{Z}$ symmetry.}
    \label{fig:15jsnippet}
\end{figure}

In the language of category theory, the 2-simplex data is a simple object of $\mathcal{C}$, and the data on the 3-simplex is an element of the space Hom$\left(a_{013} \otimes a_{123}, a_{013} \otimes a_{012}\right)$, i.e. an element of $\bigoplus_b V^b_{013,123} \otimes V_b^{023,{012}}$. In other words, if we had allowed $N_{ab}^c > 1$, we would also need to associate elements of a fusion space and a splitting space to the 3-simplices; this generalization is straightforward. The picture for each 3-simplex in Fig.~\ref{fig:3simplexu1} is projected onto a 2d plane, and then we get a graphical calculus for each 3-simplex as described in Fig.~\ref{fig:15jsnippet}.

\begin{figure}[htb]
    \centering
    \includegraphics{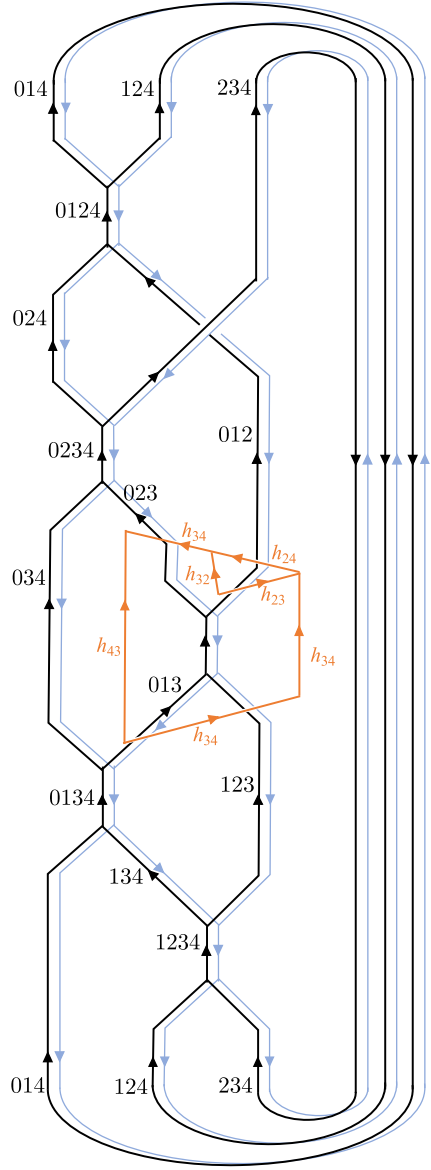}
    \caption{15j symbol for a $+$ simplex. The blue lines represent the dual of the $c$ field. The blue lines are behind the black lines of anyons, except for the the case that $c_{234}$ line crosses over the $a_{234}$ line. The 15j symbol of a $-$ simplex gives the complex conjugate for that of a $+$ simplex.}
    \label{fig:15jsymbol}
\end{figure}

We now define our path integral. Given a labeling of all the simplices, we will compute an anyon diagram for each 4-simplex based on the labeling. We multiply the results, and then sum over all labelings. Let $\Delta^4$ be a 4-simplex, and let $\epsilon(\Delta^4) = \pm$ be its orientation. Then define complex numbers $Z^{\epsilon(\Delta^4)}(\Delta^4)$ in the following diagrammatic way, shown in Fig.~\ref{fig:15jsymbol}. For each 3-simplex in $\Delta^4$, take the diagram obtained in Fig.~\ref{fig:15jsnippet} given the induced orientation on that 3-simplex. Lay out these five diagrams in a plane such that anyon lines shared between two 3-simplices are near each other. Then connect up all of the lines and symmetry defects, sliding symmetry defects along anyon lines and bending symmetry defects far from anyon lines as necessary to obtain a closed diagram. These anyon diagrams arise from projecting the boundary 3-simplices, which form a triangulation of $S^3$, into the plane. 
There are an arbitrary choice in this definition of the 15j symbol; in Fig.~\ref{fig:15jsnippet}, we could have resolved the four-fold fusion in a different channel, related by an $F$-move to our current choice. In~\cite{bulmash2020}, it was shown that the partition function does not depend on the choice in the definition. In addition, there are choices for resolving the blue $c$ lines in Fig.~\ref{fig:15jsnippet}; we could have resolved the blue lines on the left of the black anyon lines. One can easily find that the choice does not affect on the value of the 15j symbol. Also, one can notice that the angles of orange $h$ lines are freely changed in going from Fig.~\ref{fig:15jsnippet} to Fig.~\ref{fig:15jsymbol}. These changes of angles do not affect on the value of the 15j symbol either, and one can freely bend the $h$ lines to obtain Fig.~\ref{fig:15jsymbol}.

The evaluation of the $15$j symbol involves the symmetry action of Villain gauge fields on anyons defined in Sec.~\ref{subsec:nonflat}. This requires us to define a $\R$ lift $q_a\in\R$ of the fractional charge $Q_a$ that satisfies $q_a=Q_a$ mod 1.
We can regard the fractional charge of an anyon $a_{ijk}$, $q_{ijk}: \langle ijk\rangle\to \mathbb{R}$ as a 2-cochain $q\in C^2(M^4,\mathbb{R})$, satisfying $(\delta q)_{0123}=q_{123}-q_{023}+q_{013}-q_{012}=0$ mod 1 on each 3-simplex due to the charge conservation~\eqref{eq:fusioncharge}.

The $15$j symbol is now computed as
\begin{align}
\label{Z0U1}
\begin{split}
    Z^{+}(01234)&=
    e^{-2\pi i q_{012}(c+\delta h)_{234}-2\pi i (\delta q)_{0123}h_{34}}Z_0^{+}(01234), \\
    Z^{-}(01234)&=
    e^{2\pi i q_{012}(c+\delta h)_{234}+2\pi i (\delta q)_{0123}h_{34}}Z_0^{-}(01234), 
\end{split}
\end{align}
We note that the factor $e^{-2\pi i q_{012}(c+\delta h)_{234}-2\pi i (\delta q)_{0123}h_{34}}$ can be expressed as $e^{-2\pi i (q\cup(c+\delta h)-2\pi i (\delta q)\cup h)_{01234}}$ using the cup product.
$Z_0^{\pm}(01234)$ is the $15$j symbol without any symmetry defects~\cite{crane1993},
    \begin{widetext}
\begin{align}
Z^+_0(01234) = \mathcal{N}_{01234} \sum_{d,a \in \mathcal{C}} &F^{024,234,012}_{d,0234,a} R_a^{012,234} (F^{024,012,234}_{d})^{-1}_{a,0124}F^{014,124,234}_{d,0124,1234}\times \nonumber \\
&\times (F^{014,134,123}_{d})^{-1}_{1234,0134}F^{034,013,123}_{d,0134,0123} \times (F^{034,023,012}_{d})^{-1}_{0123,0234}
\label{eqn:Zplusdef}
\\
Z^-_0(01234) = \mathcal{N}_{01234} \sum_{d,a \in \mathcal{C}} &(F^{024,234,012}_{d})^{-1}_{a,0234} (R_a^{012,234})^{-1} F^{024,012,234}_{d,0124,a}\left(F^{014,124,234}_{d}\right)^{-1}_{1234,0124}\times \nonumber\\
&\times F^{014,134,123}_{d,0134,1234} \left(F^{034,013,123}_d\right)^{-1}_{0123,0134}\times F^{034,023,012}_{d,0234,0123}
\label{eqn:Zminusdef}
\end{align}
\end{widetext}
where $\{F, R\}$ are $F$ and $R$ symbols of BTC given in Appendix~\ref{sec:anyon}.
The normalization factor $\mathcal{N}_{01234}$ is given by
\begin{equation}
\mathcal{N}_{01234} = \sqrt{\frac{\prod_{\Delta^3 \in 3\text{-simplices}} d_{b_{\Delta^3}}}{\prod_{\Delta^2 \in 2\text{-simplices}}d_{a_{\Delta^2}}}} ,
\label{eqn:normalization}
\end{equation}
where the products are over $3$-simplices $\Delta^3$ and $2$-simplices $\Delta^2$ of the $4$-simplex $01234$, and $d_a$ is the quantum dimension of anyon $a$. 
We may now define the path integral. On a closed 4-manifold $M^4$, we define
\begin{widetext}
\begin{equation}
\begin{split}
Z(M^4,(h,c)) &= \sum_{\{a,b\}} {\mathcal{D}^{2(N_0-N_1)-\chi}}\frac{\prod_{\Delta^2 \in \mathcal{T}^2}d_{a_{\Delta^2}}\prod_{\Delta^4 \in \mathcal{T}^4}Z^{\epsilon(\Delta^4)}(\Delta^4)}{\prod_{\Delta^3 \in \mathcal{T}^3}d_{b_{\Delta^3}}} \\
&=\sum_{\{a,b\}} {\mathcal{D}^{2(N_0-N_1)-\chi}}\frac{\prod_{\Delta^2 \in \mathcal{T}^2}d_{a_{\Delta^2}}\prod_{\Delta^4 \in \mathcal{T}^4}Z_0^{\epsilon(\Delta^4)}(\Delta^4)}{\prod_{\Delta^3 \in \mathcal{T}^3}d_{b_{\Delta^3}}}
\prod_{\Delta^4 \in \mathcal{T}^4}e^{-2\pi i\epsilon(\Delta^4)(q\cup(c+\delta h)+\delta q\cup h)}.
\label{eqn:Zclosed}
\end{split}
\end{equation}
\end{widetext}
Here $\mathcal{T}^k$ denotes the set of $k$-simplices and $N_k = |\mathcal{T}^k|$ is the number of $k$-simplices in the triangulation. $\chi$ is the Euler characteristic of $M^4$. The sum is over all possible labelings of the anyons $\{a\}$, $\{b\}$ on the $2$- and $3$-simplices.

In Appendix~\ref{app:pachner}, we show that the above partition function is invariant under the Pachner move (i.e., re-triangulations), which guarantees the topological invariance of $Z(M^4,(h,c))$.

We note that the partition function $Z(M^4,(h,c))$ on a closed manifold is independent of the choice of $\R$ lift $q_a$ of the fractional charges, since the change of the lift $q\to q+\Lambda$ with $\Lambda\in C^2(M^4,\Z)$ shifts the partition function by
\begin{align}
    e^{-2\pi i\int (\Lambda\cup(c+\delta h)+\delta\Lambda\cup h)}=e^{-2\pi i\int \delta(\Lambda\cup h)}=1.
\end{align}
The path integral thus gives a well-defined (3+1)D invariant based on the data of U(1) fractional charge $Q_a\in\R/\Z$.

Since the input category $\mathcal{C}$ is a modular tensor category, we expect that the resulting (3+1)D path integral defines an invertible TQFT, i.e. $|Z(M^4, (h,c))| = 1$. When $(h,c) = 0$, the above path integral reduces to the Crane-Yetter path integral \cite{crane1993}, which is known to define an invertible TQFT when $\mathcal{C}$ is modular. It is generally expected that if $|Z(M^4, A = 0)| = 1$, for any background gauge field $A$, then $|Z(M^4, A)| = 1$ for $A \neq 0$ as well, although we are not aware of a rigorous proof. This is indeed borne out in a variety of examples studied in Sec.~\ref{sec:evaluation} and in closely related constructions studied in previous works~\cite{bulmash2020, tata2021anomalies}. 

Since we expect that $Z$ defines an invertible TQFT, it should be a bordism invariant, and so we expect that in general the path integral defined above evaluates to:
\begin{align}
    Z(M^4, (h,c) ) = e^{\frac{2\pi i}{8} c_- \sigma(M^4) + i \Theta C_2},
\end{align}
where $\sigma(M^4)$ is the signature of $M^4$, $c_-$ is the chiral central charge, $C_2$ is the total second Chern number of the $\U$ bundle determined from $c$. $\Theta$ defines the electromagnetic theta term, which, as discussed in Section \ref{summarySec}, is determined by the topological twist of the vison, $e^{i \Theta/2} = \theta_v^*$. 

Finally, we note that our construction above assumed that $\mathcal{C}$ is modular, however this was not really an essential ingredient. As long as we have a set of charges $Q_a$ for each anyon $a$ that is compatible with the fusion rules, we can see that the above path integral will be well-defined and topologically invariant. What is less clear, if $\mathcal{C}$ is not modular, is when such a set of charges can be defined, and whether it provides an exhaustive characterization of $\U$ symmetry fractionalization.

\section{Bosonic state sum of (3+1)D TQFT with $\U\rtimes H$ symmetry}
\label{sec:u1rtimesg}

We can consider bosonic topological phases with $\U\rtimes H$ global symmetry, by incorporating into the state sum the flat background gauge field of the symmetry group $H$ following~\cite{bulmash2020}. 

The $H$ symmetry action is classified by a $\mathbb{Z}_2$ grading corresponding to whether $\mathbf{g}\in H$ has a unitary or anti-unitary action on the category:
\begin{align} \label{antiunitaryactionmain}
s(\mathbf{g}) = \left\{
\begin{array} {ll}
+1 & \text{if $\mathbf{g}$ is unitary} \\
-1  & \text{if $\mathbf{g}$ is anti-unitary} \\
\end{array} \right.
\end{align}

The symmetry group $\U\rtimes H$ is characterized by the $H$-action on U(1), corresponding to whether the group elements of $H$ complex conjugate the U(1) elements. This is given by the map
\begin{align}
    \sigma_{\mathbf{g}}: (h,c)\to \sigma(\mathbf{g})\cdot(h,c),
\end{align}
with $\sigma(\mathbf{g})\in\{\pm 1\}$.
Then, the $\U\rtimes H$ gauge field is represented as the U(1) gauge field, defined by the pair $h\in C^1(M,\mathbb{R}), c\in Z_{\sigma}^2(M,\Z)$, together with a flat $H$ gauge field $A_{H}$. Here, $Z_{\sigma}^d(M,\Z)$ means the twisted cocycle by the $H$ action, see Appendix~\ref{app:twisted} for an explanation of twisted cohomology.
Accordingly, the U(1) gauge transformation in the presence of an $H$ gauge field also gets twisted by the $H$ action as
\begin{align}
    c\to c-\delta_{\sigma}\lambda,\quad h\to h+\lambda+\delta_{\sigma}\chi
\end{align}
with $\lambda\in C^1(M,\mathbb{Z})$, $\chi\in C^0(M,\mathbb{R})$.
Since $H$ acts on the U(1) by conjugation, in the diagrammatics the $H$ defect acts on the labels of anyons together with the U(1) gauge field $(h,c)$ when the $H$ defect crosses over the $h$ and $c$ lines.

As explained in Appendix~\ref{app:tilde}, the graphical calculus of Villain gauge fields in the presence of $H$ defects is defined based on the $\U\rtimes H$ symmetry fractionalization data. This is constructed by the U(1) fractional charge $Q_a$ satisfying the transformation law
\begin{align}
Q_{^{\mathbf{g}}a}=s(\mathbf{g})\sigma(\mathbf{g})Q_{a} \mod 1,
\label{eq:Qtransformmain}
\end{align}
and the symmetry fractionalization data $U,\eta$ for $H$ symmetry. Hence, our construction takes as input a modular tensor category $\mathcal{C}$ describing the bosonic topological order, symmetry fractionalization data $U$ and $\eta$ for $H$ on $\mathcal{C}$, and the fractional charge of anyons $Q_a\in \R/\Z$ for U(1) symmetry satisfying Eq.~\eqref{eq:Qtransformmain}. 
See Appendix~\ref{app:tilde} for the construction of $\U\rtimes H$ symmetry fractionalization data and the Villain calculus in the presence of $H$ symmetry defects.

\begin{figure}[htb]
    \centering
    \includegraphics{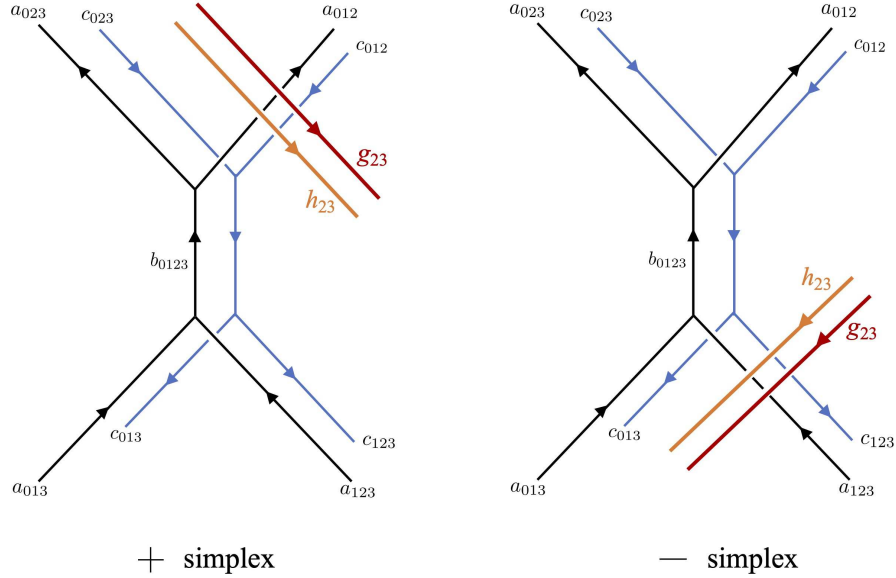}
    \caption{Graphical calculus for a 3-simplex with $+$ or $-$ orientation. The red line labeled by $\mathbf{g}_{23}\in H$ denotes the $H$-symmetry defect.}
    \label{fig:15jsnippetG}
\end{figure}

\begin{figure}[htb]
    \centering
    \includegraphics{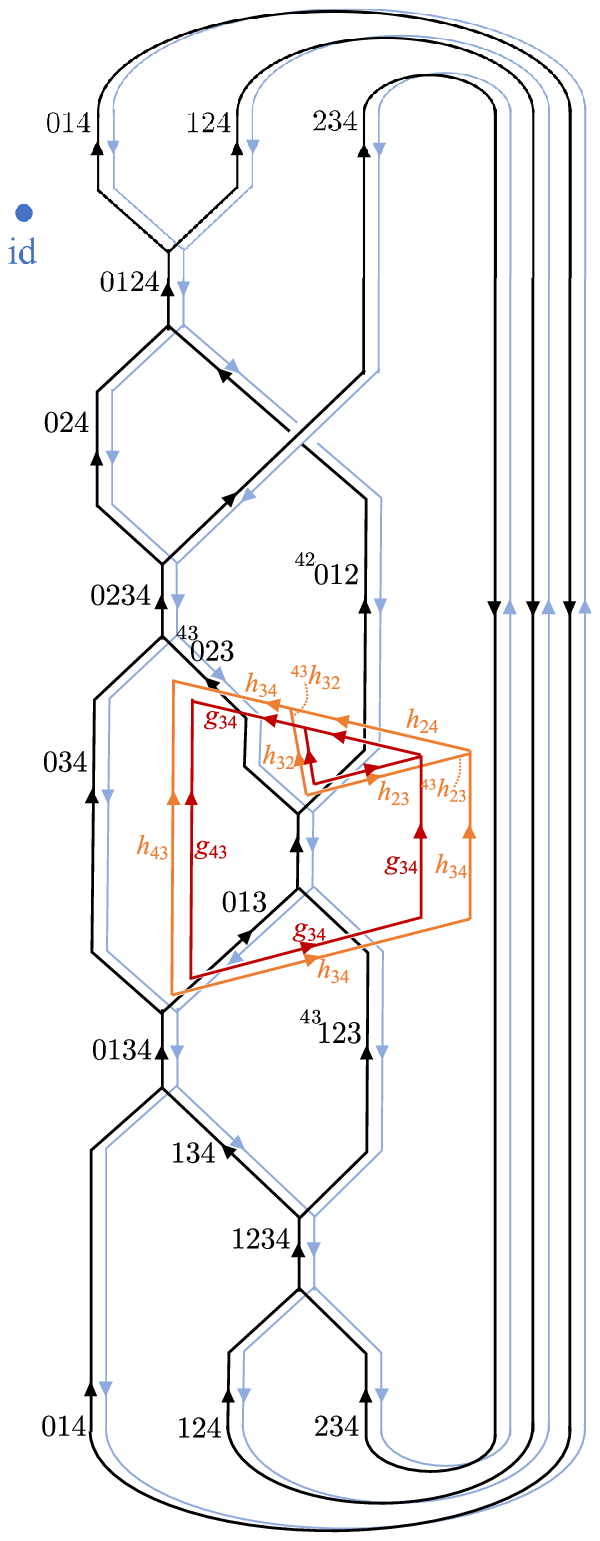}
    \caption{15j symbol for a $+$ simplex. The blue lines represent the dual of the $c$ field. The blue lines are behind the black lines of anyons, except for the the case that $c_{234}$ line crosses over the $a_{234}$ line. Here id means the identity in $H$, and the blue dot denotes the background domain of $H$ symmetry, see Appendix~\ref{app:symmfrac},~\ref{app:tilde} for the graphical calculus.}
    \label{fig:15jsymbolG}
\end{figure}

 The 15j symbol on a 4-simplex is given by connecting the diagrams on each boundary 3-simplex shown in Fig.~\ref{fig:15jsnippetG}. The resulting 15j symbol is given by Fig.~\ref{fig:15jsymbolG}. On a $+$ or $-$ 4-simplex, the 15j symbol evaluates as

\begin{align}
\begin{split}
    Z^{+}(01234)&=e^{-2\pi i ^{42}q_{012}(c+\delta_{\sigma}h)_{234}-2\pi i ^{43}(\delta_{s\sigma}q)_{0123}h_{34}}\cdot\frac{U_{34}(013,123,0123)^{s(34)}}{U_{34}(023,^{32}012,0123)^{s(34)}\eta_{012}(23,34)^{s(24)}}\cdot Z_0^{+}(01234), \\
    Z^{-}(01234)&=e^{2\pi i ^{42}q_{012}(c+\delta_{\sigma}h)_{234}+2\pi i ^{43}(\delta_{s\sigma}q)_{0123}h_{34}}\cdot\frac{U_{34}(023,^{32}012,0123)^{s(34)}\eta_{012}(23,34)^{s(24)}}{U_{34}(013,123,0123)^{s(34)}}\cdot Z_0^{-}(01234), 
\end{split}
\end{align}
where $^{lm}q_{ijk}$ means the $\mathbf{g}_{lm}$ action on the U(1) charge, 
$^{lm}q_{ijk}:=q(^{lm}a_{ijk})$.
Here, $q_{a}$ satisfies
\begin{align}
^{ji}q_{a}=s(\mathbf{g}_{ji})\sigma(\mathbf{g}_{ji})q_{a}.
\label{eq:gactioncharge}
\end{align}
which follows from Eq.~\eqref{eq:Qtransformmain}.
Then, the $\R$-lifted fractional charge $q_{ijk}:\langle ijk\rangle\to \R$ of anyons can be regarded as a 2-cochain $q\in C^2(M^4,\R)$ that satisfies $\delta_{s\sigma}q=0$ mod 1, where $\delta_{s\sigma}$ is a twisted coboundary defined by the $H$ action on $q_a$ in Eq.~\eqref{eq:gactioncharge}.
We note that the first term is neatly expressed as twisted cup product (see Appendix~\ref{app:twisted})
\begin{align}
    e^{\mp2\pi i ( q\cup_{s\sigma}(c+\delta_{\sigma}h)+ \delta_{s\sigma}q\cup_{s\sigma}h)}.
\end{align}
Then
\begin{widetext}
\begin{align}
Z^+_0(01234) = \mathcal{N}_{01234} \sum_{d,a \in \mathcal{C}} &F^{024,234,^{42}012}_{d,0234,a} R_a^{^{42}012,234} (F^{024,^{42}012,234}_{d})^{-1}_{a,0124}F^{014,124,234}_{d,0124,1234}\times \nonumber \\
&\times (F^{014,134,^{43}123}_{d})^{-1}_{1234,0134}F^{034,^{43}013,^{43}123}_{d,0134,^{43}0123} \times (F^{034,^{43}023,^{42}012}_{d})^{-1}_{^{43}0123,0234}
\label{eqn:ZplusdefG}
\\
Z^-_0(01234) = \mathcal{N}_{01234} \sum_{d,a \in \mathcal{C}} &(F^{024,234,^{42}012}_{d})^{-1}_{a,0234} (R_a^{^{42}012,234})^{-1} F^{024,^{42}012,234}_{d,0124,a}\left(F^{014,124,234}_{d}\right)^{-1}_{1234,0124}\times \nonumber\\
&\times F^{014,134,^{43}123}_{d,0134,1234} \left(F^{034,^{43}013,^{43}123}_d\right)^{-1}_{^{43}0123,0134}\times F^{034,^{43}023,^{42}012}_{d,0234,^{43}0123}
\label{eqn:ZminusdefG}
\end{align}
\end{widetext}

In the presence of the defects for anti-unitary symmetry $H$, the space-time manifold $M^4$ in general becomes unoriented. In that case, the anti-unitary symmetry defect is realized as the Poincar\'e dual of the first Stiefel-Whitney class by $A_H^*s=w_1$, reflecting that the symmetry defect reverses the orientation of the space-time manifold. See~\cite{bulmash2020, tata2021anomalies} for detailed description.

The partition function is then given by
\begin{widetext}
\begin{equation}
\begin{split}
Z(M^4,(h,c),A_{H}) &= \sum_{\{a,b\}} {\mathcal{D}^{2(N_0-N_1)-\chi}}\frac{\prod_{\Delta^2 \in \mathcal{T}^2}d_{a_{\Delta^2}}\prod_{\Delta^4 \in \mathcal{T}^4}Z^{\epsilon(\Delta^4)}(\Delta^4)}{\prod_{\Delta^3 \in \mathcal{T}^3}d_{b_{\Delta^3}}}.
\label{eqn:Zclosedtilde}
\end{split}
\end{equation}
\end{widetext}
We show the invariance of the path integral under the Pachner move in Appendix~\ref{app:pachner}, which shows that our path integral defines a topological invariant of $M^4$.

To check that the partition function is independent of the $\R$ lift of the fractional charge $Q_a$, let us study the effect of a local redefinition of the charge $q_a\to q_a+\Lambda$, where $\Lambda\in C^2(M^4,\Z)$ is supported on a single 2-simplex.
Since the partition function is invariant under $H$ gauge transformation, one can move the $H$ symmetry defect away from $\Lambda$. Then, the shift of the partition function is given by
\begin{align}
    \prod_{\Delta^4} e^{-2\pi i\epsilon(\Delta^4) \delta(\Lambda\cup h)}=1.
\end{align}
This shows that the path integral gives a well-defined (3+1)D invariant based on the data of U(1) fractional charge $Q_a\in\R/\Z$.

For the same reason discussed in the preceding section, the (3+1)D path integral $Z(M^4, (h,c), A_H)$ is expected to define an invertible TQFT with $G = \U \rtimes H$ symmetry. Therefore $Z(M^4, (h,c), A_H)$ is expected to be a smooth bordism invariant, and can be viewed as defining an element of the Pontyragin dual of the smooth bordism group for 4-manifolds equipped with background $G$ gauge fields.

\section{Construction of fermionic topological phases: fermion condensation}
\label{sec:condensation}

Our constructions described above can also be extended to describe fermionic topological phases with $\U$ symmetry, where the fermion carries charge under the $\U$, following the general procedure of \cite{tata2021anomalies,Gaiotto:2015zta}. In this case, the input into the construction is the data of a (2+1)D fermionic topological order, characterized by a super-modular category, and fermionic symmetry fractionalization \cite{bulmashSymmFrac,aasen21ferm}. 

Since we are interested in fermionic topological phases with $\U^f$ symmetry, our resulting TQFT will depend on a $\Spin^c=(\Spin\times \U)/\Z_2$ structure on our space-time 4-manifold $M^4$. More generally, the full fermionic symmetry group, including the space-time symmetry, is a group extension $\Z_2^f\to\mathcal{G}_f\to \mathcal{G}_b$, corresponding to a lift of the bosonic symmetry group $\mathcal{G}_b = \SO(d) \rtimes G_b$. To define such a lift, the fermionic theory in general requires a variant of the Spin$^c$ structure, twisted by the elements of $G_b$, which we refer to as a $\mathcal{G}_f$-structure, $\xi_{\mathcal{G}}$ (see Eq.~\eqref{Gfstructure}). A general review of  $\mathcal{G}_f$-structures is provided in Appendix of~\cite{tata2021anomalies}. The internal bosonic symmetry group $G_b= \mathcal{G}_b/\SO(d)$, and recall that the internal fermionic symmetry group $G_f$ is a $\Z_2^f$ extension of $G_b$, $\Z_2^f\to G_f\to G_b$, characterized by $[\omega_2]\in H^2(BG_b,\Z_2)$.

The construction proceeds by starting with a path integral $Z_b(M^4, A_b)$ for a bosonic topological phase, which is referred to as the bosonic shadow. $Z_b$ is defined by using the constructions in the preceding sections, with a super-modular category and symmetry fractionalization data as input. Here $A_b$ is a background gauge field for the bosonic symmetry group $G_b$. Since the super-modular category has an invisible particle, the fermion $\psi$, the resulting (3+1)D TQFT is non-invertible, and corresponds to a dynamical (3+1)D $\Z_2$ gauge theory with a single topologically non-trivial point-like particle, which is a fermion (see \cite{tata2021anomalies} for an extended discussion in the context of flat $G_b$ bundles). To obtain an invertible fermionic topological phase, i.e a fermionic SPT, we must condense the fermion to Higgs the $\Z_2$ gauge field. The fermion condensation can be done by defining the bosonic path integral $Z_b$ in the presence of fermion world-lines, and then summing over all possible configurations of wordlines. Mathematically, this is achieved by coupling the theory to a background $3$-form $\Z_2$ gauge field $f_3 \in Z^3(M^4, \Z_2)$, whose Poincare dual defines the fermion worldlines, to obtain $Z_b(M^4, A_b, f_3)$. 

Due to the Fermi statistics of the fermion, the path integral $Z_b(M^4, A_b, f_3)$ is not actually gauge invariant, and thus possesses a number of 't Hooft anomalies involving the 2-form $\Z_2$ symmetry and $0$-form $G_b$ symmetry. The fermion condensation procedure requires carefully canceling these anomalies by combining the theory with another, intrinsically fermionic theory, whose path integral is denoted $z_c(M^4, f_3, \xi_{\mathcal{G}})$. Physically this corresponds to the statement that for a topological phase with an emergent fermion, the only way to ``condense" the fermion $\psi$ is to stack another trivial topological phase with a physical (not emergent) fermion, $c$, and condense the composite $\psi c$.

Below, we briefly review the fermion condensation procedure for general $d$-dimensional theories, starting with a bosonic path integral $Z_b(M^d, A_b, f_{d-1})$, and we summarize the 't Hooft anomalies carried by the bosonic theory $Z_b(M^d, A_b, f_{d-1})$. We will see how the anomalies of the bosonic theory $Z_b$ is canceled out by stacking with the trivial fermionic theory $z_c$, and how the anomalous bosonic theory is converted to a topologically invariant fermionic theory free of anomalies by the fermion condensation procedure.

\subsection{'t Hooft anomaly of a bosonic shadow theory $Z_b(M^d, A_b, f_{d-1})$}

The global symmetry of the bosonic shadow theory is a 0-form symmetry $G_b$ with the background gauge field $A_b$ and a $\Z_2$ $(d-1)$-form symmetry generated by the Wilson line of the fermionic particle $\psi$ with the background gauge field $f_{d-1}\in Z^{d-1}(M^d,\Z_2)$.

In this paper, the bosonic symmetry group $G_b$ has the form of $G_b=\U\rtimes H$, where $H$ is a symmetry group that can act on $\U$ by charge conjugation  $\sigma: H\to\mathrm{Aut}(\U)$. The $U(1)$ part of the $G_b$ gauge field $A_b$, is characterized by the Villain fields $(h,c)$, and we use $A_H$ to denote the background $H$ gauge field. 

The extension class $[\omega_2]$ that characterizes the group extension~$\Z_2^f\to G_f\to G_b$ satisfies
\begin{align}
    A_b^*\omega_2 = 2c + A_H^* \omega_{H}\quad  \mod 2,
\end{align}
where $\omega_{H}\in H^2(BH,\Z_2)$ characterizes the group extension restricted to $H$, and $c$ is the Chern class $2c\in H^2_{\sigma}(M^d,\Z)$ twisted by the $H$ action on $\U$. Here we slightly abuse notation and view the gauge fields $A_b$ and $A_H$ as maps $A_b: M \rightarrow BG_b$, $A_H : M^d \rightarrow BH$, and $A_b^*$, $A_H^*$ denote their pullbacks. 

In the presence of background $H$ gauge fields, the U(1) Villain gauge field is realized by the gauge fields $(h,c)$ with the field strength $c+\delta h$, setting $h\in C^1_{\sigma}(M^d,\R)$,  $2c\in Z^2_{\sigma}(M^d,\Z)$ on each 2-simplex. The gauge transformation is given by
    \begin{align}
    c\to c-\delta_{\sigma}{\lambda},\quad h\to h+\lambda+\delta_{\sigma}\chi,
\end{align}
with $\lambda\in C^1(M^d,\Z), \chi\in C^0(M^d,\R)$.

Now let us explain the 't Hooft anomaly of the bosonic shadow theory $Z_b(M^d, A_b, f_{d-1})$ constructed in this paper. The 't Hooft anomaly of $Z_b(M^d, A_b, f_{d-1})$ is characterized by the $(d+1)$-dimensional response action
\begin{align}
    e^{i S_{d+1,b}}=(-1)^{\int \Sq^2(f_{d-1})+A_b^*\omega_2\cup f_{d-1}}=(-1)^{\int \Sq^2(f_{d-1})+(2c+A_{H}^*\omega_{H})\cup f_{d-1}} . 
\end{align}
The 't Hooft anomaly means the non-invariance of the partition function $Z_b(M^d, A_b, f_{d-1})$ under the gauge transformations and re-triangulations. That is, given two triangulations $T$ and $T'$ of $M^d$, we consider a triangulation of $M^d \times [0,1]$ which reduces to $T$ and $T'$ on the two boundaries. We also consider extending the definition of $A_b$ and $f_{d-1}$ to $M^d \times [0,1]$, which reduce to their values on the initial and final triangulations. Then the change in phase of $Z_b$ is given by
        \begin{align}
            Z_b( (M^d, T), A_b, f_{d-1}) = Z_b( (M^d, T'), A_b', f_{d-1}' )(-1)^{\int_{M\times[0,1]} \Sq^2(f_{d-1})+A_b^*\omega_2\cup f_{d-1}}.
        \end{align}
Here we have explicitly written the dependence on the triangulation $T$ in the path integral. Note that we could also consider a different manifold $(M^d)'$ on the right-hand side instead, in which case we consider a bordism from $M^d$ to $(M^d)'$. The 't Hooft anomaly reflects that the excitation $\psi$ behaves as a fermionic particle that carries the nontrivial symmetry fractionalization characterized by $\omega_2$ under the $G_b$ symmetry. See~\cite{tata2021anomalies} for the detailed description.

\subsection{Anomaly cancellation and fermion condensation}

We obtain a re-triangulation invariant, anomaly-free path integral, by ``stacking'' the bosonic theory $Z_b$ with another theory which transforms in exactly the same way as to cancel these anomalies. Since we want a fermionic theory after stacking, this other theory should be an intrinsically fermionic theory. An important requirement of such a fermionic theory is that it should depend on a choice of a $\mathcal{G}_f$-structure $\xi_{\mathcal{G}}$, which is required to specify how to couple the theory with $G_b$ internal symmetry to fermions. Thus, we consider the product:
        \begin{align}
        \label{}
        Z_b( (M^d, T), A_b, f_{d-1}) z_c( (M^d, T), f_{d-1}, \xi_{\mathcal{G}}) . 
        \end{align}
        Here $z_c$ is the path integral of a fermionic theory, i.e. which contains physical fermions, and will be defined in the next subsection. Note that $z_c$ depends on the $G_b$-connection $A_b$ implicitly through the $\mathcal{G}_f$ structure $\xi_{\mathcal{G}}$ which we will also define in the next subsection. When $z_c$ is defined appropriately, the product above will be re-triangulation invariant, anomaly-free, and independent of a change of $\xi_{\mathcal{G}}$ by a 1-coboundary. 

        Finally, the topologically invariant path integral for our fermionic theory can then be obtained by condensing the fermions of the bosonic shadow by summing over all inequivalent choices of $f_{d-1}$:
        \begin{align}
        \label{pathIntegral}
        Z(M^d, A_b, \xi_{\mathcal{G}}) \propto \sum_{[f_{d-1}] \in H^{d-1}(M^d, \Z_2)} Z_b((M^d,T), A_b, f_{d-1}) z_c((M^d,T),f_{d-1},\xi_{\mathcal{G}}).
        \end{align}
        For the bosonic shadow theory that we construct in Sec.~\ref{sec:shadow}, the resulting theory after fermion condensation $Z(M^d, A_b, \xi_{\mathcal{G}})$ defines an invertible fermionic TQFT with fermionic $G_f$ symmetry. As such, it is a smooth bordism invariant for 4-manifolds equipped with $G_b$ background gauge fields and $\xi_{\mathcal{G}}$ structures.

        \subsection{$\mathcal{G}_f$ structure and the definition of $z_c$}
        
        The definition of $z_c$ depends on a $\mathcal{G}_f$ structure, which is a $1$-cochain with the property that~\footnote{In fact, the precise definition of $z_c$ requires $\xi_{\mathcal{G}}$ to be a $(d-1)$-chain $\xi_{\mathcal{G}}\in C_{d-1}(M^d,\Z_2)$, while $A_b^*\omega_2$ in the RHS of Eq.~\eqref{Gfstructure} is defined as a cocycle. The precise formulation of $\xi_{\mathcal{G}}$ hence needs a map $f_{\infty}: Z^k(M^d,\Z_2)\to Z_{d-k}(M^d,\Z_2)$ that turns a cochain to a chain, and define  $\delta \xi_\mathcal{G} = w_2 + w_1^2 + f_{\infty}(A_b^* \omega_2)$, with $w_k$ the chain representative of $k^{\mathrm{th}}$ Stiefel-Whitney class. See~\cite{tata2021anomalies} for the detailed description.}
        \begin{align}
        \label{Gfstructure}
            \delta \xi_\mathcal{G} = w_2 + w_1^2 + A_b^* \omega_2 . 
        \end{align}
        As mentioned above, the $\mathcal{G}_f$-structure $\xi_{\mathcal{G}}$ allows us to specify a lift from the full bosonic symmetry (including space-time symmetries) of the field theory to the fermionic theory. 
        A detailed description of $\mathcal{G}_f$-structure can be found in Appendix of~\cite{tata2021anomalies}. The fermionic path integral $z_c$ then has the form of
        \begin{align}
            z_c( (M^d, T), f_{d-1}, \xi_{\mathcal{G}}) = \sigma( (M^d, T), f_{d-1}) (-1)^{ \int_{M^d} \xi_{\mathcal{G}}\cup f_{d-1}}. 
        \end{align} 
        Here, $\sigma( (M^d, T), f_{d-1})$ is referred to as the Gu-Wen Grassmann integral constructed in~\cite{gu2014,Gaiotto:2015zta, Kobayashi2019pin, tata2020}.
        The Grassmann integral $\sigma( (M^d, T), f_{d-1})$ is not invariant under the gauge transformation and re-triangulation, and the shift is characterized by the $(d+1)$-dimensional bulk response action
        \begin{align}
            e^{iS_{d+1,\sigma}}=(-1)^{\int \Sq^2(f_{d-1})+(w_2+w_1^2)\cup f_{d-1}}.
        \end{align}
     It is known that the combination $\Sq^2(f_{d-1})+(w_2+w_1^2)\cup f_{d-1}$ is trivial in cohomology according to the Wu formula~\cite{ManifoldAtlasWu}. This means that $S_{d+1,\sigma}$ should be regarded as an action of a trivial $(d+1)$-dimensional invertible topological phase. One can regard the Grassmann integral $\sigma( (M^d, T), f_{d-1})$ as giving the trivial boundary of the $(d+1)$-dimensional trivial invertible theory determined by $S_{d+1,\sigma}$.
     
     Meanwhile, the fermionic action $(-1)^{ \int_{M^d} \xi_{\mathcal{G}}\cup f_{d-1}}$ has an 't Hooft anomaly given by
     \begin{align}
            e^{iS_{d+1,\xi}}=(-1)^{\int (w_2+w_1^2+A_b^* \omega_2)\cup f_{d-1}}.
        \end{align}
     Altogether, we see that the 't Hooft anomaly of the combined action $Z_b( (M^d, T), A_b, f_{d-1}) z_c( (M^d, T), f_{d-1}, \xi_{\mathcal{G}})$ exactly cancels as
     \begin{align}
         e^{iS_{d+1,b}}e^{iS_{d+1,\sigma}}e^{iS_{d+1,\xi}}=1.
     \end{align}
     Thus one can gauge the $(d-2)$-form symmetry $f_{d-1}$ of the combined theory to perform the fermion condensation as described in Eq.~\ref{pathIntegral}.
     
     \subsection{Examples of $\mathcal{G}_f$ structures}
     
     Here we show the examples of $\mathcal{G}_f$ structures that correspond to the Altland-Zirnbauer class A, AI, AII, AIII. See Table~\ref{tab:AZclass} for a summary of the groups $G_f$, $G_b$, $\mathcal{G}_f$, and $\mathcal{G}_b$ for these symmetry classes. 
     \subsubsection{class A: $\Spin^c$ structure}
     
        Spin$^c$ structure can be obtained by setting $G_b=\U$, and $A_b^*\omega_2=2c$ mod 2. There is no time-reversal symmetry, so we take $M^d$ as an oriented manifold. The $\mathcal{G}_f$ structure is then given by
\begin{align}
    \delta\xi_{\mathcal{G}}=w_2+2c \quad \mod 2, 
\end{align}
which implies that $[w_2]=[2c]$ as an element of $H^2(M^d,\Z_2)$. This results in the Dirac quantization of the Spin$^c$ gauge field in Eq.~\eqref{eq:spincdirac}.

\subsubsection{class AI: $\Pin^{\tilde c}_-$ structure}

$\Pin^{\tilde c}_-$ structure can be obtained by setting $G_b=\U\rtimes \Z_2^{\mathbf{T}}$ with $\Z_2^{\mathbf{T}}$ that conjugates the U(1) charge. The symmetry extension class is given by $A_b^*\omega_2=2c$ mod 2, implying that $\mathbf{T}^2=1$. $M^d$ is in general unoriented, and the $\mathcal{G}_f$ structure is then given by
\begin{align}
    \delta\xi_{\mathcal{G}}=w_2+w_1^2+2c \quad \mod 2, 
\end{align}
which implies that $[w_2+w_1^2]=[2c]$ as an element of $H^2(M^d,\Z_2)$. The Dirac quantization of the Villain gauge field is given by
\begin{align}
        \int_C c=\frac{1}{2}\int_C (w_2+w_1^2) \quad \mod 1,
    \end{align}
    for any oriented two-dimensional cycle $C$ in the spacetime.

\subsubsection{class AII: $\Pin^{\tilde c}_+$ structure}
$\Pin^{\tilde c}_+$ structure can be obtained by setting $G_b=\U\rtimes \Z_2^{\mathbf{T}}$ with $\Z_2^{\mathbf{T}}$ that conjugates the U(1) charge. The symmetry extension class is given by $A_b^*\omega_2=2c+w_1^2$ mod 2, implying that $\mathbf{T}^2=(-1)^f$. $M^d$ is in general unoriented, and the $\mathcal{G}_f$ structure is then given by
\begin{align}
    \delta\xi_{\mathcal{G}}=w_2+2c \quad \mod 2, 
\end{align}
which implies that $[w_2]=[2c]$ as an element of $H^2(M^d,\Z_2)$. The Dirac quantization of the Villain gauge field is given by
\begin{align}
        \int_C c=\frac{1}{2}\int_C w_2 \quad \mod 1,
    \end{align}
    for any oriented two-dimensional cycle $C$ in the spacetime.
    
   \subsubsection{class AIII: $\Pin^{c}$ structure}
$\Pin^{c}$ structure can be obtained by setting $G_b=\U\times \Z_2^{\mathbf{T}}$ with $\Z_2^{\mathbf{T}}$ that does not conjugate the U(1) charge. The symmetry extension class is given by $A_b^*\omega_2=2c$ mod 2. $M^d$ is in general unoriented, and the $\mathcal{G}_f$ structure is then given by
\begin{align}
    \delta\xi_{\mathcal{G}}=w_2+2c \quad \mod 2, 
\end{align}
which implies that $[w_2]=[2c]$ as an element of $H^2(M^d,\Z_2)$. The Dirac quantization of the Villain gauge field is given by
\begin{align}
        \int_C c=\frac{1}{2}\int_C w_2 \quad \mod 1,
    \end{align}
    for any oriented two-dimensional cycle $C$ in the spacetime. 
    
    \begin{table}[t]
        \centering
        \begin{tabular}{l|l|l||l|l}
           Cartan & $G_b$ & $G_f$ & $\mathcal{G}_b$ & $\mathcal{G}_f$  \\
          \hline
          A & $\U$ & $\U^f$ & $\SO\times \U$ & $\Spin^c$ \\
          AI & $\U \rtimes \Z_2^{\bf T}$ & $\U^f \rtimes \Z_2^{\bf T}$ & $\mathrm{O} \ltimes \U$ & $[\Pin^- \ltimes \U]/\Z_2$\\
          AII & $\U \rtimes \Z_2^{\bf T}$ & $[\U^f \rtimes \Z_4^{ {\bf T},f}]/\Z_2$ & $\mathrm{O} \ltimes \U$ & $[\Pin^+ \ltimes \U]/\Z_2$ \\
          AIII & $\U \times \Z_2^{\bf T}$ &  $\U^f \times \Z_2^{\bf T}$ & $\mathrm{O} \times \U$ & $\Pin^c$ \\
        \end{tabular}
        \caption{Altland-Zirnbauer class involving U(1), and corresponding global symmetries.}
        \label{tab:AZclass}
    \end{table}

\section{Bosonic shadow for (3+1)D fermionic SPT with curved U(1) bundle}
\label{sec:shadow}

In this section, we construct a path integral $Z_b( (M^4,T) , A_b, f_3)$ that corresponds to the bosonic shadow discussed in the previous section. Compared to Sec.~\ref{sec:statesumu1},~\ref{sec:u1rtimesg}, here we take as input a super-modular category $\mathcal{C}$ with fermionic symmetry fractionalization data, and we couple the theory to a non-zero background flat $3$-form $\Z_2$ gauge field $f_3$. 


Since the $G_b=\U\rtimes H$ gauge field $A_b$ is specified by a pair of flat $H$ gauge field $A_{H}$ and the Villain gauge field $(h,c)$, we sometimes write the bosonic shadow as $Z_b( (M^4,T) , A_{H}, (h,c), f_3)$.
        
Our construction takes as input a super-modular category $\mathcal{C}$ describing the fermionic topological order, the bosonic symmetry group $G_b$ and the class $[\omega_2] \in H^2(BG_b,\Z_2)$ specifying $G_f$, symmetry fractionalization data $U$ and $\eta$ for $H$ on $\mathcal{C}$, and the fractional charges of anyons $Q_a\in \R/(2\Z)$ for U(1) symmetry reviewed in Sec.~\ref{subsubsec:spinccharge}, which satisfy

            \begin{align}
Q_{^{\mathbf{g}}a}=s(\mathbf{g})\sigma(\mathbf{g})Q_{a} \mod 2.
\label{eq:Qtransformfermionmain}
\end{align}
In Appendix~\ref{app:tilde}, we see that these data define the symmetry fractionalization data for $\mathcal{C}$.


        As reviewed in Appendix~\ref{app:symmfrac}, we require 
        \begin{align}
            U_{\bf g}(\psi,\psi;1) &= 1\\
            \eta_{\psi}({\bf g}, {\bf h}) &= \omega_2({\bf g}, {\bf h})
        \end{align}
    Subject to the first constraint, the consistency condition Eq.~\eqref{UUoverU_equals_EtaOverEtaEta} between $\eta$ and $U$ implies that $\eta_{\psi}({\bf g,h}) = \pm 1$, which turns the consistency condition for $\eta_{\psi}$ Eq.~\eqref{EtaEta_equals_EtaEta} into an (untwisted) $\Z_2$-valued cocycle condition, so the second condition makes sense.

        Given a triangulation and branching structure $T$ of $M^4$, we assign the following fixed background data:
        \begin{itemize}
            \item A cocycle $f \in Z^3(M^4,\Z_2)$ representing the background fermion line, which consists of an assignment $f_{ijkl}$ on each 3-simplex satifying $f_{0123} + f_{0124} + f_{0134} + f_{0234} + f_{1234} = 0 \text{ (mod 2)}$ on each 4-simplex $\braket{01234}$ (using the additive $\Z_2$ notation). 
            \item A flat gauge field $A_{H}$ which consists of group elements ${\bf g}_{ij} \in H$ on each edge $\braket{ij}$ satifying ${\bf g}_{ij}{\bf g}_{jk} = {\bf g}_{ik}$
            \item A 1-form Villain gauge field $h\in C^1(M^4,\R)$ which consists of group elements $h_{ij}\in\R$ on each edge $\braket{ij}$
            \item A 2-form Villain gauge field $c$ with $2c\in Z^2_{\sigma}(M^4,\Z)$ which consists of group elements $c_{ijk}\in\Z/2$ on each face $\braket{ijk}$.
        \end{itemize}
        The state sum consists of a summation over all possible assignments of the following data:
        \begin{itemize}
            \item To each 2-simplex $\braket{ijk}$, assign a simple object (anyon) $a_{ijk}  \in \mathcal{C}$
            \item To each 3-simplex $\braket{ijkl}$, assign an anyon $b_{ijkl} \in \C$ and an element of the vector space $V^{\psi^{f_{ijkl}}\times b}_{ijl,jkl} \otimes V_b^{ikl,\act{lk}{ijk}}$ .
        \end{itemize}
        
        Where it does not lead to ambiguity, we will often ignore the distinction between a simplex and the anyon or group element data assigned to it, i.e. simply write ${\bf g}_{ij}$ as $ij$, $a_{ijk}$ as $ijk$, or $b_{ijkl}$ as $ijkl$. We will always refer to the 3-form gauge field as $f_{ijkl}$.
       
\begin{figure}[htb]
    \centering
    \includegraphics{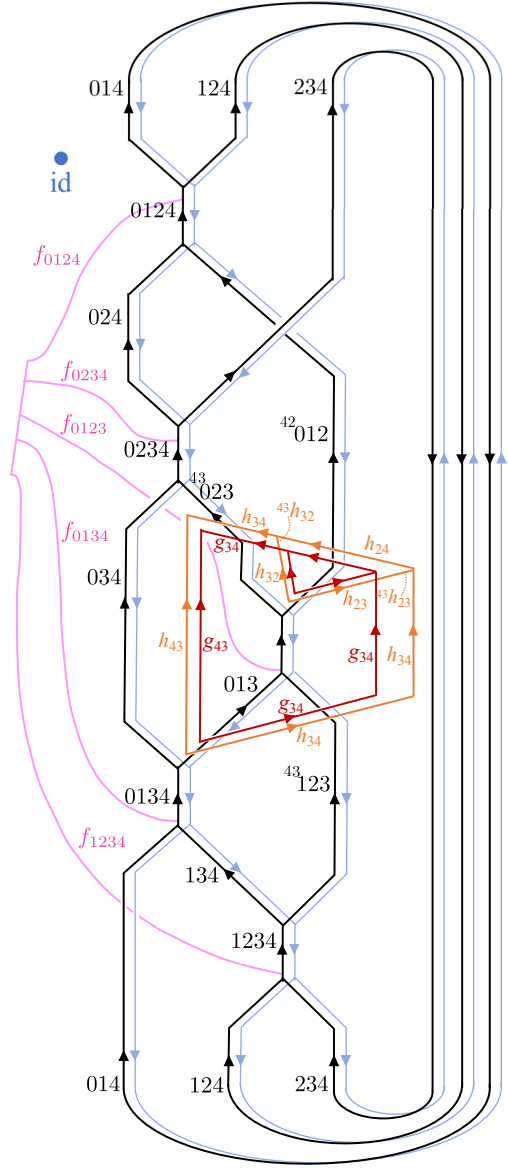}
    \caption{15j symbol of $Z_b$ for a $+$ simplex. The blue lines represent the dual of the $c$ field. The blue lines cross over the pink lines of $f$ fields, and cross
    behind the black lines of anyons, except for the the case that $c_{234}$ line crosses over the $a_{012}$ line. Here id means the identity in $H$, and the blue dot denotes the background domain of $H$ symmetry, see Appendix~\ref{app:symmfrac},~\ref{app:tilde} for the graphical calculus.}
    \label{fig:15jfermionG}
\end{figure}

       The evaluation of the $15$j symbol in Fig.\ref{fig:15jfermionG} involves the symmetry action of Villain gauge fields on anyons defined in Sec.~\ref{subsec:nonflatspinc}. This definition uses a $\R$ lift $q_a\in\R$ of the fractional charge $Q_a\in\R/(2\Z)$ that satisfies $q_a=Q_a$ mod 2.
        
       We will also need to assign to each $4$-simplex $\Delta_4$ of $M$ an orientation $\epsilon(\Delta_4) = \pm$. If $M$ is orientable, this simply amounts to choosing a global orientation of $M$. 
       We then assign an orientation-dependent amplitude $Z_b^{\epsilon(\Delta_4)}(\Delta_4)$ to each 4-simplex $\Delta_4$ of $M$. This amplitude is given diagrammatically in Fig.~\ref{fig:15jfermionG}, with a normalization factor
       \begin{equation} \label{fig:15jfermion}
           \mathcal{N}_{01234} = \sqrt{\frac{\prod_{\Delta_3 \in \text{3-simplices}} d_{b_{\Delta_3}}}{\prod_{\Delta_2 \in \text{2-simplices}} d_{a_{\Delta_2}}}} .
       \end{equation}

Now we construct a (3+1)D bosonic shadow theory $Z_b(M^4, A_{H}, (h,c), f)$ for the fermionic SPT phase based on the $\mathcal{G}_f$ structure. With the gauge $F^{\psi,\psi,a}=F^{a,\psi,\psi}=1$ for all $a\in\mathcal{C}$, we can evaluate the diagram in Fig.~\ref{fig:15jfermionG} as
 \begin{align}
\begin{split}
    Z^{+}_b(01234)&=e^{-2\pi i (^{42}q_{012}(c+\delta_{\sigma}h)_{234}+q_{\psi^{f_{0123}}}c_{034}+q_{\psi^{f_{1234}}}c_{014}+ ^{43}(\delta_{s\sigma} q)_{0123}h_{34})}Z_{0,b}^{+}(01234), \\
    Z^{-}_b(01234)&=e^{2\pi i (^{42}q_{012}(c+\delta_{\sigma}h)_{234}+q_{\psi^{f_{0123}}}c_{034}+q_{\psi^{f_{1234}}}c_{014}+ ^{43}(\delta_{s\sigma} q)_{0123}h_{34})}Z_{0,b}^{-}(01234), 
\end{split}
\end{align}
where $^{lm}q_{ijk}$ means the $\mathbf{g}_{lm}\in H$ action on the U(1) charge $q_{ijk}$.
One can regard $q_{ijk}:\braket{ijk}\to\R$ as a 2-cochain $q\in C^2(M^4,\R)$ which satisfies $\delta_{s\sigma} q=f$ mod 2. This mod 2 relation is a consequence of Eq.~\eqref{eq:fusionchargeliftedspinc} and Eq.~\eqref{eq:fermionchargelifted} on each 3-simplex.

$Z_{0,b}^{\pm}$ is the 15j symbol without U(1) gauge field $(h,c)$ expressed as
\begin{equation}
       \begin{split}
           Z_{0,b}^+(01234) =& \frac{1}{\eta_{012}(23,34)^{s(24)}} \Big( \frac{U_{34}(013,123 ; 0123 \times f_{0123})}{U_{34}(023, {^{32}}012 ; 0123 ) U_{34}(f_{0123},0123 ; 0123 \times f_{0123})} \Big)^{s(34)} \\
           &\cdot 
           \Big( F^{f_{0123} \, , \,  {^{43}}013 \, , \,  123}_{0123 \times f_{0123} \, , \,  {^{43}}013 \times f_{0123} \, , \,  0123} \,\,\,
                  \big(F^{034 \, , \, f_{0123} \, , \,  {^{43}}013 \times f_{0123}}_{0134 \times f_{0134} \, , \,  034 \times f_{0123} \, , \,  {^{43}}013} \big)^* \,\,\, 
                  R^{f_{0123} \, , \,  034}_{034 \times f_{0123}} \Big) \\
           &\cdot
           \left( F^{f_{0234} \, , \,  034 \, , \,  {^{43}}023}_{0234 \, , \,  034 \times f_{0234} \, , \,  0234 \times f_{0234}} \right) \\
           &\cdot
           \Big( F^{f_{0124} \, , \,  024 \, , \,  {^{42}}012}_{0124 \, , \,  024 \times f_{0124} \, , \,  0124 \times f_{0124}} \,\,\,
                 F^{f_{0124} \, , \,  024 \times f_{0124} \, , \,  234}_{0234 \, , \,  024 \, , \,  0234 \times f_{0124}} \,\,\,
                 F^{f_{0124} \, , \,  034 \times f_{0234} \, , \,  {^{43}}023}_{0234 \times f_{0124} \, , \,  034 \times f_{0234} \times f_{0124} \, , \,  0234} \Big) \\
           &\cdot 
           \Big( \big(F^{f_{0134} \, , \,  034 \times f_{0123} \, , \,  {^{43}}013 \times f_{0123}}_{0134 \, , \,  034 \times f_{0123} \times f_{0134} \, , \,  0134 \times f_{0134}} \big)^* \Big) \\
           &\cdot 
           \Big( \big(F^{f_{1234} \, , \,  134 \, , \,  {^{43}}123}_{1234 \, , \,  134 \times f_{1234} \, , \,  1234 \times f_{1234}} \big)^* \,\,\,
                  F^{014 \, , \,  f_{1234} \, , \,  134 \times f_{1234}}_{0134 \, , \,  014 \times f_{1234} \, , \,  134} \,\,\,
                  \big( R^{f_{1234} \, , \,  014}_{014 \times f_{1234}} \big)^* \\
                  &\quad\quad
                  \big(F^{f_{1234} \, , \,  014 \, , \,  134 \times f_{1234}}_{0134 \, , \,  014 \times f_{1234} \, , \,  0134 \times f_{1234}} \big)^* \,\,\,
                  \big(F^{f_{1234} \, , \,  034 \times f_{0123} \times f_{0134} \, , \,  {^{43}}013 \times f_{0123}}_{0134 \times f_{1234} \, , \,  034 \times f_{0123} \times f_{0134} \times f_{1234} \, , \,  0134} \big)^* \,
           \Big) \\
           &\cdot \Bigg( 
           \sum_{d,a \in \mathcal{C}} 
           F^{024 \times f_{0124} \, , \,  234 \, , \,  {^{42}}012}_{d \, , \,  0234 \times f_{0124} \, , \,  a} \,\,
           R^{{^{42}}012 \, , \,  234}_a \,\,
           \big(F^{024 \times f_{0124} \, , \,  012 \, , \,  234}_{d \, , \, 0124 \, , \,  a} \big)^* \,\,
           F^{014 \, , \,  124 \, , \,  234}_{d \, , \,  0124 \, , \,  1234} \\
           &\quad\quad\quad\quad\quad
           \big(F^{014 \, , \,  134 \times f_{1234} \, , \,  {^{43}}123}_{d \, , \,  0134 \times f_{1234} \, , \,  1234} \big)^* \,\,
           F^{034 \times f_{0234} \times f_{0124} \, , \,  {^{43}}013 \times f_{0123} \, , \,  {^{43}}123}_{d \, , \,  0134 \times f_{1234} \, , \,  {^{43}}0123 \times f_{0123}} \,\,
           \big(F^{034 \times f_{0234} \times f_{0124} \, , \,  {^{43}}023 \, , \,  {^{42}}012}_{d \, , \,  0234 \times f_{0124} \, , \,  {^{43}}0123 \times f_{0123}} \big)^*  \,\, d_d 
           \Bigg)
       \end{split}
       \end{equation}
       where we mean $f_{ijkl} = (\psi)^{f(ijkl)}$, $s(ij) = s({\bf g}_{ij})$, ${^{lk}}ijk = {^{{\bf g}_{lk}}a_{ijk}}$, etc. On a $-$ simplex, the weight is simply given by the complex conjugate, 
       \begin{equation}
           Z_b^-(01234) = (Z_b^+(01234))^*.
       \end{equation}
       The partition function is then given by
\begin{widetext}
\begin{equation}
\begin{split}
Z_b( (M^4,T) , A_{H}, (h,c), f_3) &= \sum_{\{a,b\}} {\mathcal{D}^{2(N_0-N_1)-\chi}}\frac{\prod_{\Delta^2 \in \mathcal{T}^2}d_{a_{\Delta^2}}\prod_{\Delta^4 \in \mathcal{T}^4}Z_b^{\epsilon(\Delta^4)}(\Delta^4)}{\prod_{\Delta^3 \in \mathcal{T}^3}d_{b_{\Delta^3}}} \\
&=\sum_{\{a,b\}} {\mathcal{D}^{2(N_0-N_1)-\chi}}\frac{\prod_{\Delta^2 \in \mathcal{T}^2}d_{a_{\Delta^2}}\prod_{\Delta^4 \in \mathcal{T}^4}Z_{0,b}^{\epsilon(\Delta^4)}(\Delta^4)}{\prod_{\Delta^3 \in \mathcal{T}^3}d_{b_{\Delta^3}}}
\prod_{\Delta^4 \in \mathcal{T}^4}e^{-2\pi i\epsilon(\Delta^4)(q\cup_{s\sigma}(c+\delta_{\sigma}h)+c\cup_1 f+ \delta_{s\sigma} q\cup_{s\sigma}h)}
\label{eqn:Zclosedfermion}
\end{split}
\end{equation}
\end{widetext}

In Appendix~\ref{app:pachner}, we will see that the bosonic shadow theory has the expected 't Hooft anomaly characterized by the response action
\begin{align}
    S_{5,b}=(-1)^{\int \Sq^2(f)+(2c+A_{H}^*\omega_{H})\cup f}
\end{align}

To check that the partition function is independent of the $\R$ lift of the fractional charge $Q_a$, let us study the effect of a local redefinition of the charge $q_a\to q_a+2\Lambda$, where $\Lambda\in C^2(M^4,\Z)$ is supported on a single 2-simplex.
Firstly, we move the $H$ symmetry defect away from $\Lambda$, which has the effect of shifting the partition function by some phase $e^{i\phi}$ according to the 't Hooft anomaly.
Then, we redefine the fractional charge $q_a\to q_a+2\Lambda$. the shift of the partition function under is given by
\begin{align}
    \prod_{\Delta^4} e^{-2\pi i\epsilon(\Delta^4) \delta(2\Lambda\cup h)}=1.
\end{align}
We then perform the inverse gauge transformation for the $H$ symmetry defect to the initial configuration. This shifts the partition function by $e^{-i\phi}$, since the response action $S_{5,b}$ of the 't Hooft anomaly is independent of the fractional charge $q_a$. After all, we can see that the partition function is not shifted by $q_a\to q_a+\Lambda$.
This shows that the path integral gives a well-defined (3+1)D invariant based on the data of U(1) fractional charge $Q_a\in\R/(2\Z)$.

\section{Explicit evaluation of U(1) anomaly indicators for (2+1)D TQFT}
\label{sec:evaluation}

The (3+1)D invertible TQFTs that are defined by the (2+1)D data can be viewed as defining the 't Hooft anomaly of the (2+1)D theory. One can uniquely identify the (3+1)D invertible TQFTs by computing the partition function on the generating manifolds for the appropriate bordism groups, which define anomaly indicator formulas \cite{barkeshli2019tr,lapa2019,bulmash2020,tata2021anomalies}. 

Below, we will show that the path integrals defined in the preceding sections can be given a relatively simple formula in the case where the input category $\mathcal{C}$ is Abelian. This allows us to explicitly compute anomaly indicators for a variety of physically relevant symmetry groups. 

\subsection{U(1) anomaly for (2+1)D Abelian TQFT: bosonic case}

\subsubsection{General simplification of path integral for Abelian modular $\mathcal{C}$}

For (2+1)D Abelian modular category, one can conveniently compute the path integral $Z(M^4,(h,c))$ by using the particular parameterization of $(F,R)$ given in Eq.~\eqref{eq:AbelianF} and Eq.~\eqref{eq:AbelianR}, with $\mathcal{A}=\Z_{n_1}\times\Z_{n_2}\times \dots\times \Z_{n_k}$. 

First, we note that for Abelian theories, the 15j symbol $Z_0^{\pm}(01234)$ given in Eq.~\eqref{Z0U1} simplifies to
\begin{align}
Z^+_0(01234) = &R^{012,234} F^{014,124,234} (F^{014,134,123})^{-1}F^{034,013,123} (F^{034,023,012})^{-1}
\label{eqn:Zplusabel}
\\
Z^-_0(01234) =  & (R^{012,234})^{-1} \left(F^{014,124,234}\right)^{-1} F^{014,134,123} \left(F^{034,013,123}\right)^{-1}F^{034,023,012}
\label{eqn:Zminusabel}
\end{align}
One can express the 15j symbol in terms of a quadratic function of $\{a_i\}$. To do this, we utilize the hexagon equation for BTC
\begin{align}
    \begin{split}
        R^{a,c}F^{a,c,b} R^{b,c}&=F^{c,a,b} R^{a\times b, c} F^{a,b,c},
    \end{split}
\end{align}
In the gauge that we are using for $F$ (Eq.~\eqref{eq:AbelianF}) we have $F^{a,b,c} = F^{a,c,b}$, so that the above simplifies as
\begin{align}
    \begin{split}
        F^{a,b,c}=R^{b,a}R^{c,a}(R^{b\times c,a})^{-1}.
        \label{eq:hexagonsimplify}
    \end{split}
\end{align}
One can then see that

\begin{align}
    \begin{split}
        Z_0^+(01234) &=   R^{012,234} (R^{234,014}R^{124,014}(R^{1234,014})^{-1}) ((R^{134,014})^{-1}(R^{123,014})^{-1}R^{1234,014})\\ &\times (R^{123,034}R^{013,034}(R^{0123,034})^{-1})((R^{023,034})^{-1}(R^{012,034})^{-1}R^{0123,034}) \\
        &= R^{012,234}\times (R^{234,014}R^{124,014}(R^{134,014})^{-1}(R^{123,014})^{-1}) \\ & \times (R^{123,034}R^{013,034}(R^{023,034})^{-1}(R^{012,034})^{-1}) \\
        &=\prod_i(\theta_{p_i})^{\hat{a}_i\cup \hat{a}_i+\hat{a}_i\cup_1\delta\hat{a}_i}\prod_{i<j}(M_{p_i,p_j})^{\hat{a}_i\cup \hat{a}_j+\hat{a}_j\cup_1\delta\hat{a}_i}.
    \end{split}
\end{align}
Here, we write the anyon $a\in\mathcal{A}$ as $a=\prod_{i}p_i^{a_i}$ with generators $\{p_i\}$ for $1\le i\le k$, and the assignment of anyons on 2-simplices defines a 2-cocycle $a_i\in Z^2(M^4,\Z_{n_i})$ for each $i$. $\hat{a}_i$ is the integral lift of $a_i\in Z^2(M^4,\Z_{n_i})$ that maps $x\in\Z_{n_i}$ to $x\in\Z$.
$(M_{p_i,p_j})^{n_j}=M_{p_i,1}=1$, so $M_{p_i,p_j}$ is an $n_j$-th root of unity (since $M_{ab}$ is symmetric, $M_{p_i,p_j}$ is also an $n_i$-th root of unity). Since $\delta\hat{a}_i$ is the integer multiple of $n_i$, the 15j symbol further simplifies as
\begin{align}
    \begin{split}
        Z_0^+(01234) &=   \prod_i(\theta_{p_i})^{\hat{a}_i\cup \hat{a}_i+\hat{a}_i\cup_1\delta\hat{a}_i}\prod_{i<j}(M_{p_i,p_j})^{{a}_i\cup {a}_j}.
    \end{split}
\end{align}
The partition function is then expressed as 
\begin{align}
\begin{split}
    Z(M^4,(h,c))= \sum_{\{a_j\}}&\mathcal{D}^{2(N_0-N_1)-\chi}\prod_i e^{-2\pi i\int(q(p_i^{a_i})\cup (c+\delta h)+\delta q(p_i^{a_i})\cup h)}(\theta_{p_i})^{\int(\hat{a}_i\cup \hat{a}_i+\hat{a}_i\cup_1\delta\hat{a}_i)}\prod_{i<j}(M_{p_i,p_j})^{\int {a}_i\cup {a}_j},
    \end{split}
    \label{eq:AbelianZwU1}
\end{align}
where $e^{\int S}$ means the path integral $\prod_{\Delta^4\in\mathcal{T}^4}e^{\epsilon(\Delta^4)S(\Delta^4)}$. Note that here we use the notation $q(a)$ instead of $q_a$. 

In general, the topological twist $\theta_{p}$ for an Abelian BTC is a $2n$-th root of unity when the anyon $p$ generates a $\mathbb{Z}_{n}$ group under fusion for even $n$, to further satisfy $(\theta_p)^{2n}=M_{p,p}^{n}=M_{p,p^n}=1$.
Meanwhile, $\theta_p$ further becomes an $n$-th root of unity for odd $n$, due to $(\theta_p)^{n^2}=\theta_{p^n}=1$. 
Here, for a given $a\in Z^2(M,\Z_n)$ let us introduce an operation $\mathcal{P}(a)$ known as the Pontryagin square,
\begin{align}
\mathcal{P}(a):=
    \begin{cases}
         & \hat{a}\cup \hat{a}+\hat{a}\cup_1\delta\hat{a} \in Z^4(M^4,\Z_{2n}) \quad \text{for even $n$} \\
         & a\cup a \in Z^4(M^4,\Z_{n}) \quad \text{for odd $n$},
    \end{cases}
\end{align}
where $\hat{a}$ here denotes the $\Z_{2n}$ lift of $a$.
$\mathcal{P}(a_i)$ is known to define a cohomology operation, which gives a map $\mathcal{P}: H^2(M^4,\Z_{n})\to H^4(M^4,\Z_{2n})$ for even $n$, and $\mathcal{P}: H^2(M^4,\Z_{n})\to H^4(M^4,\Z_{n})$ for odd $n$~\cite{pontryaginsquare, Aharony2013reading, kapustin2013topological}. 
Using the Pontryagin square, the above expression can be further simplified as
\begin{align}
\begin{split}
    Z(M^4,(h,c))=& \sum_{\{[a_j]\in H^2(M^4,\Z_{n_j})\}}\mathcal{D}^{2(N_0-N_1)-\chi}\prod_i{|B^2(M^4,\mathbb{Z}_{n_i})|}\prod_i e^{-2\pi i\int(q(p_i^{a_i})\cup (c+\delta h)+\delta q(p_i^{a_i})\cup h)}(\theta_{p_i})^{\mathcal{P}(a_i)}\prod_{i<j}(M_{p_i,p_j})^{\int {a}_i\cup {a}_j} \\
    =&\prod_i\sqrt{n_i}^{-\chi}\frac{|H^0(M^4,\Z_{n_i})|}{|H^1(M^4,\Z_{n_i})|}\cdot \sum_{\{[a_j]\in H^2(M^4,\Z_{n_j})\}}\prod_i e^{-2\pi i\int(q(p_i^{a_i})\cup (c+\delta h)+\delta q(p_i^{a_i})\cup h)}(\theta_{p_i})^{\mathcal{P}(a_i)}\prod_{i<j}(M_{p_i,p_j})^{\int {a}_i\cup {a}_j}
    \end{split}
    \label{eq:AbelianZwU1simplify}
\end{align}
Here, we have written the normalization factor using $\mathcal{D}^2=\prod_in_i$ and the formula
\begin{align}
    n^{N_0-N_1}|B^2(M,\mathbb{Z}_{n})|=\frac{|H^0(M,\mathbb{Z}_{n})|}{|H^1(M,\mathbb{Z}_{n})|}
    \label{eq:simplenormal}
\end{align}
The proof is by considering the chain complex $1 \xrightarrow{\delta_0} (\Z_n)^{N_0} \xrightarrow{\delta_1} (\Z_n)^{N_1} \xrightarrow{\delta_2} (\Z_n)^{N_2} \xrightarrow{\delta} \to \cdots $. Note that $|B^2(M,\Z_n)| = |\text{im}(\delta_2)|$. And by definition, $|\text{ker}(\delta_1)| /|\text{im}(\delta_0)| = |H^0(M,\Z_n)|$ and $|\text{ker}(\delta_2)| / |\text{im}(\delta_1)| = |H^1(M,\Z_n)|$. Further we have $|\text{ker}(\delta_1)|  |\text{im}(\delta_1)| = n^{N_0}$ and $|\text{ker}(\delta_2)|  |\text{im}(\delta_2)| = n^{N_1}$. Putting these all together gives the above claim.

In the following subsections, we explicitly compute the partition functions on spacetime manifolds of our interest.

\subsubsection{Partition function on $\mathbb{CP}^2$}
\label{subsec:CP2boson}

Firstly, let us evaluate the partition function of our theory on $\mathbb{CP}^2$ with U(1) gauge field.
We consider the U(1) gauge field with the nontrivial 1st Chern class $[c]\in H^2(\mathbb{CP}^2,\mathbb{Z})$.
Since the field strength is entirely given by $c$, $\delta h = 0$, and so
without loss of generality, we can take the $\mathbb{R}$ field $h=0$ by a gauge transformation. 

Since the generator of $H^2(\mathbb{CP}^2,\mathbb{Z}_{n_i})$ is given by a $\Z_{n_i}$ reduction of that of $H^2(\mathbb{CP}^2,\mathbb{Z})$, one can pick a representative $a_i$ of  $[a_i]\in H^2(\mathbb{CP}^2,\mathbb{Z}_{n_i})$ that lifts to an integral cocycle, where we can write $\mathcal{P}(a_i)=\hat{a}_i\cup\hat{a}_i$. Since $\mathcal{P}(a_i)$ is a cohomology operation, any representative of a fixed cohomology class $[a_i]$ gives the same result. Hence, we just write $\mathcal{P}(a_i)=\hat{a}_i\cup\hat{a}_i$ for $M^4=\mathbb{CP}^2$ in the following discussion.

The partition function is then expressed as
\begin{align}
\begin{split}
    Z(\mathbb{CP}^2,(h,c))= \prod_i\sqrt{n_i}^{-\chi}\frac{|H^0(\mathbb{CP}^2,\Z_{n_i})|}{|H^1(\mathbb{CP}^2,\Z_{n_i})|} \cdot \sum_{\{[a_j]\in H^2(\mathbb{CP}^2,\mathbb{Z}_{n_j})\}}& \prod_i e^{-2\pi i\int q(p_i^{a_i})\cup c}(\theta_{p_i})^{\int\hat{a}_i\cup \hat{a}_i}\prod_{i<j}(M_{p_i,p_j})^{\int {a}_i\cup {a}_j}.
    \end{split}
\end{align}

Now, with some abuse of notation, we identify the 2-form field $[a_i]\in H^2(\mathbb{CP}^2,\mathbb{Z}_{n_i})$ as a $\mathbb{Z}_{n_i}$-valued number $a_i\in \mathbb{Z}_{n_i}$, by rewriting the cohomology as $a_i [x]\in H^2(\mathbb{CP}^2,\mathbb{Z}_{n_i})$ with a generator $[x]$. Using $\chi(\mathbb{CP}^2)=3$, $|H^0(\mathbb{CP}^2,\mathbb{Z}_{n})|=n$, $|H^1(\mathbb{CP}^2,\mathbb{Z}_{n})|=1$, the normalization factor is simply given by $1/\mathcal{D}$. The partition function is then expressed as
\begin{align}
\begin{split}
    Z(\mathbb{CP}^2,(h,c))= \frac{1}{\mathcal{D}}\sum_{\{a_j\in \mathbb{Z}_{n_j}\}}&e^{-2\pi i\int q(p_i^{a_i})\cup c}(\theta_{p_i})^{\hat{a}_i^2}\prod_{i<j}(M_{p_i,p_j})^{{a}_i {a}_j}.
    \end{split}
\end{align}
To further simplify the expression, we introduce an anyon $p=\prod_i p_i^{a_i}$. We then have
\begin{align}
    \theta_{p}=\prod_i\theta_{p_i^{a_i}}\prod_{i<j}M_{p_i^{a_i},p_j^{a_j}}=\prod_i(\theta_{p_i})^{\hat{a}_i^2}\prod_{i<j}(M_{p_i,p_j})^{{a}_i{a}_j}.
\end{align}
Hence, we get
\begin{align}
\begin{split}
    Z(\mathbb{CP}^2,(h,c))= \frac{1}{\mathcal{D}}\sum_{p\in\mathcal{C}}e^{-2\pi iq(p)\cdot C}\theta_p,
    \end{split}
\end{align}
where $C\in\mathbb{Z}$ denotes the Chern number given by integrating $[c]\in H^2(\mathbb{CP}^2,\mathbb{Z})$ on the fundamental 2-cycle. Using $q_p=Q_p$ mod 1, we get the final expression
\begin{align}
\begin{split}
    Z(\mathbb{CP}^2,(h,c))= \frac{1}{\mathcal{D}}\sum_{p\in\mathcal{C}}e^{-2\pi iQ_p\cdot C}\theta_p.
    \end{split}
\end{align}
One can immediately see that this partition function becomes a phase, as also discussed in~\cite{lapa2019}. Using a vison $v\in\mathcal{C}$, we can rewrite the fractional charge as
\begin{align}
    e^{-2\pi iQ_p\cdot C}=M_{p,v^{-C}}=\frac{\theta_{p\times v^{-C}}}{\theta_{p}\theta_{v^{-C}}}.
    \label{eq:rewritecharge}
\end{align}
The partition function is then given by a phase
\begin{align}
    Z(\mathbb{CP}^2,(h,c))=(\theta_{\overline{v}})^{-C^2}\cdot \frac{1}{\mathcal{D}}\sum_{p\in\mathcal{C}}\theta_p=(\theta_{{v}})^{-C^2}\cdot e^{\frac{2\pi i}{8}c_-}.
\end{align}

\subsubsection{Partition function on $S^2\times S^2$}

Next we consider the manifold $S^2\times S^2$. By simplifying the normalization factor, we have
\begin{align}
\begin{split}
    Z(S^2\times S^2,(h,c))= \frac{1}{\mathcal{D}^2}\sum_{\{[a_j]\in H^2(S^2\times S^2,\mathbb{Z}_{n_j})\}}&
    \prod_i e^{-2\pi i\int q(p_i^{a_i})\cup c}(\theta_{p_i})^{\int\hat{a}_i\cup\hat{a}_i }\prod_{i<j}(M_{p_i,p_j})^{\int {a}_i\cup {a}_j}.
    \end{split}
\end{align}
Note that $[a_i]$ is again lifted to the integral cohomology $[\hat{a}_i]$, and then we used $\delta\hat{a}_i=0$.
We express $[a_j]\in H^2(S^2\times S^2, \mathbb{Z}_{n_j})$ using two $\mathbb{Z}_{n_j}$-valued numbers $[a_j]=\alpha_j [x]+ \beta_j [y]$, where $x$ and $y$ are fundamental 2-cycles. Then 
\begin{align}
\begin{split}
    Z(S^2\times S^2,(h,c))= \frac{1}{\mathcal{D}^2}\sum_{\{\alpha_j,\beta_j\}}&
    \prod_i e^{-2\pi i q(p_i)(\alpha_iC_{\beta}+\beta_iC_{\alpha})}(\theta_{p_i})^{2\alpha_i\beta_i}\prod_{i<j}(M_{p_i,p_j})^{\alpha_i\beta_j+\beta_i\alpha_j}.
    \end{split}
\end{align}
where we label the 1st Chern class $[c]=C_\alpha [x]+C_\beta [y]$ with $C_\alpha,C_\beta\in\mathbb{Z}$. Let $p_{\alpha}=\prod_i p_i^{\alpha_i}$, $p_{\beta}=\prod_i p_i^{\beta_i}$. Then the braiding between $p_{\alpha}$, $p_{\beta}$ is given by
\begin{align}
M_{p_{\alpha},p_{\beta}}=\prod_i(\theta_{p_i})^{2\alpha_i\beta_i}\prod_{i<j}(M_{p_i,p_j})^{\alpha_i\beta_j+\beta_i\alpha_j}.
\end{align}
We then get the final expression
\begin{align}
\begin{split}
    Z(S^2\times S^2,(h,c))=\frac{1}{\mathcal{D}}\sum_{{p}_{\alpha},p_{\beta}\in \mathcal{C}}e^{-2\pi i Q(p_{\alpha})C_\beta}e^{-2\pi i Q(p_{\beta})C_\alpha}S^*_{{p}_{\alpha},p_{\beta}}.
    \end{split}
\end{align}
To see that the partition function becomes a phase, we again rewrite it using Eq.~\eqref{eq:rewritecharge} as
\begin{align}
\begin{split}
    Z(S^2\times S^2,(h,c))&=\sum_{p_{\alpha},p_{\beta}\in \mathcal{C}}{S}^*_{\overline{p}_{\alpha},v^{C_{\beta}}}M_{{p}_{\beta},v^{-C_{\alpha}}}S_{\overline{p}_{\alpha},p_{\beta}}\\
    &= \sum_{p_{\beta}\in \mathcal{C}}M_{{p}_{\beta},v^{C_{-\alpha}}}\delta_{v^{C_{\beta}},{p}_{\beta}} \\
    &=M_{v^{C_{\beta}},v^{-C_{\alpha}}}= (\theta_{v})^{-2C_{\alpha}C_{\beta}}.
\end{split}
\end{align}

\subsubsection{$\U\times \mathbb{Z}_2^{\mathbf{T}}$ symmetry: partition function on $S^2\times\mathbb{RP}^2$}
\label{subsubsec:S2RP2}

Here we compute the partition function with symmetry group $G = \U\times \mathbb{Z}_2^{\mathbf{T}}$. We consider the case where time-reversal symmetry on unitary modular tensor category (UMTC) does not permute the anyons. In that case, the consistency equations involving the $H=\Z_2^{\mathbf T}$ symmetry data $\eta$ greatly simplifies: 
\begin{align}
    {U_{\mathbf{T}}(a,b;c)}^*U_{\mathbf{T}}(a,b;c)=\frac{\eta_c(\mathbf{T},\mathbf{T})}{\eta_a(\mathbf{T},\mathbf{T})\eta_b(\mathbf{T},\mathbf{T})}, \quad \eta_a(\mathbf{T},\mathbf{T})={\eta_a(\mathbf{T},\mathbf{T})}^*.
\end{align}
This means that $\eta_a(\mathbf{T},\mathbf{T})=\pm1$ and satisfies $\eta_a(\mathbf{T},\mathbf{T})\eta_b(\mathbf{T},\mathbf{T})=\eta_{a\times b}(\mathbf{T},\mathbf{T})$ for Abelian UMTC. The rest of the $\eta$'s are set to be trivial by suitable gauge transformations,
\begin{align}
    \eta_1(\mathbf{g},\mathbf{h})=\eta_a(1,\mathbf{g})=\eta_a(\mathbf{g},1)=1.
\end{align}
We will simply write $\eta_{a}^{\mathbf{T}}:= \eta_a(\mathbf{T},\mathbf{T})$.

Further, we have $\theta_a=\pm 1$ and $M_{a,b}=\pm 1$ for any pair of anyons, since the $\mathbf{T}$ symmetry does not permute anyons. This means that $a^2$ for some $a\in\mathcal{A}$ is transparent, since $M_{a^2,b}=(M_{a,b})^2=1$ for any $b\in\mathcal{A}$. $a^2$ hence becomes a trivial anyon in UMTC, so the group of Abelian anyons has the form of $\mathcal{A}=(\Z_2)^k$.  

When the symmetry does not permute anyons in the Abelian anyon system for $G = \U\times \mathbb{Z}_2^{\mathbf{T}}$, one can set $U_{\mathbf{g}}(a,b;c)=1$ by a suitable natural isomorphism.~\footnote{In that case, the consistency condition Eq.~\eqref{eqn:tildeFConsistency}, Eq.~\eqref{eqn:tildeURConsistency} imply that $U_{\mathbf{g}}(a,b)$ defines an element of $Z^2(B\mathcal{A},\U)$ for fixed $\mathbf{g}\in H$, which is symmetric under the exchange of $a,b$. For an Abelian discrete group $\mathcal{A}$, one can easily show that any symmetric function $U_{\mathbf{g}}(a,b)$ is trivial in cohomology $H^2(B\mathcal{A},\U)$. Hence it can be set as 1 by a suitable natural isomorphism, which is regarded as a shift of $U$ by coboundary $\delta \gamma(a,b):=(\gamma(a)\gamma(b))/\gamma(a\times b)$.} We hence consider the case where $U_{\mathbf{g}}(a,b;c)=1$, and ignore the contribution of $U$ symbols.

We consider the case with trivial U(1) holonomy $h=0$, and the nontrivial Chern number on $S^2$ given by
\begin{align}
    \int_{S^2}c=C\in\Z.
\end{align}
By simplifying the normalization factor using $\chi=2$ and $|H^0(S^2\times\mathbb{RP}^2,\mathbb{Z}_{2})|=|H^1(S^2\times\mathbb{RP}^2,\mathbb{Z}_{2})|=2$, the partition function is then obtained as
\begin{align}
\begin{split}
    Z(S^2\times\mathbb{RP}^2,(h,c))=\frac{1}{\mathcal{D}^2}
    &\sum_{\{[a_j]\in H^2(S^2\times \mathbb{RP}^2,\mathbb{Z}_2)\}}\prod_i e^{-2\pi i\int q({p_i})a_i\cup c}(\theta_{p_i})^{\int{a}_i\cup{a}_i }(\eta^{\mathbf{T}}_{p_i})^{\int a_i\cup w_1^2}\prod_{i<j}(M_{p_i,p_j})^{\int{a}_i\cup{a}_j}.
    \end{split}
\end{align}
Here we have used that $\eta_{p_{012}}(23,34)$ on a single 4-simplex with an anyon $p_{012}=\prod_{i}p_i^{a_i(012)}$ can be expressed as $\prod_i(\eta^{\mathbf{T}}_{p_i})^{a_i\cup w_1^2}$, where $w_1$ is the 1st Stiefel-Whitney class. This can be seen by $\eta_{p_{012}}(23,34)=\prod_{i}(\eta_{p_i}(23,34))^{a_i(012)}=\prod_{i}(\eta_{p_i}^{\mathbf{T}})^{a_i(012)s(\mathbf{g}_{23})s(\mathbf{g}_{34})}$, and $s(\mathbf{g}_{ij})=w_1(ij)$.
We express $[a_j]\in H^2(S^2\times \mathbb{RP}^2, \mathbb{Z}_2)$ using a $\mathbb{Z}_{2}$-valued number $\alpha_j$, $\beta_j$ in terms of $[a_j]=\alpha_j [x]+ \beta_j [y]$, where $x$ and $y$ are fundamental 2-cocycles of $S^2$ and $\mathbb{RP}^2$ respectively. Then
\begin{align}
\begin{split}
    Z(S^2\times\mathbb{RP}^2,(h,c))=\frac{1}{\mathcal{D}^2}\sum_{\{\alpha_j,\beta_j\}}\prod_i e^{-2\pi i q(p_i)\beta_i C}(\theta_{p_i})^{2\alpha_i\beta_i }(\eta^{\mathbf{T}}_{p_i})^{\alpha_i}\prod_{i<j}(M_{p_i,p_j})^{\alpha_i\beta_j+\alpha_j\beta_i}.
    \end{split}
\end{align}

Let $p_{\alpha}=\prod_i p_i^{\alpha_i}$, $p_{\beta}=\prod_i p_i^{\beta_i}$. The braiding between $p_{\alpha}$, $p_{\beta}$ is then given by
\begin{align}
M_{p_{\alpha},p_{\beta}}=\mathcal{D}\cdot S_{p_{\alpha},p_{\beta}}=\prod_i(\theta_{p_i})^{2\alpha_i\beta_i}\prod_{i<j}(M_{p_i,p_j})^{\alpha_i\beta_j+\beta_i\alpha_j}.
\end{align}
We then get the final expression
\begin{align}
\begin{split}
    Z(S^2\times\mathbb{RP}^2,(h,c))&=\frac{1}{\mathcal{D}}
    \sum_{\{p_{\alpha},p_{\beta}\in\mathcal{C}\}} e^{-2\pi i q(p_{\beta})C}\eta^{\mathbf{T}}_{p_\alpha}S_{p_\alpha,p_\beta}.
    \end{split}
\end{align}

To see that the partition function becomes a phase, we again rewrite it using Eq.~\eqref{eq:rewritecharge}, and that $\eta^{\mathbf{T}}_{p}$ can be expressed as $M_{p,w}$ with some anyon $w\in \mathcal{C}$. Then
\begin{align}
\begin{split}
    Z(S^2\times\mathbb{RP}^2,(h,c))&=\mathcal{D}\sum_{p_{\alpha},p_{\beta}\in \mathcal{C}}S_{p_{\alpha},w}S_{p_{\beta},v^{-C}}S_{p_{\alpha},p_{\beta}} \\
    &= \mathcal{D}\sum_{p_{\alpha},p_{\beta}\in \mathcal{C}}S_{p_{\alpha},w}\overline{S}_{p_{\beta},v^{-C}}S_{p_{\alpha},p_{\beta}} 
    = \mathcal{D}\sum_{p_{\alpha}\in \mathcal{C}}S_{p_{\alpha},w}\delta_{p_{\alpha},v^{C}}\\
    &=M_{v^{C},w}.
    \end{split}
\end{align}

\subsection{$\U^f$ anomaly for (2+1)D Abelian Spin$^c$ TQFT: fermionic case}

\subsubsection{General simplification of path integral for Abelian super-modular $\mathcal{C}$}

We evaluate the the bosonic shadow theory of the fermionic state sum~\eqref{eqn:Zclosedfermion} in the case that $\mathcal{C}$ is Abelian. In that case, the anyons form a group $\mathcal{A}=\mathbb{Z}_2^\psi\times\Z_{n_1}\times \Z_{n_2}\times\dots \times\Z_{n_k}$, where $\mathbb{Z}_2^\psi$ is generated by a transparent fermion $\psi$.
We fix generators $p_i$ of $\Z_{n_i}$ for $0\le i\le k$ with $p_0=\psi$ and $n_0=2$. An element $a\in\mathcal{A}$ is then expressed as $a=\prod_i p_i^{a_i}$. $(F,R)$ symbols are again given given in Eq.~\eqref{eq:AbelianF} and Eq.~\eqref{eq:AbelianR}.

We immediately see that $F^{\psi,b,c}=1, R^{\psi,b}=R^{b,\psi}=(-1)^{b_0}$ for any $b,c\in\mathcal{A}$.

Based on these $(F,R)$ symbols, the 15j symbol $Z_{0,b}^{\pm}(01234)$ is expressed in a simpler form by using Eq.~\eqref{eq:hexagonsimplify} as
\begin{align}
\label{eq:15jfermion}
\begin{split}
           Z_{0,b}^+(01234) =&  
           \Big( 
                  \big(F^{034 \, , \, f_{0123} \, , \,  013 \times f_{0123}}_{0134 \times f_{0134} \, , \,  034 \times f_{0123} \, , \,  013} \big)^* \,\,\, 
                  R^{f_{0123} \, , \,  034}_{034 \times f_{0123}}
           \cdot 
                  F^{014 \, , \,  f_{1234} \, , \,  134 \times f_{1234}}_{0134 \, , \,  014 \times f_{1234} \, , \,  134} \,\,\,
                  \big( R^{f_{1234} \, , \,  014}_{014 \times f_{1234}} \big)^*            \Big) \\
           &\cdot \Bigg( 
           F^{024 \times f_{0124} \, , \,  234 \, , \,  012}_{d \, , \,  0234 \times f_{0124} \, , \,  a} \,\,
           R^{012 \, , \,  234}_a \,\,
           \big(F^{024 \times f_{0124} \, , \,  012 \, , \,  234}_{d \, , \, 0124 \, , \,  a} \big)^* \,\,
           F^{014 \, , \,  124 \, , \,  234}_{d \, , \,  0124 \, , \,  1234} \big(F^{014 \, , \,  134 \times f_{1234} \, , \,  123}_{d \, , \,  0134 \times f_{1234} \, , \,  1234} \big)^* \,\,\\
           &\quad\quad\quad\quad\quad
           F^{034 \times f_{0234} \times f_{0124} \, , \,  013 \times f_{0123} \, , \,  123}_{d \, , \,  0134 \times f_{1234} \, , \,  0123 \times f_{0123}} \,\,
           \big(F^{034 \times f_{0234} \times f_{0124} \, , \,  023 \, , \,  012}_{d \, , \,  0234 \times f_{0124} \, , \,  0123 \times f_{0123}} \big)^* 
           \Bigg) \\
           =&   (R^{013\times f_{0123},034})^{-1}R^{013,034}
           R^{134\times f_{1234},014}(R^{134,014})^{-1} \\
           & \cdot R^{012,234}\cdot R^{124,014}R^{234,014}\cdot (R^{134\times f_{1234},014}R^{123,014})^{-1}\\
           & \cdot R^{013\times f_{0123},034\times f_{0234}\times f_{0124}}R^{123,034\times f_{0234}\times f_{0124}}
           \cdot(R^{023,034\times f_{0234}\times f_{0124}}R^{012,034\times f_{0234}\times f_{0124}})^{-1} \\
           =& (-1)^{f_{0234}(a_0)_{013}+f_{0124}(a_0)_{013}+f_{0234}(a_0)_{123}+f_{0124}(a_0)_{123}+f_{0234}(a_0)_{023}+f_{0124}(a_0)_{023}+f_{0234}(a_0)_{012}+f_{0124}(a_0)_{012}} \\
           &\cdot (-1)^{f_{0123}(f_{0234}+f_{0124})}\\
           & R^{012,234}\cdot (R^{234,014}R^{124,014}(R^{134,014})^{-1}(R^{123,014})^{-1}) \\ & \cdot (R^{123,034}R^{013,034}(R^{023,034})^{-1}(R^{012,034})^{-1}) \\
           =& R^{012,234}\cdot (R^{234,014}R^{124,014}(R^{134,014})^{-1}(R^{123,014})^{-1}) \\ & \cdot (R^{123,034}R^{013,034}(R^{023,034})^{-1}(R^{012,034})^{-1}) \\
           =&\prod_i(\theta_{p_i})^{\hat{a}_i\cup \hat{a}_i+\hat{a}_i\cup_1\delta\hat{a}_i}\prod_{i<j}(M_{p_i,p_j})^{\hat{a}_i\cup \hat{a}_j+\hat{a}_j\cup_1\delta\hat{a}_i},
       \end{split}
       \end{align}
where we used $\delta(a_0)=f$ in the fourth equation.
The partition function is then given by
\begin{align}
\begin{split}
    Z_b(M^4,(h,c),f)= \sum_{\{a_j\}}&\mathcal{D}^{2(N_0-N_1)-\chi}e^{-2\pi i\int c\cup_1f}\prod_i e^{-2\pi i\int (q(p_i^{a_i})\cup (c+\delta h)+\delta q(p_i^{a_i})\cup h)}(\theta_{p_i})^{\int(\hat{a}_i\cup \hat{a}_i+\hat{a}_i\cup_1\delta\hat{a}_i)}\prod_{i<j}(M_{p_i,p_j})^{\int {a}_i\cup {a}_j}
    \end{split}
    \label{eq:AbelianZwU1fermion}
\end{align}

\subsubsection{Partition function on $\mathbb{CP}^2$}
\label{subsec:CP2fermion}

We evaluate the partition function of our theory on $\mathbb{CP}^2$ with Spin$^c$ structure. We consider the gauge fields $h=0$ by suitable gauge transformations, and we have $\int c=C\in \Z+1/2$ with the integral taken over a fundamental 2-cycle of $\mathbb{CP}^2$. By using the similar logic to Sec.~\ref{subsec:CP2boson}, we have
\begin{align}
\begin{split}
    Z_b(\mathbb{CP}^2,(h,c),f)=& \mathcal{D}^{2(N_0-N_1)-\chi}\prod_{0<i}{|B^2(\mathbb{CP}^2,\mathbb{Z}_{n_i})|}\sum_{\{a_j\in \mathbb{Z}_{n_j}|0<j\}, }\prod_{0<i} e^{-2\pi i\int q(p_i^{a_i})\cup c}(\theta_{p_i})^{\hat{a}_i^2}\prod_{0<i<j}(M_{p_i,p_j})^{{a}_i {a}_j} \\
    &\cdot \sum_{\substack{a_0\in C^2(\mathbb{CP}^2,\Z_2),\\ \delta(a_0)=f}}(-1)^{\int 2c\cup_1f}(-1)^{\int q(\psi^{a_0})\cup 2c}(-1)^{\int a_0\cup a_0+a_0\cup_1 f}
    \end{split}
\end{align}
When $f=0$, the path integral is computed as 
\begin{align}
\begin{split}
    Z_b(\mathbb{CP}^2,(h,c),f=0)=\frac{1}{\mathcal{D}}\sum_{p\in\mathcal{C}}e^{-2\pi iq(p)\cdot C}\theta_p.
    \end{split}
\end{align}

We then get the partition function of the invertible fermionic theory coupled to the Spin$^c$ structure $\xi_{\mathcal{G}}$ by performing the fermion condensation:
\begin{equation*}
            Z(M,A_b, \xi_{\mathcal{G}}) = \frac{1}{\mathcal{N}} \sum_{f_3 \in H^3(M,\mathbb{Z}_2)} Z_b(M,A_b,f_3)z_c(M,\xi_{\mathcal{G}},f_3).
        \end{equation*}
In~\cite{tata2021anomalies}, the normalization constant is shown to be
$\mathcal{N} = \sqrt{|H^2(M^4, \Z_2)|}$. With this normalization, it is argued in~\cite{tata2021anomalies} that the partition function of the path integral $Z(M,A_b, \xi_{\mathcal{G}})$ becomes a phase for general configuration of $A_b$. 

Since $H^3(\mathbb{CP}^2,\Z_2)$ is trivial, we only need to consider the case where $f = 0$. 
Using $z_c(M,\xi_{\mathcal{G}},f_3=0)=1$, we obtain
\begin{align}
    Z(\mathbb{CP}^2,(h,c),\xi_{\mathcal{G}})=\frac{1}{\sqrt{2}}Z_b(\mathbb{CP}^2,(h,c),f_3=0)=\frac{1}{\sqrt{2}\mathcal{D}}\sum_{p\in\mathcal{C}}e^{-2\pi iQ_p\cdot C}\theta_p,
\end{align}
where we used $q_p=Q_p$ mod 2.

\subsubsection{Partition function on $S^2\times S^2$}

We evaluate the partition function of our Spin$^c$ theory on $S^2\times S^2$. Since $H^3(S^2\times S^2,\Z)=0$ and $w_2(S^2\times S^2)=0$, we can set $f_3=0$ and the Dirac quantization condition for U(1) gauge field as $\int c=\Z$. The partition function of the bosonic shadow is computed in the same fashion as the bosonic case, and given by
\begin{align}
\begin{split}
    Z_b(S^2\times S^2,(h,c),f_3=0)=\frac{1}{\mathcal{D}}\sum_{{p}_{\alpha},p_{\beta}\in \mathcal{C}}e^{-2\pi i Q(p_{\alpha})C_\beta}e^{-2\pi i Q(p_{\beta})C_\alpha}S^*_{{p}_{\alpha},p_{\beta}}.
    \end{split}
\end{align}
Since $z_c(M^4,\xi_{\mathcal{G}},f_3=0)=1$, the fermionic partition function is given by
\begin{align}
\begin{split}
    Z(S^2\times S^2,(h,c),\xi_{\mathcal{G}})&= \frac{1}{\sqrt{|H^2(M^4,\Z_2)|}}Z_b(S^2\times S^2,(h,c),f_3=0) \\
    &= \frac{1}{2\mathcal{D}}\sum_{{p}_{\alpha},p_{\beta}\in \mathcal{C}}e^{-2\pi i Q(p_{\alpha})C_\beta}e^{-2\pi i Q(p_{\beta})C_\alpha}S^*_{{p}_{\alpha},p_{\beta}}.
    \end{split}
\end{align}

\subsection{Partition function on $\mathbb{RP}^4$: class DIII, AII, AIII}

We can also compute the partition function of our (3+1)D theory in the symmetry class AII and AIII, which correspond to $\Pin_+^{c}$ and $\Pin^{c}$ structure respectively. In both cases, we can compute the path integral on $\mathbb{RP}^4$ with the U(1) gauge field turned off, $(h,c)=(0,0)$. In that case, both spacetime structure reduces to Pin$^+$ structure. The partition function on $\mathbb{RP}^4$ with Pin$^+$ structure was computed in~\cite{tata2021anomalies}, and given by
\begin{align}
Z(\mathbb{RP}^4, \xi_{\mathcal{G}}) &= \frac{1}{\sqrt{2}\mathcal{D}} \left(\sum_{x | x = {\,^{\bf T}}x} d_x \theta_x \eta^{\bf T}_x \pm i \sum_{x | x = {\,^{\bf T}}x \times \psi} d_x \theta_x \eta^{\bf T}_x\right)
\label{eqn:Z16TotalZ}
\end{align}
Here, the choice of sign in $\pm i$ depends on the choice of $\xi_{\mathcal{G}}$ structure, of which there are precisely two choices. $\eta^{\bf T}_x$ is given by
\begin{equation} \label{gaugeInvariant_T_Eta}
            \eta_a^{\bf T} := \begin{cases}
            \eta_a({\bf T},{\bf T}) & \,^{\bf T}a=a\\
            \eta_a({\bf T},{\bf T})U_{\bf T}(a,\psi;a\psi)F^{a,\psi, \psi} & \,^{\bf T}a = a \times \psi
            \end{cases} 
        \end{equation}



\section{Discussion}
\label{sec:discussions}

In this work we have studied (3+1)D path integral state sums defined in terms of the algebraic data characterizing a (2+1)D topological order and symmetry fractionalization class. While the path integrals reduce to a simple form for Abelian topological phases, they are complicated to evaluate for non-Abelian topological phases. A natural direction is to develop an alternative method of computation of the path integrals in terms of skein modules, which would allow one to perform computations more efficiently in terms of gluing handles of a handle decomposition \cite{walker2006,barkeshli2019tr,walker2021}. We expect that such techniques will allow one to prove the anomaly indicator formulas discussed here in full generality, including non-Abelian theories, and also to compute new anomaly indicators for other symmetry groups. 


In this paper, we explicitly studied symmetry groups of the form $G = \U \rtimes H$. We expect that our methods should also generalize to groups of the form $G = \U^n \rtimes H$, with any $H$ action, and a symmetric integer matrix characterizing the second Chern numbers of the $\U^n$ bundle. If we restrict to the case where $H$ does not permute the different $\U$ factors, then the technical results presented here can be straightforwardly generalized, but we have not considered the case where $H$ can permute the different $\U$ factors. 

The Villain formulation used in this work has a natural generalization beyond the case of $\U$. For example, consider the case of $\PSU(N) = \SU(N)/\Z_N$, where $\PSU(2) = \SO(3)$. In this case we can consider Villain gauge fields $(h,c)$, with $h \in\SU(N)$ and $c \in \Z_N$, which can be used to define a $\PSU(N)$ bundle with non-trivial generalized Stiefel-Whitney class $[w_2] \in H^2(M^4,\Z_N)$. We can formulate a (3+1)D path integral on such a curved $\PSU(N)$ bundle as well. When $h=0$, the construction of the (3+1)D path integral is straightforward. The symmetry action on the BTC is characterized by the mutual braiding between the $\Z_N$ 't Hooft line $c$ and the anyon line as described in Fig.~\ref{fig:Zaction}; instead of a fractional charge, one has a fractional $\PSU(N)$ spin. For example, in the case SO(3) (i.e. $N = 2$), this corresponds to the possibility of a spin-1/2 carried by the anyons. The path integral of the state sum interacting with the $\Z_N$ 't Hooft line should allow derivation of the anomaly indicators involving SO(3) symmetry, which were studied in~\cite{Chenjie2021mirror} using different methods. However, it is not clear how to generalize our Villain graphical calculus in the presence of non-zero $\SU(N)$-valued $h$ fields. In particular, it is not clear to us how to generalize the rules for $h$ defects in Fig.~\ref{fig:Uetau1} to the PSU$(N)$ case. Though the above discussion on a path integral with $h=0$ corresponds to the (3+1)D PSU$(N)$ discrete theta angle~\cite{Aharony2013reading, kapustin2013topological}, it would be interesting to study lattice realization of continuous theta angles by developing the Villain graphical calculus for the PSU$(N)$ case.

Looking further, it is a natural question whether we can define exact path integral state sums for SU$(N)$ bundles with non-vanishing second Chern class. In the Villain formulations studied here, we can think of $c$ on each 2-simplex as defining the local 1st Chern class, which then gives rise to a non-trivial second Chern class. However, in the SU$(N)$ case, the Villain formulation fails because any SU$(N)$ bundle always has vanishing 1st Chern class even though the second Chern class can be non-trivial, and it is unclear how to obtain a $\SU(N)$ gauge field on lattice with non-vanishing second Chern class. 
In addition, the $\SU(N)$ symmetry fractionalization of (2+1)D topological phases becomes trivial because of $H^2(B\SU(N),\mathcal{A})=0$. This implies that the (2+1)D topological phases cannot interact with $\SU(N)$ background gauge field in a nontrivial way, which makes hard to obtain a (3+1)D path integral for a SPT phase with $\SU(N)$ symmetry.

Finally, we note that our path integral evaluated on a 4-manifold with boundary defines a wave function, which is expected to be the ground state of a commuting projector Hamiltonian. It would be interesting to further understand this Hamiltonian perspective, generalizing the Walker-Wang type Hamiltonian constructions given in \cite{bulmash2020,williamson2017,walker2012}. In particular, there is an interesting work~\cite{Jingyuanchen2019} that defines lattice Hamiltonians for (2+1)D topological ordered phases with $\U$ symmetry based on Villain formulation of $\U$ gauge field. Also, recently there has been work showing how one can obtain exactly solvable models for (2+1)D and (3+1)D topological phases with $\U$ symmetry with a finite dimensional Hilbert space on each site \cite{wang2021exactly}; an intriguing problem is to develop the Walker-Wang type Hamiltonian construction based on our Villain formulation.

\section{Acknowledgments}

We thank Daniel Bulmash and Srivatsa Tata for discussions and collaboration on related recent work. RK thanks Yasunori Lee, Ken Shiozaki and Yuya Tanizaki for useful conversations. This work is supported by NSF CAREER (DMR- 1753240) and JQI-PFC-UMD.

\appendix

\section{Review of (2+1)D anyon systems}   
\label{sec:anyon}
\subsection{Notations of BTC}

In this appendix, we briefly review the notation that we use to describe braided tensor category (BTC). For a more comprehensive review of the notation that we use, see, e.g., Ref.~\cite{barkeshli2019}. The topologically  non-trivial quasiparticles of a (2+1)D topologically ordered state are referred to as anyons. In the category theory terminology, they correspond to isomorphism classes of simple objects of the BTC. 

A BTC $\mathcal{C}$ contains splitting spaces $V_{c}^{ab}$, and their dual fusion spaces, $V_{ab}^c$, where $a,b,c \in \mathcal{C}$ are anyons. These spaces have dimension 
$\text{dim } V_{c}^{ab} = \text{dim } V_{ab}^c = N_{ab}^c$, where the fusion coefficients $N_{ab}^c$ determine the fusion rules. In particular, the fusion rules of the anyons are written as $a \times b = \sum_c N_{ab}^c c$, so that fusion from $a \times b \to c$ is possible if and only if $N_{ab}^c \ge 1$. If $N_{ab}^c > 1$, then each fusion corresponds to a higher dimensional vector space with more possible `fusion outcomes'. 

The fusion spaces are depicted graphically as: 
\vspace{-30pt}
\begin{equation}
\left( d_{c} / d_{a}d_{b} \right) ^{1/4}
\begin{pspicture}[shift=0.5](-0.1,-0.2)(1.5,1.2)
  \small
  \psset{linewidth=0.9pt,linecolor=black,arrowscale=1.5,arrowinset=0.15}
  \psline{-<}(0.7,0)(0.7,-0.35)
  \psline(0.7,0)(0.7,-0.55)
  \psline(0.7,-0.55) (0.25,-1)
  \psline{-<}(0.7,-0.55)(0.35,-0.9)
  \psline(0.7,-0.55) (1.15,-1)	
  \psline{-<}(0.7,-0.55)(1.05,-0.9)
  \rput[tl]{0}(0.4,0){$c$}
  \rput[br]{0}(1.4,-0.95){$b$}
  \rput[bl]{0}(0,-0.95){$a$}
 \scriptsize
  \rput[bl]{0}(0.85,-0.5){$\mu$}
  \end{pspicture}
=\left\langle a,b;c,\mu \right| \in
V_{ab}^{c} ,
\label{eq:bra}
\end{equation}

\begin{equation}
\left( d_{c} / d_{a}d_{b}\right) ^{1/4}
\begin{pspicture}[shift=-0.65](-0.1,-0.2)(1.5,1.2)
  \small
  \psset{linewidth=0.9pt,linecolor=black,arrowscale=1.5,arrowinset=0.15}
  \psline{->}(0.7,0)(0.7,0.45)
  \psline(0.7,0)(0.7,0.55)
  \psline(0.7,0.55) (0.25,1)
  \psline{->}(0.7,0.55)(0.3,0.95)
  \psline(0.7,0.55) (1.15,1)	
  \psline{->}(0.7,0.55)(1.1,0.95)
  \rput[bl]{0}(0.4,0){$c$}
  \rput[br]{0}(1.4,0.8){$b$}
  \rput[bl]{0}(0,0.8){$a$}
 \scriptsize
  \rput[bl]{0}(0.85,0.35){$\mu$}
  \end{pspicture}
=\left| a,b;c,\mu \right\rangle \in
V_{c}^{ab},
\label{eq:ket}
\end{equation}
where $\mu=1,\ldots ,N_{ab}^{c}$, $d_a$ is the quantum dimension of $a$, 
and the factors $\left(\frac{d_c}{d_a d_b}\right)^{1/4}$ are a normalization convention for the diagrams. 

Diagrammatically, inner products come from connecting the fusion/splitting spaces' lines as:
\begin{equation}
  \begin{pspicture}[shift=-0.95](-0.2,-0.35)(1.2,1.75)
  \small
  \psarc[linewidth=0.9pt,linecolor=black,border=0pt] (0.8,0.7){0.4}{120}{240}
  \psarc[linewidth=0.9pt,linecolor=black,arrows=<-,arrowscale=1.4,
    arrowinset=0.15] (0.8,0.7){0.4}{165}{240}
  \psarc[linewidth=0.9pt,linecolor=black,border=0pt] (0.4,0.7){0.4}{-60}{60}
  \psarc[linewidth=0.9pt,linecolor=black,arrows=->,arrowscale=1.4,
    arrowinset=0.15] (0.4,0.7){0.4}{-60}{15}
  \psset{linewidth=0.9pt,linecolor=black,arrowscale=1.5,arrowinset=0.15}
  \psline(0.6,1.05)(0.6,1.55)
  \psline{->}(0.6,1.05)(0.6,1.45)
  \psline(0.6,-0.15)(0.6,0.35)
  \psline{->}(0.6,-0.15)(0.6,0.25)
  \rput[bl]{0}(0.07,0.55){$a$}
  \rput[bl]{0}(0.94,0.55){$b$}
  \rput[bl]{0}(0.26,1.25){$c$}
  \rput[bl]{0}(0.24,-0.05){$c'$}
 \scriptsize
  \rput[bl]{0}(0.7,1.05){$\mu$}
  \rput[bl]{0}(0.7,0.15){$\mu'$}
  \endpspicture
=\delta _{c c ^{\prime }}\delta _{\mu \mu ^{\prime }} \sqrt{\frac{d_{a}d_{b}}{d_{c}}}
  \pspicture[shift=-0.95](0.15,-0.35)(0.8,1.75)
  \small
  \psset{linewidth=0.9pt,linecolor=black,arrowscale=1.5,arrowinset=0.15}
  \psline(0.6,-0.15)(0.6,1.55)
  \psline{->}(0.6,-0.15)(0.6,0.85)
  \rput[bl]{0}(0.75,1.25){$c$}
  \end{pspicture}
  ,
\end{equation}
This is a way of phrasing topological charge conservation. In addition, we have the usual `resolution of the identity' in a UMTC, phrased diagrammatically:
\begin{equation} \label{resOfIdentity}
\begin{pspicture}[shift=-0.65](-0.1,-0.2)(1.0,1.2)
  \small
  \psset{linewidth=0.9pt,linecolor=black,arrowscale=1.5,arrowinset=0.15}
  \psline{->}(0.25,0)(0.25,0.6)
  \psline(0.25,0)(0.25,1.0)
  \psline{->}(0.7,0)(0.7,0.6)
  \psline(0.7,0)(0.7,1.0)
  \rput[br]{0}(0.15,0.5){$a$}
  \rput[bl]{0}(0.8,0.5){$b$}
 \end{pspicture}
=\sum_{c} \sqrt{\frac{d_c}{d_a d_b}}
\begin{pspicture}[shift=-0.65](-0.4,-0.2)(1.5,1.3)
  \small
 \psset{linewidth=0.9pt,linecolor=black,arrowscale=1.5,arrowinset=0.15}
  \psline{->}(0.7,0.25)(0.7,0.7)
  \psline(0.7,0.25)(0.7,0.8)
  \psline(0.7,0.8) (0.25,1.25)
  \psline{->}(0.7,0.8)(0.3,1.2)
  \psline(0.7,0.8) (1.15,1.25)	
  \psline{->}(0.7,0.8)(1.1,1.2)
  \psline{->}(0.25,-0.3)(0.6,0.15)
  \psline(0.25,-0.3)(0.7,0.25)
  \psline{->}(1.15,-0.3)(0.8,0.15)
  \psline(1.15,-0.3)(0.7,0.25)
  \rput[bl]{0}(0.4,0.5){$c$}
  \rput[br]{0}(1.4,1.05){$b$}
  \rput[bl]{0}(0,1.05){$a$}
  \rput[bl]{0}(0,-0.2){$a$}
  \rput[br]{0}(1.4,-0.2){$b$}
  \end{pspicture},
\end{equation}
implicitly assuming $N_{ab}^c \leq 1$ for all $a,b,c$. 

We denote $\bar{a}$ as the topological charge conjugate of $a$, for which
$N_{a \bar{a}}^1 = 1$, i.e.
\begin{align}
a \times \bar{a} = 1 +\cdots
\end{align}
Here $1$ refers to the identity particle, i.e. the vacuum topological sector, which physically describes all 
local, topologically trivial bosonic excitations. 

The $F$-symbols are defined as the following basis transformation between the splitting
spaces of $4$ anyons:
\begin{equation}
\begin{pspicture}[shift=*](0,-0.45)(1.8,1.8)
  \small
  \psset{linewidth=0.9pt,linecolor=black,arrowscale=1.5,arrowinset=0.15}
  \psline(0.2,1.5)(1,0.5)
  \psline(1,0.5)(1,0)
  \psline(1.8,1.5) (1,0.5)
  \psline(0.6,1) (1,1.5)
  \psline{->}(0.6,1)(0.3,1.375)
  \psline{->}(0.6,1)(0.9,1.375)
  \psline{->}(1,0.5)(1.7,1.375)
  \psline{->}(1,0.5)(0.7,0.875)
  \psline{->}(1,0)(1,0.375)
  \rput[bl]{0}(0.05,1.6){$a$}
  \rput[bl]{0}(0.95,1.6){$b$}
  \rput[bl]{0}(1.75,1.6){${c}$}
  \rput[bl]{0}(0.5,0.5){$e$}
  \rput[bl]{0}(0.9,-0.3){$d$}
 \scriptsize
  \rput[bl]{0}(0.3,0.8){$\alpha$}
  \rput[bl]{0}(0.7,0.25){$\beta$}
\end{pspicture}
= \sum_{f,\mu,\nu} \left[F_d^{abc}\right]_{(e,\alpha,\beta)(f,\mu,\nu)}
 \begin{pspicture}[shift=-1.0](0,-0.45)(1.8,1.8)
  \small
  \psset{linewidth=0.9pt,linecolor=black,arrowscale=1.5,arrowinset=0.15}
  \psline(0.2,1.5)(1,0.5)
  \psline(1,0.5)(1,0)
  \psline(1.8,1.5) (1,0.5)
  \psline(1.4,1) (1,1.5)
  \psline{->}(0.6,1)(0.3,1.375)
  \psline{->}(1.4,1)(1.1,1.375)
  \psline{->}(1,0.5)(1.7,1.375)
  \psline{->}(1,0.5)(1.3,0.875)
  \psline{->}(1,0)(1,0.375)
  \rput[bl]{0}(0.05,1.6){$a$}
  \rput[bl]{0}(0.95,1.6){$b$}
  \rput[bl]{0}(1.75,1.6){${c}$}
  \rput[bl]{0}(1.25,0.45){$f$}
  \rput[bl]{0}(0.9,-0.3){$d$}
 \scriptsize
  \rput[bl]{0}(1.5,0.8){$\mu$}
  \rput[bl]{0}(0.7,0.25){$\nu$}
  \end{pspicture}
.
\end{equation}

To describe topological phases, these are required to be unitary transformations, i.e.

\begin{eqnarray}
\left[ \left( F_{d}^{abc}\right) ^{-1}\right] _{\left( f,\mu
,\nu \right) \left( e,\alpha ,\beta \right) }
&= \left[ \left( F_{d}^{abc}\right) ^{\dagger }\right]
  _{\left( f,\mu ,\nu \right) \left( e,\alpha ,\beta \right) }
  \nonumber \\
&= \left[ F_{d}^{abc}\right] _{\left( e,\alpha ,\beta \right) \left( f,\mu
,\nu \right) }^{\ast }
.
\end{eqnarray}

The $R$-symbols define the braiding properties of the anyons, and are defined via the the following
diagram:
\begin{equation}
\begin{pspicture}[shift=-0.65](-0.1,-0.2)(1.5,1.2)
  \small
  \psset{linewidth=0.9pt,linecolor=black,arrowscale=1.5,arrowinset=0.15}
  \psline{->}(0.7,0)(0.7,0.43)
  \psline(0.7,0)(0.7,0.5)
 \psarc(0.8,0.6732051){0.2}{120}{240}
 \psarc(0.6,0.6732051){0.2}{-60}{35}
  \psline (0.6134,0.896410)(0.267,1.09641)
  \psline{->}(0.6134,0.896410)(0.35359,1.04641)
  \psline(0.7,0.846410) (1.1330,1.096410)	
  \psline{->}(0.7,0.846410)(1.04641,1.04641)
  \rput[bl]{0}(0.4,0){$c$}
  \rput[br]{0}(1.35,0.85){$b$}
  \rput[bl]{0}(0.05,0.85){$a$}
 \scriptsize
  \rput[bl]{0}(0.82,0.35){$\mu$}
  \end{pspicture}
=\sum\limits_{\nu }\left[ R_{c}^{ab}\right] _{\mu \nu}
\begin{pspicture}[shift=-0.65](-0.1,-0.2)(1.5,1.2)
  \small
  \psset{linewidth=0.9pt,linecolor=black,arrowscale=1.5,arrowinset=0.15}
  \psline{->}(0.7,0)(0.7,0.45)
  \psline(0.7,0)(0.7,0.55)
  \psline(0.7,0.55) (0.25,1)
  \psline{->}(0.7,0.55)(0.3,0.95)
  \psline(0.7,0.55) (1.15,1)	
  \psline{->}(0.7,0.55)(1.1,0.95)
  \rput[bl]{0}(0.4,0){$c$}
  \rput[br]{0}(1.4,0.8){$b$}
  \rput[bl]{0}(0,0.8){$a$}
 \scriptsize
  \rput[bl]{0}(0.82,0.37){$\nu$}
  \end{pspicture}
  .
\end{equation}
Under a basis transformation, $\Gamma^{ab}_c : V^{ab}_c \rightarrow V^{ab}_c$, the $F$ and $R$ symbols change:
\begin{align} \label{eq:vertexBasisTransformation}
  F^{abc}_{def} &\rightarrow \check{F}^{abc}_d = \Gamma^{ab}_e \Gamma^{ec}_d F^{abc}_{def} [\Gamma^{bc}_f]^\dagger [\Gamma^{af}_d]^\dagger
  \nonumber \\
  R^{ab}_c & \rightarrow \check{R}^{ab}_c = \Gamma^{ba}_c R^{ab}_c [\Gamma^{ab}_c]^\dagger .
\end{align}
  where we have suppressed splitting space indices and dropped brackets on the $F$-symbol for shorthand.
  These basis transformations are referred to as vertex basis gauge transformations. Physical quantities correspond to gauge-invariant combinations
  of the data. 


The topological twist $\theta_a$ is defined via the diagram:
\begin{equation}
\theta _{a}=\theta _{\bar{a}}
=\sum\limits_{c,\mu } \frac{d_{c}}{d_{a}}\left[ R_{c}^{aa}\right] _{\mu \mu }
= \frac{1}{d_{a}}
\begin{pspicture}[shift=-0.5](-1.3,-0.6)(1.3,0.6)
\small
  \psset{linewidth=0.9pt,linecolor=black,arrowscale=1.5,arrowinset=0.15}
  \psarc[linewidth=0.9pt,linecolor=black] (0.7071,0.0){0.5}{-135}{135}
  \psarc[linewidth=0.9pt,linecolor=black] (-0.7071,0.0){0.5}{45}{315}
  \psline(-0.3536,0.3536)(0.3536,-0.3536)
  \psline[border=2.3pt](-0.3536,-0.3536)(0.3536,0.3536)
  \psline[border=2.3pt]{->}(-0.3536,-0.3536)(0.0,0.0)
  \rput[bl]{0}(-0.2,-0.5){$a$}
  \end{pspicture}
.
\end{equation}
Finally, the modular, or topological, $S$-matrix, is defined as
\begin{equation}
S_{ab} =\mathcal{D}^{-1}\sum
\limits_{c}N_{\bar{a} b}^{c}\frac{\theta _{c}}{\theta _{{a}}\theta _{b}}d_{c}
=\frac{1}{\mathcal{D}}
\begin{pspicture}[shift=-0.4](0.0,0.2)(2.6,1.3)
\small
  \psarc[linewidth=0.9pt,linecolor=black,arrows=<-,arrowscale=1.5,arrowinset=0.15] (1.6,0.7){0.5}{167}{373}
  \psarc[linewidth=0.9pt,linecolor=black,border=3pt,arrows=<-,arrowscale=1.5,arrowinset=0.15] (0.9,0.7){0.5}{167}{373}
  \psarc[linewidth=0.9pt,linecolor=black] (0.9,0.7){0.5}{0}{180}
  \psarc[linewidth=0.9pt,linecolor=black,border=3pt] (1.6,0.7){0.5}{45}{150}
  \psarc[linewidth=0.9pt,linecolor=black] (1.6,0.7){0.5}{0}{50}
  \psarc[linewidth=0.9pt,linecolor=black] (1.6,0.7){0.5}{145}{180}
  \rput[bl]{0}(0.1,0.45){$a$}
  \rput[bl]{0}(0.8,0.45){$b$}
  \end{pspicture}
,
\label{eqn:mtcs}
\end{equation}
where $\mathcal{D} = \sqrt{\sum_a d_a^2}$.

We also denote by $\mathcal{A}$ the Abelian group corresponding to fusion of Abelian anyons, for which each $a \in \mathcal{A}$ satisfies $d_a = 1$ and $a \times b$ has a unique fusion product for any $b \in \mathcal{C}$.

 The double braid, or mutual statistics, of anyons $a$ and $b$ is defined as
 \begin{align}
     M_{ab}=\frac{S^*_{ab}S_{00}}{S_{0a}S_{0b}}
     \label{doubleBraid}
 \end{align}
 and is a phase if either $a$ or $b$ is an Abelian anyon.

Also, we will eventually make use of the `ribbon identity' 
\begin{equation} \label{ribbonId}
R^{a b}_c R^{b a}_c = \frac{\theta_c}{\theta_a \theta_b} \mathbbm{1}
\end{equation}

\subsection{Abelian BTC}
\label{subsec:Abelian}

When all anyons $a \in\mathcal{C}$ have quantum dimension $d_a=1$, $\mathcal{C}$ is referred to as an Abelian BTC. The anyons form a group $\mathcal{A}$ under fusion for Abelian $\mathcal{C}$. In that case, the mutual statistics $M_{a,b}$ gives a bilinear form $M: \mathcal{A}\times \mathcal{A}\to \U$
under fusion, since $M_{a\times b,c}=M_{a,c}M_{b,c}$ and $M_{a, b\times c}=M_{a,b}M_{a,c}$. For any $a\in\mathcal{A}$ we have
\begin{align}
    M_{a,a}=\theta_a^2.
    \label{eq:homogenious}
\end{align}

The topological twists of anyons then become a quadratic function characterized by the bilinear form $M_{a,b}$,
\begin{align}
    \theta_{a\times b}=\theta_a\theta_b\cdot M_{a,b}.
    \label{eq:quadtheta}
\end{align}
For a given choice of a bilinear form $M_{a,b}$ and topological twists $\theta$ of anyons that satisfy Eq.~\eqref{eq:homogenious},~\eqref{eq:quadtheta},
it is known that there is a unique equivalence class of $(F, R)$. The representative for the data $(F,R)$ is given following~\cite{Quinn}. Firstly, we fix a decomposition of the set of anyons as $\mathcal{A}=\Z_{n_1}\times \Z_{n_2}\times\dots \times\Z_{n_k}$, and fix generators $p_i$ of $\Z_{n_i}$. Then an element $a\in\mathcal{A}$ is expressed as $a=\prod_i p_i^{a_i}$. $(F,R)$ symbols are then given by
\begin{align}
        F^{a,b,c}=\prod_i
        \begin{cases}
            1 & \text{if}\ b_i+c_i < n_i, \\
             (\theta_{p_i})^{n_ia_i} & \text{if}\ b_i+c_i \ge n_i\\
        \end{cases}
        \label{eq:AbelianF}
\end{align}

\begin{align}
R^{a,b}=\prod_i(\theta_{p_i})^{a_ib_i}\prod_{i<j}(M_{p_i,p_j})^{a_ib_j}.
\label{eq:AbelianR}
\end{align}
  
\subsection{ Modular and super-modular tensor categories} \label{sec:superModularCategories}

        Physically realizable bosonic topological orders are described by modular tensor categories, which have the property that the $S$-matrix is unitary. This means that braiding is non-degenerate, that is, every anyon $a$ can be detected by its non-trivial mutual statistics with some other anyon $b$.
        
        Fermionic topological phases have local fermions, and locality requires that these fermions have trivial mutual statistics with all other excitations. One way to keep track of the anyon fusion and braiding properties in a fermionic topological phase is to use a super-modular tensor category $\mathcal{C}$. A super-modular tensor category is a unitary braided fusion category where there are exactly two transparent anyons: the identity $1$ and a fermion $\psi$.\footnote{Mathematically speaking, $\mathcal{C}$ is super-modular if its M{\"u}ger center is equivalent to the unitary symmetric fusion category sVec of super-vector spaces.} That is, $\psi$ has $\theta_{\psi}=-1$ and trivial mutual braiding with all other particles. As such, the braiding is degenerate.
        
        In a super-modular tensor category, the anyons (simple objects) \it as a set \rm form the structure $\{1,a,b, ...\} \times \{1,\psi\}$.
        The $S$ matrix factorizes as
        \begin{align}
          S = \tilde{S} \otimes \frac{1}{\sqrt{2}} \left(\begin{matrix} 1 & 1 \\ 1 & 1 \end{matrix} \right) ,
        \end{align}
        where $\tilde{S}$ is unitary.

        The $S$ and $T$ matrices do not form a representation of $SL(2;\Z)$. Rather, one has a representation of a subgroup of $SL(2;\Z)$ which keeps the spin structure invariant. Collecting the Hilbert spaces on the 2-torus for \textit{all} spin structures realizes a representation of the ``metaplectic group" $Mp_1(\Z)$, which is the non-trivial $\Z_2$ extension of $SL(2;\Z)$~\cite{delmastro2021}. 
        
        There is a convenient, canonical gauge-fixing available for a super-modular category. In particular,
        \begin{align}
            F^{\psi, \psi, a} &\rightarrow \Gamma^{\psi \psi}_1 F^{\psi, \psi, a} (\Gamma^{\psi a})^{\ast} (\Gamma^{\psi, \psi \times a})^{\ast}\\
            F^{a,\psi,\psi} &\rightarrow \Gamma^{\psi,a} \Gamma^{a\times \psi, \psi} F^{a,\psi,\psi} \left(\Gamma^{\psi \psi}_1\right)^{\ast}
        \end{align}
        under a vertex basis transformation $\Gamma$. One can check that this is sufficient gauge freedom to gauge fix $F^{\psi,\psi,a}=1$ for all $a \in \mathcal{C}$, and likewise gauge-fix $\Gamma^{a,\psi}$ such that $F^{a, \psi,\psi}=1$ for all $a \in \mathcal{C}$. Diagrammatically, in this canonical gauge, fermion lines may be bent freely and slid past each other.

\section{Review of symmetry fractionalization}
\label{app:symmfrac}

In this appendix, we provide a review of symmetry fractionalization of braided tensor categories (BTC), summarizing results from \cite{barkeshli2019,bulmashSymmFrac,aasen21ferm}. We first recall the topological symmetries of BTC, and then describe the symmetry fractionalization of bosonic and fermionic topological phases. Parts of the review provided here are taken directly from the review provided in Ref. \cite{tata2021anomalies}; we include it here to make the paper more self-contained. 

\subsection{ Topological symmetry and braided auto-equivalence}

An important property of a BTC $\mathcal{C}$ is the group of ``topological symmetries,'' which are related to ``braided auto-equivalences'' in the mathematical literature, although the former contains anti-unitary symmetries as well. 

The topological symmetries consist of the invertible maps
\begin{align}
\varphi: \mathcal{C} \rightarrow \mathcal{C} .
\end{align}
We consider the different $\varphi$ modulo equivalences known as natural isomorphisms of the form
\begin{align}
    \Upsilon (\ket{a,b;c}) = \frac{\gamma_a\gamma_b}{\gamma_c}\ket{a,b;c}
\end{align}
The equivalence classes $[\varphi]$ then constitute a group, which we denote as Aut$(\mathcal{C})$ ~\cite{barkeshli2019}.\footnote{Ref. \cite{barkeshli2019} also included a parity grading, which we do not include here.}

The symmetry maps can be classified according to a $\mathbb{Z}_2$ grading corresponding to whether $\varphi$ has a unitary or anti-unitary action on the category:
\begin{align} \label{defnOfS_anti-unitary}
s(\varphi) = \left\{
\begin{array} {ll}
+1 & \text{if $\varphi$ is unitary} \\
*  & \text{if $\varphi$ is anti-unitary} \\
\end{array} \right.
\end{align}
where $*$ refers to complex conjugation. 
Thus the topological symmetry group can be decomposed as
\begin{align}
\text{Aut}(\mathcal{C}) = \text{Aut}_{0}(\mathcal{C}) \sqcup \text{Aut}_{1}(\mathcal{C}),
\end{align}
where Aut$_{0}(\mathcal{C})$ is the subgroup corresponding to topological symmetries that are unitary (this is referred to in the mathematical literature as the group of ``braided auto-equivalences''), and $\text{Aut}_{1}(\mathcal{C})$ are the anti-unitary ones.

The maps $\varphi$ may permute the topological charges:
\begin{align}
\varphi(a) = a' \in \mathcal{C}, 
\end{align}
subject to the constraint that 
\begin{align}
N_{a'b'}^{c'} &= N_{ab}^c
\nonumber \\
S_{a'b'} &= S_{ab}^{s(\varphi)},
\nonumber \\
\theta_{a'} &= \theta_a^{s(\varphi)},
\end{align}
The maps $\varphi$ have a corresponding action on the $F$- and $R$- symbols of the theory, as well as on the fusion and splitting spaces, reviewed below.

\subsection{ Global symmetry and symmetry fractionalization: bosonic case}
\label{globsym}

We now consider a bosonic system which has a global symmetry group $G$. In the bosonic case, the global symmetry acts on the anyons and the topological state space through the action of a group homomorphism
\begin{align}
[\rho] : G \rightarrow \text{Aut}(\mathcal{C}) . 
\end{align}
We use the notation $[\rho_{\bf g}] \in \text{Aut}(\mathcal{C})$ for a specific element ${\bf g} \in G$. The square brackets indicate the equivalence class of symmetry maps related by natural isomorphisms, which we define below. $\rho_{\bf g}$ is thus a representative symmetry map of the equivalence class $[\rho_{\bf g}]$. We use the notation
\begin{align}
\,^{\bf g}a \equiv \rho_{\bf g}(a). 
\end{align}
We associate a $\mathbb{Z}_2$ grading $s({\bf g})$ by defining
\begin{equation}
s({\bf g}) \equiv s( \rho_{\bf g})
\end{equation}
where we think of the complex conjugation $*$ as representing the $-1$ element of $\Z_2 = \{\pm 1\}$.

Each $\rho_{\bf g}$ has an action on the fusion/splitting spaces:
\begin{align}
\rho_{\bf g} : V_{ab}^c \rightarrow V_{\,^{\bf g}a \,^{\bf g}b}^{\,^{\bf g}c} .
\end{align}
This map is unitary if $s({\bf g}) = 1$ and anti-unitary if $s({\bf g}) = *$. We choose a basis $\ket{a,b;c,\mu}$ for $V_{ab}^c$ and write the action of $\rho_{\bf g}$ on the basis states as
\begin{align}
\rho_{\bf g} |a,b;c, \mu\rangle = \sum_{\nu} [U_{\bf g}(\,^{\bf g}a ,
\,^{\bf g}b ; \,^{\bf g}c )]_{\mu\nu} \ket{\,^{\bf g} a, \,^{\bf g} b; \,^{\bf g}c,\nu},
  \label{eqn:rhoStates}
 \end{align}
 and the action of $\rho_{\bf g}$ defined on the rest of the fusion/splitting spaces by (anti-)linearity.
 Here $U_{\bf g}(\,^{\bf g}a , \,^{\bf g}b ; \,^{\bf g}c ) $ is a $N_{ab}^c \times N_{ab}^c$ matrix.

\begin{figure}
    \centering
    \subfigure[\label{fig:symmFrac}]{\includegraphics[width=0.4\linewidth]{symfrac.eps}} \hspace{0.1\linewidth}
    \subfigure[\label{fig:graphicalCalculus_defs}]{\includegraphics[width=0.8\linewidth]{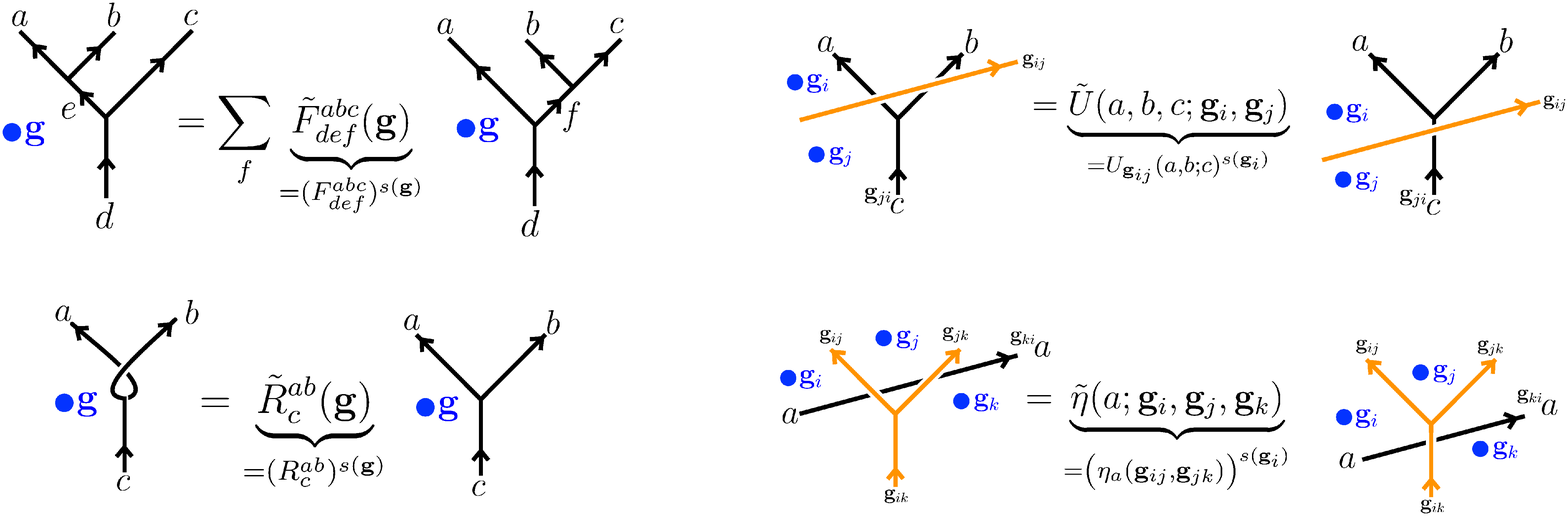}}
    \caption{(a) Anyon lines (black) passing through branch sheets (orange) and graphical definitions of the $U$ and $\eta$ symbols. (b) ``Tilded'' graphical calculus conventions. Orange lines are symmetry domain walls and black lines are anyons. The directions of the arrow on the ${\bf g}_{ij}={\bf g}_i{\bf g}_{j}^{-1}$ lines correspond to the direction in which the ${\bf g}_{ij}$ twists are applied. The figure is taken from~\cite{tata2021anomalies}.}
\end{figure}

In the presence of domain walls of $G$, a graphical calculus can be developed for the action of symmetry domain walls on anyon data. The basic pictures defining the graphical calculus are given in Fig.~\ref{fig:symmFrac}. However, if $G$ contains anti-unitary symmetries, then the anti-linearity of $\rho_{\bf g}$ is difficult to track within the usual BTC graphical calculus. We will therefore use a modified graphical calculus described in Refs.~\cite{bulmash2020,barkeshli2019rel}.

A key point from those works is that in the presence of anti-unitary symmetry domain walls, the local orientation of space reverses across the domain wall. Recall that the particular orientation chosen in space is important in defining the BTC data. Reflecting a BTC diagram (say, without any $G$ defects) across the vertical axis corresponds to a Hermitian conjugation of $F$- and $R$-symbols. The fact that space should reverse orientation after an anti-unitary domain-wall sweep means roughly that $F$- and $R$-symbols should be complex conjugated after sweeping the domain wall across.

The proposal to deal with such situations was to imagine that regions of space are labeled by group elements ${\bf g} \in G$ in which fusion spaces and their duals gain an extra label by group elements. For example, the space $V^{ab}_c$ is replaced with $\tilde{V}^{ab}_{c}({\bf g})$. Graphically this is depicted as
\begin{equation}
\left( d_{c} / d_{a}d_{b}\right) ^{1/4}
\begin{pspicture}[shift=-0.65](-0.1,-0.2)(1.5,1.2)
  \small
  \psset{linewidth=0.9pt,linecolor=black,arrowscale=1.5,arrowinset=0.15}
  \psline{->}(0.7,0)(0.7,0.45)
  \psline(0.7,0)(0.7,0.55)
  \psline(0.7,0.55) (0.25,1)
  \psline{->}(0.7,0.55)(0.3,0.95)
  \psline(0.7,0.55) (1.15,1)	
  \psline{->}(0.7,0.55)(1.1,0.95)
  \rput[bl]{0}(0.4,0){$c$}
  \rput[br]{0}(1.4,0.8){$b$}
  \rput[bl]{0}(0,0.8){$a$}
  \rput[bl]{0}(0.2,0.35){$\textcolor{blue}{{\bf g}}$}
 \scriptsize
  \rput[bl]{0}(0.85,0.35){$\mu$}
  \end{pspicture}
=\left| a,b;c,\mu \right\rangle_{{\bf g}} \in
\tilde{V}_{c}^{ab}({\bf g}),
\label{eq:ketWithG}
\end{equation}
where the tildes are meant to distinguish these spaces from the previous ones, and similarly for the dual vertices. In a region of space labeled by ${\bf g}$ we will have $\tilde{F}$ and $\tilde{R}$ symbols $\tilde{F}^{abc}_{def}({\bf g}), \tilde{R}^{ab}_c({\bf g})$. Similarly, there is now a ``tilded'' symmetry action
\begin{equation}
    \tilde{\rho}_{\bf h} : V_c^{ab}({\bf g}) \to V_{\,^{\bf h}c}^{\,^{\bf h}a \,^{\bf h}b}({\bf hg})
\end{equation}
which defines a ``tilded'' $U$ symbol
\begin{equation}
    \tilde{\rho}_{\bf h}\ket{a,b;c}_{\bf g} = \tilde{U}(\,^{\bf h}a,\,^{\bf h}b,\,^{\bf h}c;{\bf hg},{\bf g}) \ket{\,^{\bf h}a,\,^{\bf h}b;\,^{\bf h}c}_{\bf hg}
\end{equation}
where the ${\bf g}$ subscript on the state indicates that the state is in $V_c^{ab}({\bf g})$. Importantly, $\tilde{\rho}_{\bf h}$ is always unitary. The anti-unitarity of any symmetry action is tracked by the group element labels. The graphical representations of these tilded data are shown in Fig.~\ref{fig:graphicalCalculus_defs}.

The unitary or anti-unitarity of the group element labels play the role of the local orientation of space. This manifests itself in relating the $\tilde{F},\tilde{R},\tilde{U}$ to the original $F,R$ symbols
\begin{equation}
\begin{split}
\tilde{F}^{abc}_{def}({\bf g}) &= (F^{abc}_{def})^{s({\bf g})} \\
\tilde{R}^{ab}_c({\bf g})      &= (R^{ab}_{c})^{s({\bf g})}\\
\tilde{U}(a,b,c;{\bf g}_i,{\bf g}_j) &= \left(U_{{\bf g}_{ij}}(a,b;c)\right)^{s({\bf g}_i)},
\end{split}
\label{eqn:tildeToUntilded}
\end{equation}
where ${\bf g}_{ij}={\bf g}_i{\bf g}_j^{-1}$.

Symmetry fractionalization is then specified by a set of phases $\eta_{a}({\bf g,h})$, which satisfy certain consistency relations which we will discuss shortly. The data $\{U,\eta\}$ characterize a \textit{symmetry fractionalization class} and give us information about how the group symmetries fractionalize onto the different anyons. The $U_{\bf g}(a,b;c)$ relates the symmetry action on fusion vertices $c \to a,b$ to the fusion vertices $^{\bf g}c \to  \,^{\bf g}a, \,^{\bf g}b$, while the data $\eta_a({\bf g}, {\bf h})$ characterize the difference in phase obtained when acting ``locally'' on an anyon $a$ by ${\bf g}$ and ${\bf h}$ separately, as compared with acting on $a$ by the product ${\bf gh}$. The $\eta$ symbols also have a ``tilded'' version, graphically defined in Fig.~\ref{fig:graphicalCalculus_defs} and related to the usual symbols by
\begin{equation}
    \tilde{\eta}(a;{\bf g}_i,{\bf g}_j,{\bf g}_k) = \eta_a({\bf g}_{ij},{\bf g}_{jk})^{s({\bf g}_i)}.
\end{equation}

\begin{figure}[h!]
    \centering
    \includegraphics[width=\linewidth]{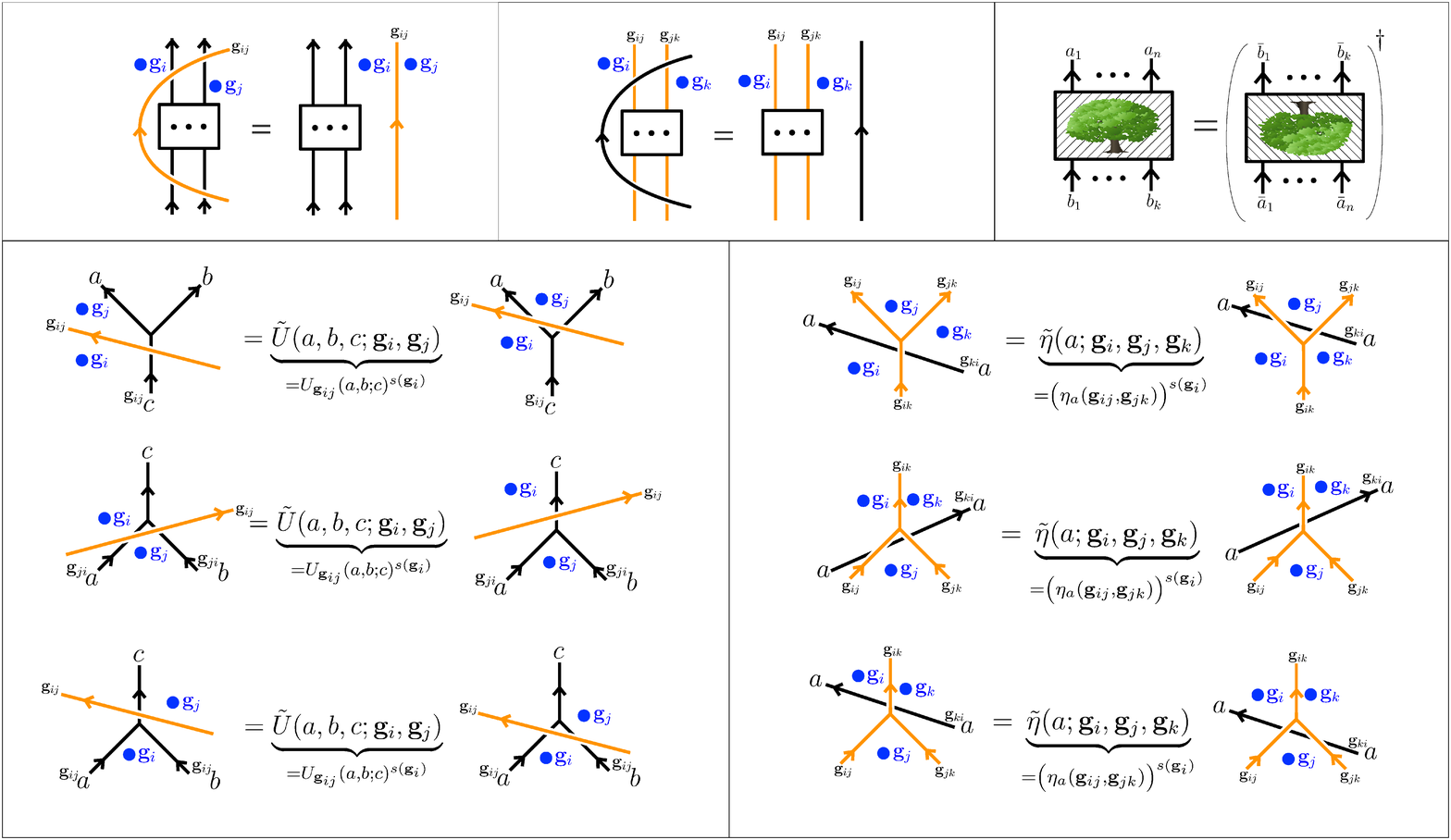}
    \caption{Other moves in the graphical calculus can be derived using isotopy invariance of sliding vertical strands (top-left and top-middle) and noting that reflecting the diagrams vertically corresponds to Hermitian conjugation (top-right). The figure is taken from~\cite{tata2021anomalies}.}
    \label{fig:graphicalCalculus_more}
\end{figure}

The diagrammatic rules and definitions in Fig.~\ref{fig:graphicalCalculus_defs} can be used to derive more rules as in Fig.~\ref{fig:graphicalCalculus_more}. One can derive them by demanding isotopy invariance under sliding vertical strands and noting that reflecting a partial diagram vertically amounts to Hermitian conjugation of relevant matrices.

Lastly, there are several consistency conditions that need to be imposed on the $U,\eta,F,R$ symbols in order for diagrammatic evaluations to be consistent under different orders of moves. In the case of $N_{ab}^c \le 1$ they can be written as
\begin{widetext}
\begin{align}
\tilde{F}^{fcd}_{egl}({\bf g})\tilde{F}^{abl}_{efk}({\bf g}) &= \sum_h \tilde{F}^{abc}_{gfh}({\bf g})\tilde{F}^{ahd}_{egk}({\bf g})\tilde{F}^{bcd}_{khl}({\bf g}) \label{eqn:tildePentagon}\\
\tilde{R}^{ca}_e({\bf g})\tilde{F}^{acb}_{deg}({\bf g})\tilde{R}^{cb}_g({\bf g}) &= \sum_f \tilde{F}^{cab}_{def}({\bf g})\tilde{R}^{cf}_d({\bf g})\tilde{F}^{abc}_{dfg}({\bf g})\\
(\tilde{R}^{ac}_e({\bf g}))^{-1}\tilde{F}^{acb}_{deg}({\bf g})(\tilde{R}^{bc}_g({\bf g}))^{-1} &= \sum_f \tilde{F}^{cab}_{def}({\bf g})(\tilde{R}^{fc}_d({\bf g}))^{-1}\tilde{F}^{abc}_{dfg}({\bf g})\\
\tilde{U}(\acts{g}{12}{a},\acts{g}{12}{b} ;\acts{g}{12}{e},{\bf g}_1,{\bf g}_2)\tilde{U}(\acts{g}{12}{e},\acts{g}{12}{c} ;\acts{g}{12}{d},{\bf g}_1,{\bf g}_2)&\tilde{F}^{\acts{g}{12}{a}\act{12}{b}\act{12}{c}}_{\act{12}{d}\act{12}{e}\act{12}{f}}({\bf g}_1) \times \nonumber \\
\times \tilde{U}^{-1}(\acts{g}{12}{b},\acts{g}{12}{c} ;\acts{g}{12}{f},{\bf g}_1,{\bf g}_2)&\tilde{U}^{-1}(\acts{g}{12}{a},\acts{g}{12}{f} ;\acts{g}{12}{d},{\bf g}_1,{\bf g}_2) = \tilde{F}^{abc}_{def}({\bf g}_2) \label{eqn:tildeFConsistency}
\\ 
  \tilde{U}(\acts{g}{12}{b},\acts{g}{12}{a} ;\acts{g}{12}{c},{\bf g}_1,{\bf g}_2)\tilde{R}^{\act{12}{a}\act{12}{b}}_{\act{12}{c}}({\bf g}_1)\tilde{U}^{-1}(\acts{g}{12}{a},\acts{g}{12}{b} ;\acts{g}{12}{c},{\bf g}_1,{\bf g}_2) &= \tilde{R}^{ab}_c({\bf g}_2) \label{eqn:tildeURConsistency}
\\
\tilde{U}(\acts{g}{21}{a},\acts{g}{21}{b} ;\acts{g}{21}{c} ,{\bf g}_2,{\bf g}_3)\tilde{U}(a,b;c,{\bf g}_1, {\bf g}_2) &= \tilde{U}(a, b;c, {\bf g}_1, {\bf g}_3)\frac{\eta_c({\bf g}_1, {\bf g}_2, {\bf g}_3)}{\eta_a({\bf g}_1, {\bf g}_2, {\bf g}_3)\eta_b({\bf g}_1, {\bf g}_2, {\bf g}_3)}
\label{UUoverU_equals_EtaOverEtaEta}
\\
  \tilde{\eta}_{{}^{{\bf g}_{21}}\!a}({\bf g}_2,{\bf g}_3,{\bf g}_4) \tilde{\eta}_a({\bf g}_1, {\bf g}_3, {\bf g}_4)
                                                             &= \tilde{\eta}_a({\bf g}_1,{\bf g}_2,{\bf g}_3) \tilde{\eta}_a({\bf g}_1, {\bf g}_2, {\bf g}_4)
                                                              \label{EtaEta_equals_EtaEta}
\end{align}
\end{widetext}
The top three are just the standard pentagon and hexagon equations from BTCs without symmetry. The next two ensure the symmetry action is compatible with the $F$- and $R$-symbols. The next ensures that the symmetry action and symmetry fractionalization are consistent with each other, and the last one is a generalized associativity condition for the $\eta$ symbols.

These data are subject to an additional class of gauge transformations, referred to as symmetry action gauge transformations, which arise by changing $\rho$ by a natural isomorphism: \cite{barkeshli2019}
\begin{align}
  U_{\bf g}(a,b;c) &\rightarrow \frac{\gamma_{a}({\bf g}) \gamma_b({\bf g})}{ \gamma_c({\bf g}) } U_{\bf g}(a,b;c)
\nonumber \\
  \eta_a({\bf g}, {\bf h}) & \rightarrow \frac{\gamma_a({\bf g h}) }{(\gamma_{\,^{\bf g} a}({\bf h}))^{s({\bf g})} \gamma_a({\bf g}) } \eta_a({\bf g}, {\bf h})
  \label{eq:naturaliso}
\end{align}
These symmetry gauge transformations may also be performed in the tilded notation:
\begin{equation}
    \tilde{\gamma}(a;{\bf g}_i, {\bf g}_j) = \gamma_a({\bf g}_{ij})^{s({\bf g}_i)}
\end{equation}
In this paper we will always fix the gauge
\begin{align}
  \eta_1({\bf g},{\bf h})=\eta_a({\bf 1},{\bf g}) = \eta_a({\bf g},{\bf 1})&=1
                                                                             \nonumber \\
  U_{\bf g}(1,b;c)=U_{\bf g}(a,1;c)&=1.
  \end{align}
  
One can show that symmetry fractionalization forms a torsor over $\mathcal{H}^2_{\rho}(G, \mathcal{A})$ in the bosonic case. That is, different possible patterns of symmetry fractionalization can be related to each other by elements of $\mathcal{H}^2_{\rho}(G, \mathcal{A})$. In particular, given an element $[\mathfrak{t}] \in \mathcal{H}^2_{\rho}(G, \mathcal{A})$, we can change the symmetry fractionalization class as
\begin{align}
\eta_a({\bf g}, {\bf h}) \rightarrow \eta_a({\bf g}, {\bf h}) M_{a \mathfrak{t}({\bf g},{\bf h})},
\end{align}
where $\mathfrak{t}({\bf g},{\bf h}) \in \mathcal{A}$ is a representative 2-cocycle for the cohomology class $[\mathfrak{t}]$ and $M_{ab}$ is the mutual braiding in Eq.~\eqref{doubleBraid}.

\subsection{ Fermionic symmetries}
\label{fermSymFracSec}

        We presently sketch a realization of fermionic symmetry in fermionic topological phases where $\C$ is super-modular, summarizing results from ~\cite{bulmashSymmFrac, aasen21ferm}.
        
        To extend the above formalism to construct symmetry actions of fermionic symmetry group $G_f$ on the super-modular tensor category $\C$, rather than directly plugging $G_f$ into the above formalism, we must instead consider $G_f$ as a central extension of $G_b$ and construct a symmetry action of $G_b$ on $\C$, taking into account certain additional constraints. 
        
        Analogously to the bosonic case, the $G_b$ action of $\mathcal{C}$ is again realized as a topological symmetry of BTC. One can immediately see that $G_b$ cannot permute $\psi$; this follows from the super-modularity of $\mathcal{C}$, where the topological symmetry has to preserve the transparent fermion $\psi$.
        
        In the fermionic case, we need to account for the locality of the fermions and for the fact that $G_f$ is present and may be a non-trivial extension of $G_b$. We discuss how to do this below.
        
        Locality of the fermion imposes a key constraint 
        \begin{equation}
            U_{\bf g}(\psi, \psi;1)=1.
            \label{eqn:Upsi1}
        \end{equation}
        Physically, this arises from the idea that symmetry localization separates the action of a symmetry into a product of its local action on anyons and its action on the topological state space. In the case of the fermion, the symmetry transformation rules of the local fermion operators are set entirely by the local Hilbert space, that is, the action of the symmetry on states containing only fermions must be determined entirely by the local action of the symmetry. Thus the action on the topological state space given by $U_{\bf g}(\psi,\psi;1)$ must be trivial. 
        This also means that there is a constraint on symmetry gauge transformations
        \begin{equation}
            \gamma_{\psi}({\bf g})=1.
            \label{eqn:gammaPsiConstraint}
        \end{equation}
        This is because a symmetry gauge transformation changes the local action of the symmetry operators on anyons but, as discussed above, the local action of symmetries on fermions is fixed from the outset by the microscopic Hilbert space because fermions are local excitations. Therefore, such gauge transformations are not allowed. 

        Finally, we must include the fact that $G_f$ may be a non-trivial central extension of $G_b$. We incorporate this by demanding that
        \begin{eqnarray}
        \eta_{\psi}({\bf g,h}) = \omega_2({\bf g,h})
        \label{eqn:etaPsiOmega2Constraint}
        \end{eqnarray}
        where $\omega_2 \in Z^2(G_b,\Z_2)$ specifies the central extension. Eq.~\eqref{eqn:etaPsiOmega2Constraint} is gauge-invariant under the restricted set of gauge transformations which obey Eq.~\eqref{eqn:gammaPsiConstraint}. 
        
        The constraint Eq.~\eqref{eqn:etaPsiOmega2Constraint} requires some physical and mathematical explanation. Physically, the fermion cannot carry fractional quantum numbers of $G_f$ since it is a local excitation. However, if ${\bf g,h}\in G_b$ and, if considered in $G_f$, ${\bf gh} = (-1)^f {\bf k}$ for some ${\bf k} \in G_b$, from the perspective of $G_b$ the fermion appears to carry a fractional quantum number because it picks up a minus sign from the action of $(-1)^f$. Hence we can reinterpret minus signs from fermion parity, that is, $\omega_2$, as symmetry fractionalization of $G_b$ on the fermion.

        Mathematically, our requirement that $\eta_{\psi}$ must be a $\Z_2$ cocycle is actually necessary, as follows.  Since $\rho_{\bf g}$ cannot permute $\psi$, the consistency equation Eq.~\eqref{EtaEta_equals_EtaEta} for $\eta_{\psi}$ becomes a 2-cocycle condition. Furthermore, using the constraint Eq.~\eqref{eqn:Upsi1} the $U$-$\eta$ consistency condition Eq.~\eqref{UUoverU_equals_EtaOverEtaEta} forces $\eta_{\psi}({\bf g,h})\in \Z_2$. Hence $\eta_{\psi} \in Z^2(BG_b,\Z_2)$ and it makes sense to set it equal to $\omega_2$. 
        
        \subsection{Locality-respecting natural isomorphisms in the fermionic case}
        While we defined the global symmetry of the bosonic topological phases as $[\rho] : G \rightarrow \text{Aut}(\mathcal{C})$, the formulation of global symmetries in the fermionic case gets modified due to the additional constraints Eqs.~\eqref{eqn:Upsi1},~\eqref{eqn:gammaPsiConstraint}.
        
        Given any element of $\mathrm{Aut}(\mathcal{C})$, one can always find a representative that satisfies $U(\psi,\psi;1)=+1$, since one can modify generic $U(\psi,\psi;1)$ to unit by a natural isomorphism with $\gamma_\psi=U(\psi,\psi;1)^{-\frac{1}{2}}$. 
        Then, let us consider the equivalence classes of topological symmetries satisfying Eq.~\eqref{eqn:Upsi1}, modulo natural isomorphisms that preserve Eq.~\eqref{eqn:Upsi1}. The group of the equivalence classes is again given by Aut($\mathcal{C}$). Since we set $U(\psi,\psi;1)=1$, the natural isomorphism mentioned here satisfies $\gamma_{\psi}=\pm 1$ to preserve this constraint.
        
        However, since we want the natural isomorphisms of a fermionic system to satisfy the locality constraint Eq.~\eqref{eqn:gammaPsiConstraint},
        we instead have to take a equivalence class of topological symmetries Aut$_{LR}(\mathcal{C})$ modulo the natural isomorphisms that respect the locality $\gamma_{\psi}=+1$. Accordingly, in a fermionic system we must specify a map
        \begin{align}
            [\rho]: G_b\to\mathrm{Aut}_{LR}(\mathcal{C}),
        \end{align}
        where 
        \begin{align}
            \rho_{\bf gh} = \kappa_{\bf g, h} \circ \rho_{\bf g} \circ \rho_{\bf h},
        \end{align}
        where $\kappa_{\bf g,h}$ is a natural isomorphism. 
        $\mathrm{Aut}_{LR}(\mathcal{C})$ may or may not be isomorphic to $\mathrm{Aut}(\mathcal{C})$. To study the difference of these groups, we consider a natural isomorphism that satisfies $\gamma_{\psi}=-1$, given by
        \begin{align}
    \Upsilon_{\psi} (\ket{a,b;c}) = \frac{\gamma_a\gamma_b}{\gamma_c}\ket{a,b;c}
\end{align}
with $\gamma_a=1$ for all $a\neq \psi$ and $\gamma_{\psi}=-1$. Though we have $\gamma_{\psi}=-1$ for $\Upsilon_{\psi}$, it is still possible that the natural isomorphism $\Upsilon_{\psi}$ preserves the locality. This is because one can modify $\gamma_a\to \gamma_a\zeta_a$ that satisfies $\zeta_a \zeta_b=\zeta_c$ when $N^{ab}_c\neq 0$ which leaves $\Upsilon_{\psi}$ invariant; if one can find such phases $\{\zeta_a\}$ with $\zeta_{\psi}=-1$, $\Upsilon_{\psi}$ reduces to a locality-respecting natural isomorphism with $\gamma_{\psi}=+1$.

If $\Upsilon_{\psi}$ respects the locality, it means that any natural isomorphisms with $\gamma_{\psi}=-1$ respect the locality, and hence $\mathrm{Aut}_{LR}(\mathcal{C})\cong \mathrm{Aut}(\mathcal{C})$. Otherwise we have $\mathrm{Aut}_{LR}(\mathcal{C})/\Z_2\cong \mathrm{Aut}(\mathcal{C})$, since $[\Upsilon_{\psi}]\in\mathrm{Aut}_{LR}(\mathcal{C})$ gives a nontrivial group element. We thus have checked that the following three statements are equivalent with each other:
\begin{itemize}
    \item $\Upsilon_{\psi}$ respects the locality
    \item $\mathrm{Aut}_{LR}(\mathcal{C})\cong \mathrm{Aut}(\mathcal{C})$
    \item There exists a set of phases $\{\zeta_a\}$ that satisfies $\zeta_{\psi}=-1$, and $\zeta_a \zeta_b=\zeta_c$ when $N^{ab}_c\neq 0$.
\end{itemize}

Note that the natural isomorphism $\kappa_{\bf g, h}$ may in some cases correspond to $\Upsilon_\psi$. This means that if $\Upsilon_\psi$ is locality-violating, then there can be cases where $[\kappa_{\bf g,h}] = [\Upsilon_\psi]$ is non-trivial. In these cases, the map $[\rho]$ is not a faithful group homomorphism \cite{aasen21ferm,bulmashSymmFrac}. In general, when the obstructions to defining fermionic symmetry fractionalization on the super-modular $\mathcal{C}$ vanish, we have
\begin{align}
    [\kappa_{\bf g, h}] = [\Upsilon_\psi]^{(1-\omega_2({\bf g,h}))/2}
\end{align}

Once we specify the map $[\rho]: G_b\to\mathrm{Aut}_{LR}(\mathcal{C})$, the symmetry fractionalization of super-modular $\mathcal{C}$ is given in the same fashion as Appendix~\ref{globsym}, under the constraints Eqs.~\eqref{eqn:Upsi1},~\eqref{eqn:gammaPsiConstraint},~\eqref{eqn:etaPsiOmega2Constraint}.

Finally, let us refer to an important consequence of the above discussion about the U(1) symmetry fractionalization. Since the fractional charge $e^{\pi i Q_a}$ defined in Eq.~\eqref{eq:chargedefspinc} satisfies $e^{i\pi Q_\psi}=-1$ and $e^{i\pi Q_a}e^{i\pi Q_b}=e^{i\pi Q_c}$ for $N^{ab}_c\neq 0$, the fractional charge exists if and only if $\Upsilon_{\psi}$ respects the locality. 
 
\section{Graphical calculus of Villain symmetry defects in the presence of $H$ symmetry defects}
\label{app:tilde}

In the main text, we consider the global symmetry $\U\rtimes H$ by incorporating the U(1) Villain symmetry defects and $H$ symmetry defects in the graphical calculus.
In this appendix, we provide the diagrammatic calculus of U(1) Villain symmetry defects in the presence of $H$ defects.
We first describe a symmetry fractionalization data $(U,\eta)$ for flat $\U\rtimes H$ background gauge field, and then provide the Villain calculus for curved $\U\rtimes H$ gauge fields.

\subsection{Symmetry fractionalization data for $\U\rtimes H$ symmetry}

Here we describe the $(U,\eta)$ symbols for $\U\rtimes H$ symmetry fractionalization for the bosonic case where $\mathcal{C}$ is modular. 

Let us recall the symmetry action of $H$ on the category and U(1) gauge fields. The $H$ symmetry action on BTC is classified by a $\mathbb{Z}_2$ grading corresponding to whether $\mathbf{g}\in H$ has a unitary or anti-unitary action on the category:
\begin{align} \label{antiunitaryactionapp}
s(\mathbf{g}) = \left\{
\begin{array} {ll}
+1 & \text{if $\mathbf{g}$ is unitary} \\
-1  & \text{if $\mathbf{g}$ is anti-unitary} \\
\end{array} \right.
\end{align}

In addition, we allow $H$ to act on $\U$ by the complex conjugation. The $\mathbf{g}\in H$ symmetry action $\sigma_{\mathbf{g}}$ on $\U$ gauge fields is classified by the $\Z_2$ grading of $H$:
\begin{align}
    \sigma_{\mathbf{g}}: x\to \sigma(\mathbf{g})\cdot x
\end{align} 
with $x\in\U$ and $\sigma(\mathbf{g})\in\{\pm 1\}$.

We write an element of $\U\rtimes H$ as $(x,\mathbf{g})\in\U\times H$, with the multiplication law $(x,\mathbf{g})\cdot (y,\mathbf{h})=(x+^\mathbf{g}y,\mathbf{gh})$ with $^\mathbf{g}y:=\sigma(\mathbf{g})\cdot y $.

We can characterize symmetry fractional in general as follows. First, we can pick a certain reference symmetry fractionalization class where the $\U$ acts completely trivially. In this case, we have
\begin{align}
    \eta_a^{\text{ref}} ( (x,{\bf g}), (y, {\bf h})) &= \eta_a'({\bf g}, {\bf h})
    \nonumber \\
    U^{\text{ref}}_{(x,{\bf g})}(a,b;c) &= U'_{\bf g} (a,b;c) ,
\end{align}
where $\{U'_{\mathbf{g}}(a,b;c), \eta'_a(\mathbf{g},\mathbf{h})\}$ characterize some arbitrary reference symmetry fractionalization pattern for $H$ alone. 

Then, any generic symmetry fractionalization data $\{U,\eta\}$ for $\U\rtimes H$ symmetry can be expressed as
\begin{align}
    \begin{split}
        \eta_a((x,\mathbf{g}),(y,\mathbf{h})) &= \eta'_a(\mathbf{g},\mathbf{h})\cdot M_{a,\mathfrak{t}((x,\mathbf{g}),(y,\mathbf{h}))}\\
        U_{(x,\mathbf{g})}(a,b;c)&= U'_{\mathbf{g}}(a,b;c)
        \label{eq:UetaGdiscgeneral}
    \end{split}
\end{align}
with $[\mathfrak{t}]\in \mathcal{H}^2_{\rho}(\U\rtimes H,\mathcal{A})$. 

We conjecture that the most general form of $\mathfrak{t}$ that covers all the representatives of $\mathcal{H}^2_{\rho}(\U\rtimes H,\mathcal{A})$ is given by\footnote{In principle, one can compute $\mathcal{H}^2_{\rho}(\U\rtimes H,\mathcal{A})$ by the Lyndon-Hochschild-Serre spectral sequence, which basically evaluates the cohomology from the data of $\mathcal{H}^p_{\rho}(H,\mathcal{H}^q(\U,\mathcal{A}))$ with $p+q=2$. Note that since $\mathcal{H}^1(\U, M)$ vanishes for any discrete group $M$, 
$\mathcal{H}^2_{\rho}(\U\rtimes H,\mathcal{A})$ does not get any contributions from $(p,q)=(1,1)$. Consequently we expect the cocycles to split into separate cocycles for the $\U$ and $H$ pieces, apart from the $H$ action on the $\U$ elements. }
\begin{align}
    \mathfrak{t}((x,\mathbf{g}),(y,\mathbf{h}))= v^{x+^\mathbf{g}y-[x+^\mathbf{g}y]}\mathfrak{t}'(\mathbf{g},\mathbf{h})
    \label{eq:visonGdisc}
\end{align}
where $v\in\mathcal{A}$ is a vison for U(1) symmetry fractionalization, and $[x + \,^\mathbf{g}y]$ means the sum of the two elements in $\U=\R/\Z$ taken mod 1. $\mathfrak{t}'$ controls the $H$ symmetry fractionalization where we have $\delta_{\rho}\mathfrak{t}'=0$, and hence $[\mathfrak{t}']\in \mathcal{H}^2_{\rho}(H,\mathcal{A})$.

We next show that $\mathfrak{t}$ given in Eq.~\eqref{eq:visonGdisc} gives the element of $Z^2_{\rho}(\U\rtimes H,\mathcal{A})$ if and only if the vison $v$ transforms under $\mathbf{g}\in H$ as
\begin{align}
    ^{\mathbf{g}}v=v^{\sigma(\mathbf{g})}.
    \label{eq:visonconstraint}
\end{align}
This is seen by the cocycle condition of $\mathfrak{t}$ that reduces to the following equation due to $\delta_{\rho}\mathfrak{t}'=0$,
\begin{align}
    (^{\mathbf{g}_{21}}v)^{\overline{\delta}_{\sigma}x(234)} v^{\overline{\delta}_{\sigma} x(124)}=v^{\overline{\delta}_{\sigma} x(123)}v^{\overline{\delta}_{\sigma} x(134)}
    \label{eq:consistencyvison}
\end{align}
where we introduced a twisted coboundary operation for $\omega\in C^{d}(M,\U)$ as
\begin{align}
    \overline{\delta}_{\sigma}\omega_{(01\dots d+1)}=\sigma({\mathbf{g}_{01}})\omega_{(1\dots d+1)}+\sum_{i=1}^{d+1}(-1)^{i}\cdot \omega_{(0\dots\hat{i}\dots d+1)}
\end{align}
In particular, $\overline{\delta}_{\sigma}x(ijk)=x_{ij}+^{\mathbf{g}_{ij}}x_{jk}-[x_{ij}+^{\mathbf{g}_{ij}}x_{jk}]$ since $x_{ik}=[x_{ij}+^{\mathbf{g}_{ij}}x_{jk}]$.
Meanwhile, since the above twisted coboundary satisfies $\overline{\delta}_{\sigma}\overline{\delta}_{\sigma}=0$, we have
\begin{align}
    v^{\sigma(\mathbf{g}_{21})\overline{\delta}_{\sigma}x(234)} v^{\overline{\delta}_{\sigma} x(124)}=v^{\overline{\delta}_{\sigma} x(123)}v^{\overline{\delta}_{\sigma} x(134)}
    \label{eq:coboundaryvison}
\end{align}
Comparing Eq.~\eqref{eq:consistencyvison} and Eq.~\eqref{eq:coboundaryvison} shows Eq.~\eqref{eq:visonconstraint}.

Let us now consider the fractional charge $e^{2\pi iQ_a}:= M_{a,v}$. Since we have $M_{^{\mathbf{g}}a,v}=M_{a,^{\mathbf{g}}v}^{s(\mathbf{g})}=M_{a,v}^{s(\mathbf{g})\sigma(\mathbf{g})}$, the fractional charge transforms as
\begin{align}
Q_{^{\mathbf{g}}a}=s(\mathbf{g})\sigma(\mathbf{g})Q_{a} \mod 1.
\label{eq:Qtransform}
\end{align}
By rewriting the $\eta(\mathbf g, \mathbf h)=\eta'(\mathbf g, \mathbf h)\cdot M_{a,\mathfrak{t}'(\mathbf g, \mathbf h)}$ and $U_{\mathbf{g}}(a,b;c)=U'_{\mathbf{g}}(a,b;c)$, the fractionalization data in Eq.~\eqref{eq:UetaGdiscgeneral} is expressed as
\begin{align}
     \begin{split}
        \eta_a((x,\mathbf{g}),(y,\mathbf{h})) &= e^{2\pi i Q_a(x+^\mathbf{g}y-[x+^\mathbf{g}y])}\eta_a(\mathbf{g},\mathbf{h})\\
        U_{(x,\mathbf{g})}(a,b;c)&= U_{\mathbf{g}}(a,b;c)
        \label{eq:UetaGdisc}
    \end{split}
\end{align}

In summary, we have constructed a symmetry fractionalization data for $\U\rtimes H$ symmetry, based on the data $\{U_{\mathbf{g}}(a,b;c), \eta_a(\mathbf{g},\mathbf{h})\}$ for $H$ symmetry and fractional U(1) charges $Q_a$, satisfying the property Eq.~\eqref{eq:Qtransform}. This is expected to be a most general form of the symmetry fractionalization data, and we take these data as the input for the construction of (3+1)D path integral given in Sec.~\ref{sec:u1rtimesg} of the main text.

The generalization to the fermionic case where $\mathcal{C}$ is super-modular is straightforward, and we can again construct $U,\eta$ symbols for $G_b=\U\rtimes H$ based on those for $H$ symmetry and fractional U(1) charges $Q_a$, satisfying the property Eq.~\eqref{eq:Qtransform}. In the fermionic case, the symmetry fractionalization data is given by
\begin{align}
    \begin{split}
        \eta_a((x,\mathbf{g}),(y,\mathbf{h})) &= e^{2\pi i Q_a(x+^\mathbf{g}y-[x+^\mathbf{g}y]_{1/2})}\eta_a(\mathbf{g},\mathbf{h})\\
        U_{(x,\mathbf{g})}(a,b;c)&= U_{\mathbf{g}}(a,b;c)
        \label{eq:UetaGdiscfermion}
    \end{split}
\end{align}
with the fractional charge satisfying
\begin{align}
Q_{^{\mathbf{g}}a}=s(\mathbf{g})\sigma(\mathbf{g})Q_{a} \mod 2.
\label{eq:Qtransformfermion}
\end{align}

\subsection{Villain calculus for $\U\rtimes H$ symmetry}
\begin{figure}[htb]
    \centering
    \includegraphics{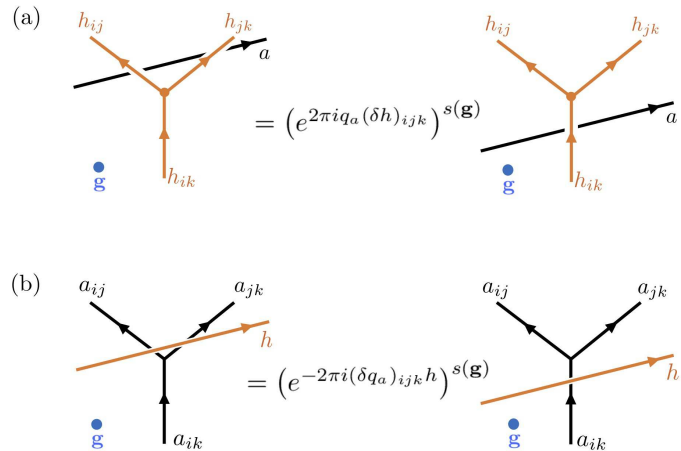}
    \caption{The graphical calculus for the $\R$ symmetry defects in the presence of $H$ defects. The blue dot denotes the $H$ domain in the background.}
    \label{fig:Uetau1tilde}
\end{figure}

\begin{figure}[htb]
    \centering
    \includegraphics{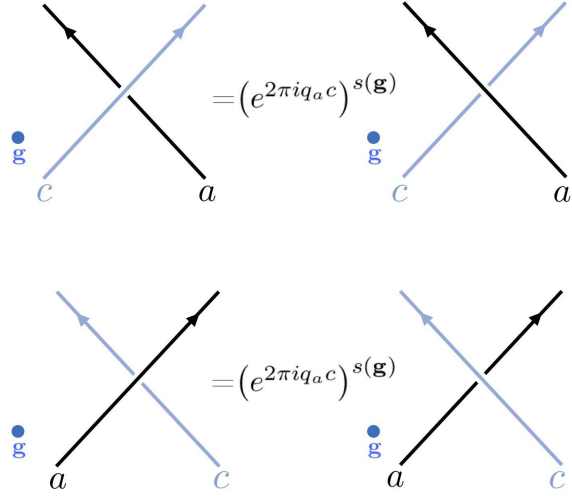}
    \caption{The action of the $\Z$ 1-form symmetry on anyons.}
    \label{fig:Zactiontilde}
\end{figure}

Now we provide the diagrammatic calculus of U(1) Villain symmetry defects in the presence of $H$ defects.
The $H$ action on Villain gauge fields is described by the $\Z_2$ grading as
\begin{align} \label{conjugateactionapp}
\sigma(\mathbf{g}) = \left\{
\begin{array} {ll}
+1 & \text{if $\sigma_{\mathbf{g}}: (h,c)\to (h,c)$} \\
-1  & \text{if $\sigma_{\mathbf{g}}: (h,c)\to(-h,-c)$} \\
\end{array} \right.
\end{align}
The calculus involving the Villain symmetry defects can be defined based on $\U\rtimes H$ symmetry fractionalization data provided in the previous subsection. We also use the $\R$ lift $q_a$ of the charge $Q_a\in\U$ to define the calculus.
In the presence of $H$ defects, the diagram is defined by setting a label of the symmetry domain $\mathbf{g}\in H$ represented by a blue dot, as we have seen in Appendix~\ref{app:symmfrac}. The calculus of Villain symmetry defects are shown in Figs.~\ref{fig:Uetau1tilde},~\ref{fig:Zactiontilde}.

\begin{figure}[htb]
    \centering
    \includegraphics{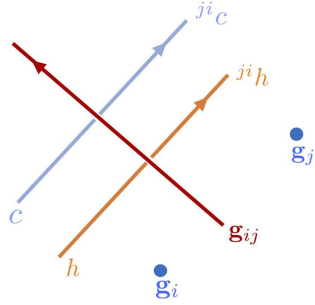}
    \caption{The action of $H$ symmetry defects on Villain symmetry defects $(h,c)$.}
    \label{fig:GactionU1}
\end{figure}

Since $H$ can act on $\U$ by complex conjugation, $H$ elements can change the label of $\U$ Villain symmetry defects by crossing over them, as shown in Fig.~\ref{fig:GactionU1}.

\begin{figure}[htb]
    \centering
    \includegraphics{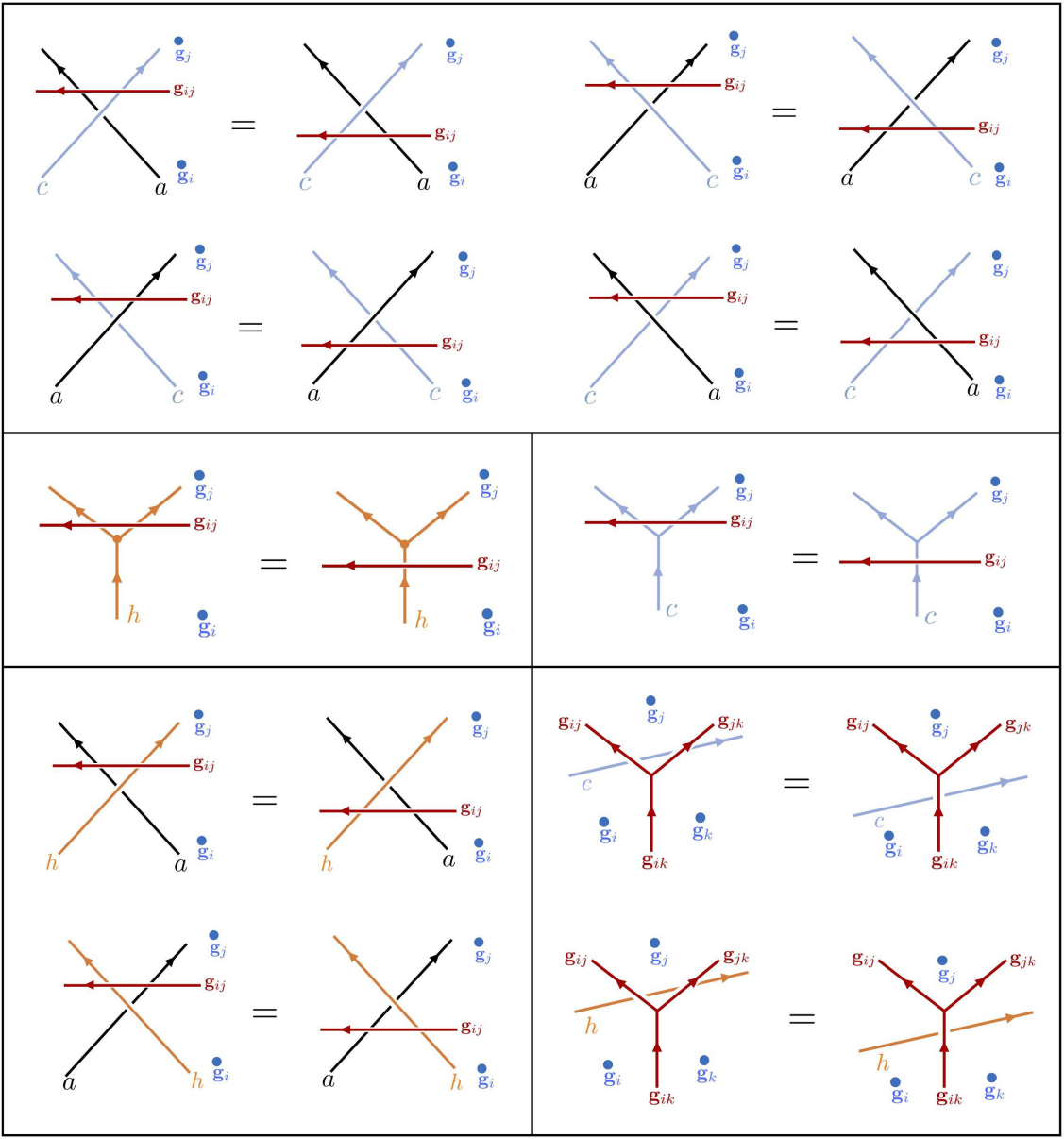}
    \caption{One can freely move the $H$ symmetry defects above the Villain symmetry defects as summarized in this table.}
    \label{fig:freemoves}
\end{figure}

Then, the rest of the calculus involving $H$ and U(1) defects are shown in Fig.~\ref{fig:freemoves}. These diagrammatics let the $H$ defects freely move above the $h$ defects without producing phases, as long as it does not cause $U$ and $\eta$ moves between anyons shown in Fig.~\ref{fig:graphicalCalculus_defs}.

Finally, freely moving the $H$ defect $\mathbf{g}_{ij}$ on Figs.~\ref{fig:Uetau1tilde},~\ref{fig:Zactiontilde} puts the consistency conditions that the lifted fractional charges $q_a$ need to satisfy:
\begin{align}
    \begin{split}
        \left(e^{2\pi iq_a(\delta h)}\right)^{s(\mathbf{g}_i)} &= \left(e^{2\pi iq_{^{ji}a}(\delta h)\cdot\sigma(\mathbf{g}_{ji})}\right)^{s(\mathbf{g}_j)} \\
        \left(e^{2\pi i(\delta q_a)h}\right)^{s(\mathbf{g}_i)} &= \left(e^{2\pi i(\delta q_{^{ji}a})h\cdot\sigma(\mathbf{g}_{ji})}\right)^{s(\mathbf{g}_j)} \\
        \left(e^{2\pi iq_a c}\right)^{s(\mathbf{g}_i)} &= \left(e^{2\pi iq_{^{ji}a}c\cdot\sigma(\mathbf{g}_{ji})}\right)^{s(\mathbf{g}_j)}
    \end{split}
\end{align}
These consistency equations are summarized in a single equation
\begin{align}
    q(^{ji}a)=s(\mathbf{g}_{ji})\sigma(\mathbf{g}_{ji})q(a),
\end{align}
which is a mod 1 relation in the bosonic case, while mod 2 in the fermionic case. This equation follows from Eq.~\eqref{eq:Qtransform}, Eq.~\eqref{eq:Qtransformfermion}.

\section{Demonstration of Spin$^c$ Gauss-Milgram formula}
\label{app:gauss}

In this appendix, we explicitly compute the indicator formula that computes $c_-$ mod 1 for a given Spin$^c$ topological order, using several non-Abelian fractional quantum Hall states. The formula is given by
\begin{align}
            e^{i\Theta_1}=e^{-2\pi i c_-}= \frac{Z\left(\mathbb{CP}^2,C=\frac{3}{2}\right)}{Z\left(\mathbb{CP}^2,C=\frac{1}{2}\right)^9},
    \end{align}
    with 
        \begin{align}
        Z(\mathbb{CP}^2,C)=\frac{1}{\sqrt{2}\mathcal{D}}\sum_{a\in\mathcal{C}}d_a^2e^{-2\pi iQ_a\cdot C}\theta_a.
        \label{eq:CP2app}
    \end{align}
    To demonstrate this formula, we consider a series of Hall states with non-Abelian anyons known as the Read-Rezayi states $\mathrm{RR}_{k,M}$~\cite{readrezayi1999}. They are labeled by a pair of integers $(k,M)$ and carry the chiral central charge $c_-=\frac{3k}{k+2}$, and describe fermionic topological orders for odd $M$. For example, the case with $(k=2,M=1)$ corresponds to the Moore-Read state with $c_-=1/2$ mod 1~\cite{Moore1991}. Below we list the anyon data and results of the computations for several fermionic Read-Rezayi states. The anyon data of these Hall states are described in~\cite{Bonderson07b,BarkeshliWen2009,BarkeshliWen2010}.
    
    \subsection{Moore-Read state: $(k,M)=(2,1)$}
    
    The Moore-Read state $\mathrm{RR}_{2,1}$ is given by taking the subset of anyons in $\mathrm{Ising}\times \U_{8}$ as 
    \begin{align}
        \{(1, [2n]_8), (\sigma, [2n+1]_8), (\psi, [2n]_8)\},
    \end{align}
    with $n\in\Z$. The Ising anyons are written as $\{1,\sigma,\psi\}$ and anyons in $\U_k$ are written as $[j]_k$ with $j\in\Z$.
    The quantum dimensions are given by $d_I=1, d_\sigma=\sqrt{2}, d_\psi=1$, and $\mathcal{D}=4$.
    The fractional charge $Q_a\in\R/(2\Z)$ is characterized by anyons of $\U_{8}$; the anyon that carries $[j]_8$ in $\mathrm{RR}_{2,1}$ has the fractional charge $Q_a=j/4$ mod 2. The category is super-modular, and the anyon $(\psi,[4]_8)$ becomes a transparent fermion with $Q=1$ mod 2.
    
    The spins $\{h_a\}$ of anyons with $\theta_a=e^{2\pi i h_a}$ are given in Table~\ref{tab:spin21}. The partition functions in Eq.~\eqref{eq:CP2app} are then given by
    \begin{align}
        Z\left(\mathbb{CP}^2,C=\frac{1}{2}\right)=e^{\frac{\pi i}{4}}, \quad  Z\left(\mathbb{CP}^2,C=\frac{3}{2}\right)=e^{\frac{5\pi i}{4}}.
    \end{align}
    The indicator is obtained as
    \begin{align}
            e^{-2\pi i c_-}= \frac{Z\left(\mathbb{CP}^2,C=\frac{3}{2}\right)}{Z\left(\mathbb{CP}^2,C=\frac{1}{2}\right)^9}=-1,
    \end{align}
    which correctly predicts $c_-=\frac{1}{2}$ mod 1.
    
    \begin{table}[t]
	\centering
	\begin{tabular} {c ||c |c |c| c| c| c| c| c}
	\ & [0]$_8$ & [1]$_8$ & [2]$_8$ & [3]$_8$ & [4]$_8$ & [5]$_8$ & [6]$_8$ & [7]$_8$ \\
	\hline
	$1$ & 0 & \ & $\frac{1}{4}$ & \ & 0 & \ & $\frac{1}{4}$ & \ \\
	$\sigma$ &\ & $\frac{1}{8}$& \ & $\frac{5}{8}$ & \ & $\frac{5}{8}$ & \ & $\frac{1}{8}$ \\
	$\psi$ &$\frac{1}{2}$ & \ & $\frac{3}{4}$ & \ & $\frac{1}{2}$ & \ & $\frac{3}{4}$ &\ \\
      \end{tabular}
      \caption{Spins of anyons in $\mathrm{RR}_{2,1}$.}
      \label{tab:spin21}
\end{table}

\subsection{Read-Rezayi state with $(k,M)=(3,1)$}

The Read-Rezayi state $\mathrm{RR}_{3,1}$ is given by $\mathrm{Fib}\times \Z_5^{(1)}\times \Z_2^{(1)}$, where $\Z_n^{(1)}$ is an Abelian topological order with $\mathcal{A}=\Z_n$, $\theta_a=e^{\frac{2\pi i }{N}a^2}$ and $M_{a,b}=e^{\frac{4\pi i}{N}ab}$ for $a,b\in\Z_n$. $\mathrm{Fib}=\{1,\phi\}$ denotes the Fibonacci category, which has a single nontrivial anyon $\phi$ with the fusion rule $\phi^2=1+\phi$, and $d_{\phi}=\frac{1+\sqrt{5}}{2}$, $\theta_{\phi}=e^{\frac{4\pi i}{5}}$. $\mathrm{RR}_{3,1}$ is super-modular, with the transparent fermion given by the nontrivial anyon $[1]_2$ of $\Z_2^{(1)}$.

For an anyon of $\mathrm{RR}_{3,1}$ labeled as $(a,[j]_5,[k]_2)$ with $a\in\mathrm{Fib}$, $j\in\Z_5, k\in\Z_2$, the fractional charge depends on $j,k$ and given by $Q_{j,k}=-\frac{4}{5}j+k$ mod 2.
The partition functions in Eq.~\eqref{eq:CP2app} are  given by
    \begin{align}
        Z\left(\mathbb{CP}^2,C=\frac{1}{2}\right)=e^{\frac{2\pi i }{8}c_{\mathrm{Fib}}}e^{\frac{-2\pi i}{5}}, \quad  Z\left(\mathbb{CP}^2,C=\frac{3}{2}\right)=e^{\frac{2\pi i }{8}c_{\mathrm{Fib}}}e^{\frac{2\pi i}{5}},
    \end{align}
where $c_{\mathrm{Fib}}=\frac{14}{5}$ is the chiral central charge of the Fibonacci category.
    The indicator is obtained as
    \begin{align}
            e^{-2\pi i c_-}= \frac{Z\left(\mathbb{CP}^2,C=\frac{3}{2}\right)}{Z\left(\mathbb{CP}^2,C=\frac{1}{2}\right)^9}=e^{-2\pi i c_{\mathrm{Fib}}},
    \end{align}
    which correctly predicts $c_-=c_{\mathrm{Fib}}=\frac{4}{5}$ mod 1.
    
    \subsection{Read-Rezayi state with $(k,M)=(4,1)$}
The Read-Rezayi state $\mathrm{RR}_{4,1}$ is described as follows. First, we consider a topological order $\mathrm{Pf}_4$ whose anyons correspond to the chiral primaries of $\Z_4$ parafermion CFT. $\mathrm{Pf}_4$ is defined by first preparing the subset of $\SU(2)_4\times \U_{-8}$ specified as
\begin{align}
    (j,[m]_8): 2j+m = 0 \mod 2,
\end{align}
where the anyons are labeled by $j\in\{0,\frac{1}{2}, 1, \frac{3}{2},2\}$, $m\in\Z_8$. We then obtain $\mathrm{Pf}_4$ by condensing a diagonal boson $(j=2,[4]_8)$ of this sub-category of $\SU(2)_4\times \U_{-8}$. For simplicity, we denote the anyon $(j,[m]_8)\in \mathrm{Pf}_4$ as $\Phi^{2j}_m$. $\mathrm{Pf}_4$ has 10 anyons, whose spins and quantum dimensions are summarized in Table~\ref{tab:pfdata}. Note that we identify the anyons as $\Phi^{2j}_m=\Phi^{2j+4}_{m+4}$ after condensing $\Phi^4_4$.
    \begin{table}[t]
	\centering
	\begin{tabular} {c ||c |c |c| c| c| c| c| c | c | c}
	\ & $\Phi^0_0$ & $\Phi^0_2$ & $\Phi^0_4$ & $\Phi^0_6$ & $\Phi^1_1$ & $\Phi^1_3$ & $\Phi^1_5$ & $\Phi^1_7$ & $\Phi^2_0$ & $\Phi^2_2$ \\
	\hline
	$h_a$ & 0 & $\frac{3}{4}$ & 0 & $\frac{3}{4}$ & $\frac{1}{16}$ & $\frac{9}{16}$ & $\frac{9}{16}$ & $\frac{1}{16}$ &$\frac{1}{3}$ & $\frac{1}{12}$ \\
	$d_a$ &1 & 1& 1 & 1 & $\sqrt{3}$ & $\sqrt{3}$ &$\sqrt{3}$ & $\sqrt{3}$ & 2 & 2 \\
      \end{tabular}
      \caption{Spins and quantum dimensions of anyons in $\mathrm{Pf}_4$.}
      \label{tab:pfdata}
\end{table}

$\mathrm{RR}_{4,1}$ is then constructed by first preparing $\mathrm{Pf}_4\times \U_{24}$ specified as
\begin{align}
    (\Phi^{\Lambda}_{\lambda},[\lambda]_{24}): \Lambda+\lambda = 0 \mod 2,
\end{align}
and then condensing a boson $(\Phi^0_4,12)$. The fractional charges of anyons are determined by $\U_{24}$ part; the charge of $ (\Phi^{\Lambda}_{\lambda},[\lambda]_{24})$ is given by $Q=\lambda/6$ mod 2.
The resulting category $\mathrm{RR}_{4,1}$ is super-modular, and has a transparent fermion $(\Phi^0_2,6)$ with $Q=1$. $\mathrm{RR}_{4,1}$ has 30 anyons, and the data of these anyons are summarized in Table~\ref{tab:41data}.

   \begin{table}[t]
	\centering
	\begin{tabular} {c ||c |c |c| c| c| c|| c| c | c | c | c | c}
	\ & ($\Phi^0_0$,0) & ($\Phi^0_2$,2) & ($\Phi^0_4$,4) & ($\Phi^0_6$,6) & ($\Phi^0_0$,8) & ($\Phi^0_2$,10) & ($\Phi^0_2$,6) & ($\Phi^0_4$,8) & ($\Phi^0_6$,10) & ($\Phi^0_4$,0) & ($\Phi^0_6$,2) & ($\Phi^0_0$,4) \\
	\hline
	$h_a$ & 0 & 5/6 & 1/3 & 1/2 & 1/3 & 5/6 & 1/2 & 1/3 & 5/6 & 0 & 5/6 & 1/3  \\
	$Q_a$ &0 & 1/3& 2/3 & 1 & 4/3 & 5/3 & 1 & 4/3 & 5/3 & 0 & 1/3 & 2/3 \\
	$d_a$ &1 & 1& 1 & 1 & 1 & 1 &1 & 1 & 1 & 1& 1& 1 \\
      \end{tabular}

      \begin{tabular} {c ||c |c |c| c| c| c|| c| c | c | c | c | c}
	\ & ($\Phi^1_1$,1) & ($\Phi^1_3$,3) & ($\Phi^1_5$,5) & ($\Phi^1_7$,7) & ($\Phi^1_1$,9) & ($\Phi^1_3$,11) & ($\Phi^1_3$,7) & ($\Phi^1_5$,9) & ($\Phi^1_7$,11) & ($\Phi^1_5$,1) & ($\Phi^1_7$,3) & ($\Phi^1_1$,5) \\
	\hline
	$h_a$ & 1/12 & 3/4 & 1/12 & 1/12 & 3/4 & 1/12 & 7/12 & 1/4 & 7/12 & 7/12 & 1/4 & 7/12  \\
	$Q_a$ &1/6 & 1/2& 5/6 & 7/6 & 3/2 & 11/6 & 7/6 & 3/2 & 11/6 & 1/6 & 1/2 & 5/6 \\
	$d_a$ &$\sqrt{3}$ & $\sqrt{3}$& $\sqrt{3}$ & $\sqrt{3}$ & $\sqrt{3}$ & $\sqrt{3}$ &$\sqrt{3}$ & $\sqrt{3}$ & $\sqrt{3}$ & $\sqrt{3}$& $\sqrt{3}$& $\sqrt{3}$ \\
      \end{tabular}
      
           \begin{tabular} {c ||c |c |c|| c| c| c}
	\ & ($\Phi^2_0$,0) & ($\Phi^2_2$,2) & ($\Phi^2_0$,4) & ($\Phi^2_2$,6) & ($\Phi^2_0$,8) & ($\Phi^2_2$,10)  \\
	\hline
	$h_a$ & 1/3 & 1/6 & 2/3 & 5/6 & 2/3 & 5/6   \\
	$Q_a$ &0 & 1/3& 2/3 & 1 & 4/3 & 5/3  \\
	$d_a$ &2 & 2& 2 & 2 & 2 &2  \\
      \end{tabular}
      
      \caption{Spins, fractional charges and quantum dimensions of 30 anyons in $\mathrm{RR}_{4,1}$. For each table, the anyons in first half of the table are obtained by fusing the transparent fermion $(\Phi^0_2,6)$ to the anyons in the latter half.}
      \label{tab:41data}
\end{table}

The partition functions in Eq.~\eqref{eq:CP2app} are  given by
    \begin{align}
        Z\left(\mathbb{CP}^2,C=\frac{1}{2}\right)=e^{\frac{\pi i}{3}}, \quad  Z\left(\mathbb{CP}^2,C=\frac{3}{2}\right)=-1.
    \end{align}
       The indicator is obtained as
    \begin{align}
            e^{-2\pi i c_-}= \frac{Z\left(\mathbb{CP}^2,C=\frac{3}{2}\right)}{Z\left(\mathbb{CP}^2,C=\frac{1}{2}\right)^9}=1,
    \end{align}
    which correctly predicts $c_-=0$ mod 1.

\section{Twisted cohomology}
\label{app:twisted}
Here we explain the definition of twisted cohomology $H_{\rho}^*(M,X)$, where $X$ is an Abelian group with $G$ action $\rho: G\to\mathrm{Aut}(X)$. For a given configuration of the $G$ gauge field $g\in Z^1(M,G)$, the twisted coboundary is defined for $\omega\in C^d(M,X)$ as
\begin{align}
    \delta_{\rho}\omega_{(01\dots d+1)}=\sum_{i=0}^{d}(-1)^{i}\cdot \omega_{(0\dots\hat{i}\dots d+1)} + 
    (-1)^{d+1}\cdot\rho_{g_{d+1,d}}[\omega_{(0\dots d)}],
\end{align}
where $\hat{i}$ means skipping over $i$.
One can see that $\delta_{\rho}\delta_{\rho}=0$, so it defines a cohomology twisted by the $G$ action $H_{\rho}^d(M,X):=Z_{\rho}^d(M,X)/B_{\rho}^d(M,X)$.
We also use a twisted version of cup product
\begin{align}
    -\cup_{\rho}-:C^k(M,\Z_2)\times C^l(M,\Z_2)\to C^{k+l}(M,\Z_2),
\end{align}
whose explicit form is written as
\begin{align}
    (\alpha\cup_\rho\beta)_{(0,\dots, k+l)} = \rho_{g_{k+l,k}}[\alpha_{(0,\dots,k)}]\cdot\beta_{(k,\dots,k+l)}.
    \label{eq:cupdef}
\end{align}
The cup product satisfies the twisted Leibniz rule at the cochain level,
\begin{align}
    \delta_{\rho}(\alpha\cup_\rho\beta)=(\delta_{\rho}\alpha)\cup_\rho\beta + (-1)^k\cdot \alpha\cup_\rho(\delta_{\rho}\beta).
\end{align}

\section{Invariance of the partition function under the Pachner move}
\label{app:pachner}

\subsection{Bosonic case}

Here, we demonstrate that our bosonic path integral is invariant under the 3-3 Pachner move following~\cite{cui2019,bulmash2020, tata2021anomalies}. Other Pachner moves can be checked analogously. 

Given five vertices $0,1,2,3,4,5$, a 3-3 Pachner move takes the 4-simplices $\langle 01234\rangle,\langle 01245\rangle,\langle02345\rangle$ to the 4-simplices $\langle 01235\rangle,\langle01345\rangle,\langle12345\rangle$. In doing so, we note that the orientations of the 4-simplices involved are identical (say $+$ orientation) except for the 4-simplex $\langle 01234\rangle$, whose orientation is instead given by $s(45)$.
To see this, we note that the orientation of 4-simplices is assigned based on a chain representative of the first Stiefel-Whitney class $w_1$, which is regarded as an orientation-reversing symmetry defect. As explained in \cite{tata2021anomalies},
the chain representative $w_1\in Z_3(M^4,\Z_2)$ is obtained from $A_H^*s\in Z^1(M^4,\Z_2)$ by using a so-called $f_{\infty}$ map \cite{Thorngrenthesis}. According to the $f_{\infty}$ map, the chain representative $w_1$ is locally given by $\partial(\braket{01234})$ when $s(45)=1$, otherwise 0. This has the effect of shifting the orientation of $\braket{01234}$ by $s(45)$ relative to the other 4-simplices. See~\cite{tata2021anomalies} for detailed explanations.

By the Pachner move, the 2-simplex $\langle 024\rangle$ is replaced by the 2-simplex $\langle135\rangle$ and the 3-simplices $\langle0124\rangle,\langle0234\rangle,\langle0245\rangle$ are replaced by the 3-simplices $\langle0135\rangle,\langle1235\rangle,\langle1345\rangle$. Our aim is to show that
\begin{widetext}
\begin{align}
\sum_{024,0124,0234,0245}\frac{d_{024}}{d_{0124}d_{0234}d_{0245}}&(Z^+(01234))^{s(45)}Z^+(01245)Z^+(02345) \nonumber \\
&= \sum_{135,0135,1235,1345}\frac{d_{135}}{d_{0135}d_{1235}d_{1345}} Z^+(01235)Z^+(01345)Z^+(12345) \label{eqn:Pachner33}
\end{align}
\end{widetext}
with fixed labels on all other simplices involved.

We contract all of the symmetry defects in the diagrams and examine the symmetry fractionalization factors that arise. These are given by the symmetry fractionalization factors of $U$ and $\eta$ for the symmetry $H$, and the U(1) fractional charges $q$ of anyons. 
Firstly, let us contract the $H$ symmetry factors involving $\eta$ and $U$, leaving the U(1) symmetry factors in the diagram.
On the left-hand-side (LHS) of Eq.~\eqref{eqn:Pachner33}, the product of the symmetry factors involving $\eta$ and $U$ are evaluated as
\begin{widetext}
\begin{align}
\begin{split}
\text{Symm}_{\text{LHS}}^{H} &= \left(\frac{U_{34}(013,123,0123)^{s(34)}}{U_{34}(023,{}^{32}012,0123)^{s(34)}\eta_{012}(23,34)^{s(24)}}\right)^{s(45)} \\
& \times \frac{U_{45}(014,124,0124)^{s(45)}}{U_{45}(024,{}^{42}012,0124)^{s(45)}\eta_{012}(24,45)^{s(25)}} \\
&\times \frac{U_{45}(024,234,0234)^{s(45)}}{U_{45}(034,{}^{43}023,0234)^{s(45)}\eta_{023}(34,45)^{s(35)}},
\label{eqn:PachnerLHS}
\end{split}
\end{align}
\end{widetext}
where the exponent $s(45)$ on the first row corresponds to the the orientation of $\langle 01234\rangle$. 
Meanwhile, the corresponding object on the right-hand-side (RHS) is given by

\begin{widetext}
\begin{align}
\begin{split}
\text{Symm}_{\text{RHS}}^{H} &= \frac{U_{35}(013,123,0123)^{s(35)}}{U_{35}(023,{}^{32}012,0123)^{s(35)}\eta_{012}(23,35)^{s(25)}} \\
& \times \frac{U_{45}(014,134,0134)^{s(45)}}{U_{45}(034,{}^{42}013,0134)^{s(45)}\eta_{012}(34,45)^{s(35)}} \\
&\times \frac{U_{45}(124,234,1234)^{s(45)}}{U_{45}(134,{}^{43}123,0234)^{s(45)}\eta_{123}(34,45)^{s(35)}}.
\label{eqn:PachnerRHS}
\end{split}
\end{align}
\end{widetext}

The strategy is that after contracting all the symmetry factors, the evaluation of the Pachner move reduces to the known case for the 15j symbols without symmetry defects~\cite{crane1993}, which is expressed as

\begin{figure}[htb]
\includegraphics{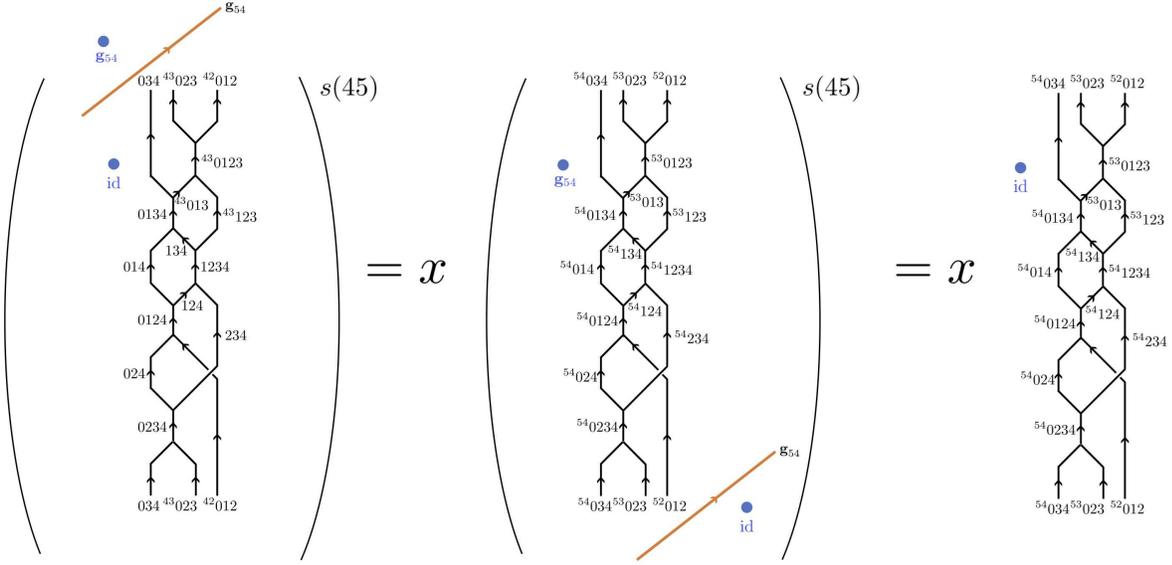}
\caption{Sweeping a ${\bf g}_{54}$ domain wall through the diagram for the 4-simplex $\langle01234\rangle$. This is a graphical calculus version of replacing all of the $F$- and $R$-symbols in the evaluation of the diagram in Eq.~\eqref{eqn:Zplusdef} with their ${\bf g}_{54}$-transformed versions using consistency conditions. 
Although not indicated in this figure, we also have the U(1) gauge field $(h,c)$ in the diagram, which is possibly conjugated by the action of $\sigma(\mathbf{g}_{54})$.
A trace is implied, i.e. open lines at the top are implied to be connected to their corresponding lines at the bottom. }
\label{fig:Pachner_sweep}
\end{figure}

\begin{widetext}
\begin{align}
\sum_{024,0124,0234,0245}\frac{d_{024}}{d_{0124}d_{0234}d_{0245}}&Z_0^+(01234)Z_0^+(01245)Z_0^+(02345) \nonumber \\
&= \sum_{135,0135,1235,1345}\frac{d_{135}}{d_{0135}d_{1235}d_{1345}} Z_0^+(01235)Z_0^+(01345)Z_0^+(12345) \label{eqn:PachnerZ0}
\end{align}
\end{widetext}
where $Z_0^+$ is the 15j symbol without symmetry factors given in Eq.~\eqref{eqn:Zplusdef}. However, following~\cite{bulmash2020}, there are two problems. First is that the diagram on $\langle01234\rangle$ is twisted by $s(45)$ relative to the above Eq.~\eqref{eqn:PachnerZ0}. Second, the anyon lines in Eq.~\eqref{eqn:PachnerZ0} on $\langle01234\rangle$ turns out to differ from the lines on the rest of the diagrams up to group multiplication by $\mathbf{g}_{45}$. To fix these issues, we ``sweep'' a $\mathbf{g}_{45}$ symmetry defect across the entire $\langle01234\rangle$ diagram $Z^+(01234)$, as schematically shown in Fig.~\ref{fig:Pachner_sweep}.

Sweeping $\mathbf{g}_{45}$ symmetry defect across the $\langle01234\rangle$ diagram has three effects. Firstly, it gives the extra factor
\begin{widetext}
\begin{align}
x= \frac{U_{54}(^{54}{024},^{54}{234},^{54}{0234})}{U_{54}(^{54}{034},^{53}{023},^{54}{0234})} &\times 
\frac{U_{54}(^{54}{014},^{54}{124},^{54}{0124})}{U_{54}(^{54}{024},^{53}{012},^{54}{0124})} \times 
\frac{U_{54}(^{53}{023},^{52}{012},^{53}{0123})}{U_{54}(^{53}{013},^{53}{123},^{53}{0123})} \times \nonumber \\
& \times 
\frac{U_{54}(^{54}{034},^{53}{013},^{54}{0134})}{U_{54}(^{54}{014},^{53}{134},^{54}{0134})} \times 
\frac{U_{54}(^{54}{134},^{53}{123},^{54}{1234})}{U_{54}(^{54}{124},^{54}{234},^{54}{1234})}
\end{align}
\end{widetext}
that will multiply Symm$_{\mathrm{LHS}}^H$. Then, the $\mathbf{g}_{45}$ sweep changes the orientation of the swept 4-simplex $\langle01234\rangle$ by $s(45)$.
Finally, since $\mathbf{g}_{45}$ can act by the charge conjugation $\sigma(\mathbf{g}_{45})$ on the U(1) gauge field, it transforms the U(1) gauge field in the $Z^+(01234)$ as $(h,c)\to(\sigma(\mathbf{g}_{54})h,\sigma(\mathbf{g}_{54})c)$. 

It is shown in~\cite{bulmash2020} that the symmetry factor involving $\eta$ and $U$ are invariant under the Pachner move, i.e., $\frac{\text{Symm}_{\text{RHS}}^{H}\cdot x}{\text{Symm}_{\text{RHS}}^{H} }=1$, by the repeated use of consistency conditions between $\eta$ and $U$. 

After the $\mathbf{g}_{54}$ sweep described above, the U(1) symmetry factors  before and after the Pachner move are given by
\begin{widetext}
\begin{align}
\begin{split}
\text{Symm}_{\text{LHS}}^{\U} &=e^{-2\pi i^{54}(q\cup_{s\sigma}(c+\delta_{\sigma}h)+ \delta_{s\sigma} q\cup_{s\sigma}h)_{01234}}e^{-2\pi i(q\cup_{s\sigma}(c+\delta_{\sigma}h)+ \delta_{s\sigma} q\cup_{s\sigma}h)_{01245}}e^{-2\pi i(q\cup_{s\sigma}(c+\delta_{\sigma}h)+ \delta_{s\sigma} q\cup_{s\sigma}h)_{02345}} \\
\text{Symm}_{\text{RHS}}^{\U} &=e^{-2\pi i(q\cup_{s\sigma}(c+\delta_{\sigma}h)+ \delta_{s\sigma} q\cup_{s\sigma}h)_{01235}}e^{-2\pi i(q\cup_{s\sigma}(c+\delta_{\sigma}h)+ \delta_{s\sigma} q\cup_{s\sigma}h)_{01345}}e^{-2\pi i(q\cup_{s\sigma}(c+\delta_{\sigma}h)+ \delta_{s\sigma} q\cup_{s\sigma}h)_{12345}}
\label{eqn:PachnerU(1)}
\end{split}
\end{align}
\end{widetext}
Then, these symmetry factors are compared as
\begin{align}
\begin{split}
    \frac{\text{Symm}_{\text{LHS}}^{\U}}{\text{Symm}_{\text{RHS}}^{\U}}&=e^{2\pi i(\delta_{s\sigma,\sigma}(q\cup_{s\sigma}(c+\delta_{\sigma}h)+ \delta_{s\sigma}q\cup_{s\sigma}h))} \\
    &= e^{2\pi i (\delta_{s\sigma} q\cup_{s\sigma}(c+\delta_{\sigma}h)-\delta_{s\sigma} q\cup_{s\sigma}\delta_{\sigma}h)} = 1,
    \end{split}
\end{align}
where $\delta_{s\sigma,\sigma}$ is a twisted coboundary operation based on the $H$ action that acts on $q$ as $s\sigma$, and acts on $(h,c)$ as $\sigma$.
Hence, all the symmetry factors are identical before and after the Pachner move, and the evaluation reduces to the known case for the 15j symbols without symmetry defects~\cite{crane1993, bulmash2020}. We have thus shown that the path integral is invariant under the 3-3 Pachner move. 

\subsection{Fermionic case}
One can show the effect of the 3-3 Pachner move on the path integral $Z_b(M^4,A_{H},(h,c),f)$ for the bosonic shadow of the fermionic SPT phase.
The bosonic shadow theory is expected to have an 't Hooft anomaly characterized by a (4+1)D response action
\begin{align}
    S_{5,b}=(-1)^{\int \Sq^2(f)+(2c+A_{H}^*\omega_{H})\cup f}
    \label{eq:anomalyshadow}
\end{align}
Let us check this anomaly by computing the Pachner move. We will show that
\begin{widetext}
\begin{align}
\sum_{024,0124,0234,0245}&\frac{d_{024}}{d_{0124}d_{0234}d_{0245}}(Z_b^+(01234))^{s(45)}Z_b^+(01245)Z_b^+(02345) \nonumber \\
&= \sum_{135,0135,1235,1345}\frac{d_{135}}{d_{0135}d_{1235}d_{1345}} Z_b^+(01235)Z_b^+(01345)Z_b^+(12345)(-1)^{(\Sq^2(f)+2c\cup f+f\cup A_{H}^*\omega_{H})_{012345}} \label{eqn:Pachner33fermion}
\end{align}
\end{widetext}
This can be checked by the same logic as the bosonic case in the previous subsection, following the steps:
\begin{enumerate}
    \item Sweep the $\mathbf{g}_{54}$ symmetry defect over the $Z_b^+(01234)$ diagram. This sweep has the effect of changing the orientation on $\langle 01234\rangle$ by $s(34)$, acting on U(1) gauge fields by the charge conjugation $(h,c)\to (^{54}h,^{54}c)$, and emitting the symmetry factor
    \begin{equation}
\begin{split}
x = &\frac{U_{54}({^{54}}014,{^{54}}124;{^{54}}0124 \times f_{0124})}
{U_{54}({^{54}}024,{^{52}}012;{^{54}}0124) U_{54}(f_{0124},{^{54}}0124 \times f_{0124};{^{54}}0124)} \\
\times &\frac{U_{54}({^{54}}024,{^{54}}234;{^{54}}0234 \times f_{0234}) U_{54}(f_{0234},{^{54}}0234 \times f_{0234};{^{54}}0234)}{U_{54}({^{54}}034,{^{53}}023;{^{54}}0234)} \\
\times &\frac{U_{54}({^{53}}023,{^{52}}012;{^{53}}0123) U_{54}(f_{0123},{^{53}}0123; {^{53}}0123 \times f_{0123})}{U_{54}({^{53}}013,{^{53}}123;{^{53}}0123 \times f_{0123})} \\
\times &\frac{U_{54}({^{54}}034,{^{53}}013;{^{54}}0134) U_{54}(f_{0134},{^{54}}0134; {^{54}}0134 \times f_{0134})}{U_{54}({^{53}}014,{^{54}}134;{^{54}}0134 \times f_{0134})} \\
\times &\frac{U_{54}({^{54}}134,{^{53}}123;{^{54}}1234) U_{54}(f_{1234},{^{54}}1234; {^{54}}1234 \times f_{1234})}{U_{54}({^{54}}124,{^{54}}234;{^{54}}1234 \times f_{1234})},
\end{split}
\end{equation}
that will multiply the LHS of Eq.~\eqref{eqn:Pachner33fermion}.

\item Contract the symmetry defects of the symmetry $H$ in Eq.~\eqref{eqn:Pachner33fermion}. With the above extra symmetry factor $x$ counted, the symmetry factor from the LHS evaluates as
\begin{align}
\begin{split}
x\cdot \text{Symm}_{\text{LHS}}^{H} = x\cdot&\bigg(\frac{1}{\eta_{012}(23,34)^{s(24)}} \Big( \frac{U_{34}(013,123 ; 0123 \times f_{0123})}{U_{34}(023, {^{32}}012 ; 0123 ) U_{34}(f_{0123},0123 ; 0123 \times f_{0123})} \Big)^{s(34)} \bigg)^{s(45)} \\
\times &\frac{1}{\eta_{012}(24,45)^{s(25)}} \Big( \frac{U_{45}(014,124 ; 0124 \times f_{0124})}{U_{45}(024, {^{42}}012 ; 0124 ) U_{45}(f_{0124},0124 ; 0124 \times f_{0124})} \Big)^{s(45)} \\
\times &\frac{1}{\eta_{023}(34,45)^{s(35)}} \Big( \frac{U_{45}(024,234 ; 0234 \times f_{0234})}{U_{45}(034, {^{43}}023 ; 0234 ) U_{45}(f_{0234},0234 ; 0234 \times f_{0234})} \Big)^{s(45)},
\end{split}
\end{align}
while the corresponding object for the RHS is given by
\begin{equation}
\begin{split}
\text{symm}^{H}_\text{RHS} = &\frac{1}{\eta_{012}(23,35)^{s(25)}} \Big( \frac{U_{35}(013,123 ; 0123 \times f_{0123})}{U_{35}(023, {^{32}}012 ; 0123 ) U_{35}(f_{0123},0123 ; 0123 \times f_{0123})} \Big)^{s(35)} \\
\times &\frac{1}{\eta_{013}(34,45)^{s(35)}} \Big( \frac{U_{45}(014,134 ; 0134 \times f_{0134})}{U_{45}(034, {^{43}}013 ; 0134 ) U_{45}(f_{0134},0134 ; 0134 \times f_{0134})} \Big)^{s(45)} \\
\times &\frac{1}{\eta_{123}(34,45)^{s(35)}} \Big( \frac{U_{45}(124,234 ; 1234 \times f_{1234})}{U_{45}(134, {^{43}}123 ; 1234 ) U_{45}(f_{1234},1234 ; 1234 \times f_{1234})} \Big)^{s(45)}.
\end{split}
\end{equation}
In~\cite{tata2021anomalies}, we show that
\begin{align}
    \frac{\text{symm}^{H}_\text{LHS} \cdot x}{\text{symm}^{H}_\text{RHS}} = \eta_{f_{0123}}({\bf g}_{34},{\bf g}_{45})=(-1)^{(f\cup A_{H}^*\omega_{H})_{012345}}.
\end{align}
\item Contract the symmetry defects of the U(1) symmetry in Eq.~\eqref{eqn:Pachner33fermion}. The symmetry factor from the LHS evaluates as

\begin{widetext}
\begin{align}
\begin{split}
\text{Symm}_{\text{LHS}}^{\U} =&e^{-2\pi i^{54}(q\cup_{s\sigma}(c+\delta_{\sigma}h)+c\cup_1 f+ \delta_{s\sigma} q\cup_{s\sigma}h)_{01234}}\\
&e^{-2\pi i(q\cup_{s\sigma}(c+\delta_{\sigma}h)+c\cup_1f+ \delta_{s\sigma} q\cup_{s\sigma}h)_{01245}} \\
&e^{-2\pi i(q\cup_{s\sigma}(c+\delta_{\sigma}h)+c\cup_1 f+\omega+ \delta_{s\sigma} q\cup_{s\sigma}h)_{02345}} 
\label{eqn:PachnerU(1)fermionLHS}
\end{split}
\end{align}
\end{widetext}

\begin{widetext}
\begin{align}
\begin{split}
\text{Symm}_{\text{RHS}}^{\U} =&e^{-2\pi i(q\cup_{s\sigma}(c+\delta_{\sigma}h)+c\cup_1 f+ \delta_{s\sigma} q\cup_{s\sigma}h)_{01235}}\\
&e^{-2\pi i(q\cup_{s\sigma}(c+\delta_{\sigma}h)+c\cup_1 f+ \delta_{s\sigma} q\cup_{s\sigma}h)_{01345}}\\
&e^{-2\pi i(q\cup_{s\sigma}(c+\delta_{\sigma}h)+c\cup_1 f+ \delta_{s\sigma} q\cup_{s\sigma}h)_{12345}}
\label{eqn:PachnerU(1)fermionRHS}
\end{split}
\end{align}
\end{widetext}
Then, these symmetry factors are compared as
\begin{align}
\begin{split}
    \frac{\text{Symm}_{\text{LHS}}^{\U}}{\text{Symm}_{\text{RHS}}^{\U}}&=e^{2\pi i(\delta_{s\sigma,\sigma}(q\cup_{s\sigma}(c+\delta_{\sigma}h)+c\cup_1 f+ \delta_{s\sigma} q\cup_{s\sigma}h))} \\
    &= e^{2\pi i (\delta_{s\sigma} q\cup_{s\sigma}(c+\delta_{\sigma}h)+\delta(c\cup_1 f)-\delta_{s\sigma} q\cup_{s\sigma}\delta_{\sigma} h)} \\
    &= e^{2\pi i (\delta_{s\sigma} q\cup_{s\sigma}c+c\cup  f+f\cup c)} \\
    &= e^{2\pi i( c\cup f)_{012345}},
    \end{split}
\end{align}
where we used $\delta_{s\sigma} q=f$ mod 2 in the last row.
The mod 2 relation $\delta_{s\sigma} q=f$ corresponds to the charge conservation on a 3-simplex in the 15j symbol, which is required because $q_{\psi^{f_{0123}}}=f_{0123}$ mod 2 on each 3-simplex $\braket{0123}$.

\item Now we have contracted all the symmetry factors for 0-form symmetries $H$ and U(1). We finally evaluate the effect of the Pachner move for the 15j symbol without symmetry factors $Z_{0,b}^+$ given in Eq.~\eqref{eq:15jfermion}.
It is shown in~\cite{tata2021anomalies} that
\begin{widetext}
\begin{align}
\sum_{024,0124,0234,0245}&\frac{d_{024}}{d_{0124}d_{0234}d_{0245}}(Z_{0,b}^+(01234))Z_{0,b}^+(01245)Z_{0,b}^+(02345) \nonumber \\
&= \sum_{135,0135,1235,1345}\frac{d_{135}}{d_{0135}d_{1235}d_{1345}} Z_{0,b}^+(01235)Z_{0,b}^+(01345)Z_{0,b}^+(12345)\cdot (-1)^{\Sq^2(f)_{012345}}.
\label{eqn:Pachner33Z0b}
\end{align}
\end{widetext}
By combining the ratio between the LHS and RHS on each step, we obtain the 't Hooft anomaly shown in Eq.~\eqref{eq:anomalyshadow}.

\end{enumerate}

\bibliography{TI}
        
\end{document}